\documentclass[
reprint,
 amsmath,amssymb,
 aps,
 prc,
floatfix,
]{revtex4-2}
\usepackage{comment}
\usepackage{graphicx}
\usepackage{dcolumn}
\usepackage{bm}
\usepackage{color}
\usepackage{xcolor}
\usepackage{mathtools}
\usepackage{url}        
\usepackage{hyperref}
\usepackage{lineno}
\usepackage{amsmath}

\begin{document}
\preprint{APS/123-QED}
\title{When positive and negative pairs differ in femtoscopy: residual Coulomb and isospin effects}

\author{Yevheniia Khyzhniak}
\author{Michael Annan Lisa}
\affiliation{%
 Physics Department, The Ohio State University
}%

\date{\today}

\begin{abstract}
We study charge-dependent modifications of identical-pion and identical-kaon
femtoscopic correlation functions from two sources: the residual Coulomb field
of the charged source and isospin-related hadronic dynamics. The residual
Coulomb effect is modeled with a modified Retière--Lisa blast-wave source,
where the same emitted particles are propagated with positive and negative
charge signs through an effective residual field.
The residual Coulomb field produces a small but systematic positive-to-negative
splitting of the correlation functions, strongest at low $k_T$ and sensitive to the effective charge, spatial
width, and expansion velocity of the residual source. It also modifies the
height and shape of the correlation function, complicating interpretation of the fitted radii.
UrQMD 3.4 cascade calculations for Au+Au
collisions at $\sqrt{s_{NN}}=7.7~\mathrm{GeV}$ show that charge-dependent
splittings can also appear even without a residual Coulomb field.
For pions, this difference is mainly driven by the initial isospin composition,
while for kaons it is strongly affected by different $K^+$ and $K^-$
production mechanisms and subsequent hadronic evolution. These results show that residual Coulomb and isospin-related effects
can compete, and neither can be interpreted reliably without constraining the
other within the same model framework.
\end{abstract}

\maketitle


\section{Introduction}

Particle femtoscopy has been one of the standard tools for studying the
space-time structure of particle production for several decades~\cite{Lisa:2005dd, Heinz:1999rw}. Its
main idea is simple: particles emitted close to each other in space and time can
remain correlated at small relative momentum~\cite{Kopylov:1972qw, Pratt:1986cc, Bertsch:1988db}. By measuring this
correlation, one can learn about the size and dynamics of the region from which
the particles were emitted~\cite{Lisa:2005dd, Heinz:1999rw, Akkelin:1995gh}.

The same correlation-function technique can be used to address different physics
questions depending on the particles in the pair.
For identical particles, such as two pions or two kaons, the correlation is driven
mainly by quantum-statistical symmetrization of the two-particle wave
function~\cite{Pratt:1986cc,Bertsch:1988db,Lisa:2005dd}.
Identical-particle femtoscopy is therefore useful for studying the size, shape,
emission duration, and collective expansion of the source at kinetic
freeze-out~\cite{Retiere:2003kf,Akkelin:1995gh,Lisa:2005dd}.

For non-identical particles, the story is different. Since the particles are not
identical, there is no quantum-statistical symmetrization of the pair wave
function. Instead, the correlation at small relative momentum is mainly shaped
by final-state interactions, such as Coulomb and strong interactions between
the two particles~\cite{Lednicky:1981su, Lednicky:2005tb}.
These correlations can be used to study whether two particle species are emitted
from the same region of the source, or whether one species
is emitted earlier, later, or from a shifted position relative to the
other~\cite{Voloshin:1997jh,Lednicky:1995vk}.
In this way, non-identical-particle femtoscopy gives access to relative
space-time emission asymmetries~\cite{Lednicky:2005tb,STAR:2003cqe}.

Femtoscopy can also be used as a complementary tool to study hadron--hadron interactions themselves~\cite{Koonin:1977fh, ALICE:2020mfd, Fabbietti:2020bfg, STAR:2015kha}. This is especially important when the interaction is not well known experimentally~\cite{ALICE:2020mfd, Kamiya:2019uiw}, or when one wants to search for near-threshold structures such as weakly bound states~\cite{STAR:2014dcy, STAR:2018uho, ALICE:2019eol}, near-threshold poles~\cite{Kamiya:2019uiw, ALICE:2021cpv}, or resonances~\cite{STAR:2006ykx, ALICE:2021cpv}. However, such signals can be small and can be modified by several competing
effects~\cite{Mihaylov:2018rva,Kisiel:2014mma,ALICE:2018ysd}.
It is therefore important to understand which physical effects can distort the
measured correlation function before interpreting a structure as evidence for a
new state or for a particular interaction scenario.

\begin{figure}
    \centering
    \includegraphics[width=0.5\linewidth]{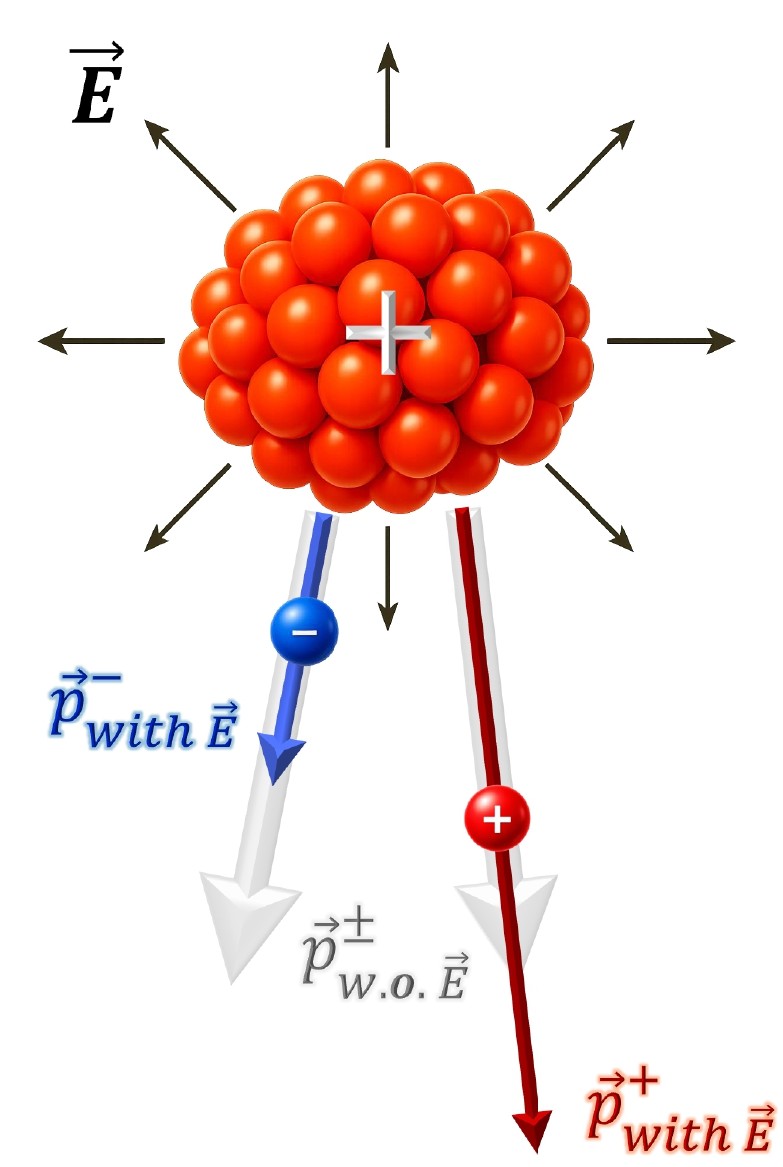}
    \caption{
Schematic illustration of the influence of the residual Coulomb field on emitted charged particles. 
The positively charged residual source, represented by the cluster of red spheres, generates an outward radial Coulomb electric field, $\vec{E}$, indicated by the black arrows. 
The gray semi-transparent vectors show the momenta that positively and negatively charged particles would have in the absence of the residual Coulomb interaction, $\vec{p}^{\,\pm}_{\mathrm{w.o.}\,\vec{E}}$. 
Under the influence of the positively charged residual source, negatively charged particles are attracted toward the source, leading to a reduction of their final momentum, shown by the blue vector $\vec{p}^{\,-}_{\mathrm{with}\,\vec{E}}$. 
In contrast, positively charged particles are repelled by the source, resulting in an increased final momentum, shown by the red vector $\vec{p}^{\,+}_{\mathrm{with}\,\vec{E}}$.
}
    \label{fig:residual_coulomb_schematic}
\end{figure}

Since femtoscopy has been used for many years, one might expect that all relevant effects have already been identified and fully accounted for. In practice, however, some ingredients of the correlation analysis, such as the source function, final-state  (including Coulomb) interactions, residual contributions, 
are still model dependent or treated approximately~\cite{Mihaylov:2018rva, Fabbietti:2020bfg, Maj:2009ue}. For charged particles, Coulomb interactions are not just a technical effect, but an integral component of the observed femtoscopic signal. In most femtoscopic analyses, the
Coulomb interaction is treated as a two-body interaction between the particles
in the measured pair~\cite{Bowler:1991vx, Sinyukov:1998fc}. This 
is essential, especially at small
relative momentum.

There is, however, another possible Coulomb effect that is less commonly considered. After particles are emitted, they may still move through the electric field produced by the remaining charged matter in the collision system~\cite{Barz:1996gr, Barz:1997es, Shoppa:1998sw, HADES:2019lek}. This residual Coulomb field can change the final momenta of the
emitted particles before they reach the detector. Since the remaining
source carries a positive net charge, the effect is charge dependent:
positively charged particles are pushed away from the residual source, while
negatively charged particles are pulled toward it. A schematic representation of this effect is shown in
Fig.~\ref{fig:residual_coulomb_schematic}. Even if this
momentum change is small for each individual particle, it can matter for
femtoscopy, because femtoscopic correlations are measured precisely at small
relative momentum.

This effect can be viewed as a third-body Coulomb effect, where the third body
is the residual charged source rather than the other particle in the measured
pair. Previous studies suggested that such a field can lead to different
apparent HBT radii for positively and negatively charged particles, and such
differences have also been investigated experimentally
~\cite{Barz:1996gr,Barz:1997es,Shoppa:1998sw,HADES:2019lek}. However, an
important point is that a difference in the fitted radii does not necessarily
mean that positive and negative particles were emitted from different
homogeneity regions. In femtoscopy, HBT radii are usually interpreted as
homogeneity lengths of the source, i.e. the regions that emit particles with
similar momenta~\cite{Akkelin:1995gh,Lisa:2005dd,Heinz:1999rw}. A residual
Coulomb field can change this interpretation, because it modifies the particle
momenta after emission and therefore distorts the measured correlation function
itself. In that case, the radius obtained from a Gaussian fit is an effective
fit parameter, not a direct measure of the original spatial homogeneity length.
For this reason, the present work focuses first on the distortions of the
correlation functions themselves, and only then on how these distortions are
reflected in the fitted femtoscopic parameters.

Another effect that can influence charge-dependent correlations is isospin.
In neutron-rich collisions, a larger neutron excess in the dense collision zone
tends to enhance $\pi^{-}$ production relative to $\pi^{+}$ production, making
the $\pi^{-}/\pi^{+}$ ratio a useful, although model-dependent, probe of the
high-density neutron-to-proton ratio and the nuclear symmetry energy
~\cite{FOPI:2006ifg,SpiRIT:2021gtq,E-0895:2003oas}.
Isospin may also affect the momentum dependence of pion emission: high-transverse-momentum
pions are expected to be more sensitive to the dense stage of the collision, so
the $\pi^{-}/\pi^{+}$ ratio can depend on pion momentum as well as on the
space-time region from which the pions are emitted~\cite{SpiRIT:2021gtq,Li:2008gp}.

In this paper, we study how identical-pion and identical-kaon correlation
functions can be modified by two effects that are not always included explicitly
in standard femtoscopic treatments: the Coulomb field of the residual charged
source and isospin-related contributions. To do this, we use a modified
blast-wave model and UrQMD to construct controlled samples of emitted particles
and to compare the resulting correlation functions with and without these
additional effects.

\section{Model for residual Coulomb effects}

\subsection{Blast-wave freeze-out source}

To study the effect of a residual Coulomb field, one first needs a realistic
freeze-out distribution for the emitted particles. In a full heavy-ion
collision, this distribution is shaped by many ingredients: the initial
geometry, pressure gradients, collective expansion, resonance decays, hadronic
rescattering, and the finite duration of particle emission~\cite{Lisa:2005dd, Retiere:2003kf}. For the
present study, however, the goal is not to reproduce every detail of the
collision dynamics. Instead, we need a source that contains the main features
relevant for femtoscopy, while still remaining simple enough that the effect of
the residual Coulomb field can be isolated and understood.

The blast-wave model provides such a source. It combines two essential pieces of
freeze-out physics: a finite spatial distribution of emitted particles and a
collective velocity field~\cite{Schnedermann:1993ws, Retiere:2003kf}. The finite geometry determines the
approximate size and shape of the emitting region, while the flow field
introduces correlations between position and momentum. These position--momentum
correlations are especially important in femtoscopy, because the measured radii
correspond to homogeneity lengths, i.e. to the regions of the source that emit
particles with similar momenta, rather than to the full geometric size of the
system~\cite{Akkelin:1995gh}. A source model without collective flow would therefore miss one
of the central features of particle production in heavy-ion collisions.

Another useful feature of the blast-wave approach is that its parameters have a
clear physical meaning. The temperature controls the amount of thermal smearing,
the transverse flow parameters control the strength and anisotropy of the
collective expansion, the geometric radii set the transverse size of the source,
and the proper-time parameters describe when particles are emitted~\cite{Retiere:2003kf}. This
makes the model flexible enough to generate pion and kaon sources with realistic
space-time and momentum correlations, but transparent enough that changes in the
input parameters can be interpreted directly.

In this paper, we study how identical-pion and identical-kaon correlation
functions can be modified by two effects that are not always included explicitly
in standard femtoscopic treatments: the Coulomb field of the residual charged
source and isospin-related contributions. To do this, we use a modified
blast-wave model and UrQMD to construct controlled samples of emitted particles.
The main emphasis is placed on the charge-dependent distortion of the
correlation functions themselves; the fitted radii and $\lambda$ parameters are
then treated as effective quantities that summarize selected features of these
distortions.

\subsection{Retière--Lisa blast-wave parametrization}

In this work, the freeze-out source is generated using the Retière--Lisa
blast-wave (RLBW) parametrization~\cite{Retiere:2003kf}.
This version of the
blast-wave model is especially useful for femtoscopic studies because it was
constructed to describe not only single-particle spectra and elliptic flow, but
also the azimuthal dependence of femtoscopic radii. It therefore contains the
main ingredients needed here: a finite transverse geometry, collective
transverse expansion, possible elliptic deformation, and a finite emission
time.

A simple radially symmetric blast-wave source would already provide a useful
first approximation to an expanding system. However, heavy-ion collisions are
not always azimuthally symmetric. The initial overlap region can be spatially
anisotropic, and this anisotropy is converted by pressure gradients into an
anisotropic flow field~\cite{Ollitrault:1992bk, Voloshin:2008dg}. The RLBW parametrization allows these
effects to be included in a controlled way: the transverse source can have
different radii in the $x$ and $y$ directions, and the strength of the
transverse flow can also depend on the azimuthal angle.

The emission point is described by the transverse source coordinates
$(r,\phi_s)$, the space-time rapidity $\eta_s$, and the proper time $\tau$.
Here, $r$ denotes the transverse distance of the emission point from the center
of the source, while $\phi_s$ gives its azimuthal direction in the transverse
plane. The subscript $s$ indicates that $\phi_s$ and $\eta_s$ are source, or
space-time, coordinates rather than momentum-space variables. The coordinate
$\eta_s$ describes the longitudinal position of the emission point in
boost-invariant coordinates, while $\tau$ is the proper emission time. These
variables are converted to Cartesian space-time coordinates as
\begin{equation}
x = r\cos\phi_s,
\qquad
y = r\sin\phi_s,
\end{equation}
\begin{equation}
t = \tau \cosh\eta_s,
\qquad
z = \tau \sinh\eta_s .
\label{eq:coord}
\end{equation}
The source rapidity is sampled in a finite interval,
\begin{equation}
-\eta_{\max} < \eta_s < \eta_{\max}.
\end{equation}
Here $\eta_{\max}$ controls how far the source extends in the longitudinal
direction. The blast-wave picture used here assumes an approximately
boost-invariant central region, where the emission probability is taken to be
uniform in the space-time rapidity $\eta_s$. Such a region cannot be infinite
in a real collision, so $\eta_{\max}$ provides a finite cutoff. Physically, its
value is related to the collision energy and to the rapidity interval over
which the produced matter can be treated as approximately boost invariant.

The longitudinal coordinate specifies where along the beam direction the
particle is emitted. We also need to specify when the emission takes place. In
the blast-wave model this is done through the proper time $\tau$, which
describes the freeze-out time in the local boost-invariant coordinates. The
proper time is sampled from a Gaussian distribution centered at $\tau_0$ with
width $\Delta \tau$,
\begin{equation}
P(\tau)
\propto
\exp
\left[
-\frac{(\tau-\tau_0)^2}{2(\Delta\tau)^2}
\right],
\qquad
\tau > 0 .
\end{equation}
The parameter $\tau_0$ sets the average freeze-out proper time, while
$\Delta\tau$ controls the duration of particle emission. The condition
$\tau>0$ is imposed to avoid unphysical emission times.

After specifying the longitudinal position and the emission time, we now turn
to the transverse structure of the source. The transverse geometry is described
by an ellipse with characteristic radii $R_x$ and $R_y$. For a given source
azimuthal angle $\phi_s$, the distance from the origin to the edge of the
source is
\begin{equation}
R(\phi_s)
=
\frac{R_x R_y}
{
\sqrt{
R_y^2\cos^2\phi_s
+
R_x^2\sin^2\phi_s
}
}.
\end{equation}
This expression follows directly from the equation of an ellipse. For an
ellipse aligned with the coordinate axes,
\begin{equation}
\frac{x^2}{R_x^2}
+
\frac{y^2}{R_y^2}
=
1 ,
\end{equation}
where $R_x$ and $R_y$ are the semi-axes in the $x$ and $y$ directions,
respectively. A point on the boundary at azimuthal angle $\phi_s$ can be
written as
\begin{equation}
x = R(\phi_s)\cos\phi_s,
\qquad
y = R(\phi_s)\sin\phi_s .
\end{equation}
Substituting these expressions into the ellipse equation gives
\begin{equation}
R^2(\phi_s)
\left(
\frac{\cos^2\phi_s}{R_x^2}
+
\frac{\sin^2\phi_s}{R_y^2}
\right)
=
1 ,
\end{equation}
which leads to the expression above. Thus, $R(\phi_s)$ represents the distance
from the center of the source to the edge of the ellipse in the direction
$\phi_s$.

It is useful to define the dimensionless elliptical radius
\begin{equation}
\tilde r
=
\frac{r}{R(\phi_s)} .
\end{equation}
Points with $\tilde r < 1$ lie inside the ellipse, while $\tilde r = 1$
corresponds to the transverse boundary. Equivalently, in Cartesian coordinates,
\begin{equation}
\tilde r
=
\sqrt{
\frac{x^2}{R_x^2}
+
\frac{y^2}{R_y^2}
}.
\end{equation}
The default transverse radii are of order $10~\mathrm{fm}$, as listed in
Table~\ref{tab:default_parameters}. These values should be understood as
geometric freeze-out parameters of the blast-wave source, not as directly
measured femtoscopic radii. The measured HBT radii correspond to homogeneity
lengths and can be smaller than the full geometric size of the emitting system
because of collective flow~\cite{Akkelin:1995gh}. Since the initial nuclear radius of a heavy
nucleus is already of order $5$--$7~\mathrm{fm}$~\cite{DeVries:1987atn,Hofstadter:1956qs} and the system expands
before kinetic freeze-out, a transverse freeze-out scale of order
$10$--$13~\mathrm{fm}$~\cite{Retiere:2003kf, Batyuk:2017smw} is a reasonable baseline for a central or
semi-central heavy-ion collision.

The difference between $R_x$ and $R_y$ introduces a moderate elliptic
deformation of the source. For the default values, the geometric anisotropy is
of order
\begin{equation}
\frac{R_y^2-R_x^2}{R_y^2+R_x^2}
\approx -0.14 ,
\end{equation}
where the sign depends on the convention used for the $x$ and $y$ axes.
Such an anisotropic freeze-out geometry is motivated by azimuthally sensitive
femtoscopy measurements, which show that the pion-emitting source can retain a
finite final-state eccentricity at kinetic freeze-out~\cite{STAR:2026skb, STAR:2014shf}.
Thus, the default source is extended and mildly anisotropic, providing a
realistic baseline for studying how the residual Coulomb field modifies the
final correlation functions.

In the Monte Carlo implementation, the transverse coordinate is first proposed
using this elliptical polar parametrization. The source azimuthal angle
$\phi_s$ is sampled uniformly in the interval $[0,2\pi)$, and the
dimensionless radial coordinate is sampled as
\begin{equation}
\tilde r = \sqrt{U},
\end{equation}
where $U$ is uniformly distributed between 0 and 1. The square root accounts for
the radial phase-space factor in the transverse-coordinate proposal. The
physical transverse radius is then obtained from
\begin{equation}
r = \tilde r R(\phi_s).
\end{equation}
This step defines only the coordinate-space proposal distribution. The proposed emission point is later combined with a proposed momentum, and the full phase-space point $(x,p)$ is accepted or rejected using the sampling weight defined below and in the Monte Carlo implementation section.

The transverse density profile is controlled by a surface function $\Omega(\tilde r)$, which determines how sharply the source falls near the elliptical boundary. In the RLBW parametrization this profile is written in a Fermi-function form, %
\begin{equation} 
\Omega(\tilde r) = \frac{1} { 1+\exp\left(\frac{\tilde r-1}{a_s}\right) }, \qquad a_s > 0 . 
\end{equation}
The parameter $a_s$ controls the diffuseness of the surface. For small $\tilde r$, the argument of the exponential is negative and $\Omega(\tilde r)$ is close to one, so the density factor is nearly constant well inside the source. Near $\tilde r=1$, the function decreases smoothly. In the formal RLBW parametrization, values $\tilde r>1$ describe the diffuse tail outside the nominal boundary and are suppressed by $\Omega(\tilde r)$. 

In the present implementation, the default calculation uses the hard-edge case, $a_s=0$. Since the Fermi-function form above is defined for $a_s>0$, the hard-edge case is implemented separately as
\begin{equation}
\Omega(\tilde r) =
\begin{cases}
1, & \tilde r \leq 1, \\
0, & \tilde r > 1 .
\end{cases}
\end{equation}
Thus, in the default setup, emission points are restricted to the interior of the ellipse. When $a_s$ is varied, the variation should be interpreted as a test of sensitivity to smoothing the density near the source boundary within the implemented proposal region, rather than as a full sampling of an extended tail outside the ellipse. This parameter is later varied to study how the surface profile affects the residual-Coulomb distortion.

The collective transverse expansion is described by a flow rapidity
$\rho(\tilde r,\phi_s)$. In the RLBW parametrization, the transverse
boost direction is not necessarily the same as the radial direction
$\phi_s$. For an elliptically deformed source, the boost direction is taken to
be normal to the surface of constant density. This introduces
position--momentum correlations that depend on both the radial position and the
azimuthal angle. The boost angle is
\begin{equation}
\phi_b
=
\operatorname{atan2}
\left(
R_x^2 \sin\phi_s,
R_y^2 \cos\phi_s
\right).
\end{equation}
The transverse flow rapidity is then written as
\begin{equation}
\rho(\tilde r,\phi_s)
=
\tilde r^{\,n}
\left[
\rho_0
+
\rho_2
\cos\left(2\phi_b\right)
\right].
\end{equation}
The parameter $\rho_0$ sets the average strength of the transverse flow, while
$\rho_2$ controls the second-harmonic modulation of the flow field. The
exponent $n$ controls how quickly the flow grows with radius. The standard
linear RLBW choice corresponds to $n=1$, which is used as the default
setting in this study.

With this definition, the local fluid four-velocity is
\begin{equation}
u^\mu
=
\left(
\cosh\eta_s \cosh\rho,\,
\cos\phi_b \sinh\rho,\,
\sin\phi_b \sinh\rho,\,
\sinh\eta_s \cosh\rho
\right).
\end{equation}
Using the metric convention $(+,-,-,-)$, this four-velocity satisfies
\begin{equation}
u^\mu u_\mu = 1 .
\end{equation}

Having specified the space-time structure and the collective flow field of the
source, we now describe how particle momenta are sampled from this freeze-out
configuration.

For a particle with four-momentum
\begin{equation}
p^\mu = (E,p_x,p_y,p_z),
\end{equation}
the thermal part of the emission probability depends on the scalar product
\begin{equation}
p_\mu u^\mu
=
E u^0
-
p_x u^x
-
p_y u^y
-
p_z u^z .
\end{equation}
The particle momentum is written in terms of the transverse momentum $p_T$, the
momentum azimuth $\phi_p$, and the momentum rapidity $Y$:
\begin{equation}
p_x = p_T\cos\phi_p,
\qquad
p_y = p_T\sin\phi_p,
\end{equation}
\begin{equation}
m_T = \sqrt{m^2+p_T^2},
\end{equation}
\begin{equation}
p_z = m_T\sinh Y,
\qquad
E = m_T\cosh Y .
\end{equation}

The sampled emission weight also contains the Cooper--Frye flux factor
$p^\mu d^3\Sigma_\mu$. In a freeze-out model, particles are emitted from a
three-dimensional hypersurface in space-time, and the number of emitted
particles depends on the flux of particles through this surface. For the
constant-proper-time hypersurface used in the blast-wave parametrization, this
factor reduces to
\begin{equation}
p^\mu d^3\Sigma_\mu
\propto
m_T\cosh(\eta_s-Y),
\end{equation}
where $\eta_s$ is the space-time rapidity of the emission point and $Y$ is the
momentum rapidity of the emitted particle. This term accounts for the
longitudinal part of the freeze-out hypersurface and weights the emission
according to the difference between the particle rapidity and the source
rapidity.

The full blast-wave weight used in the accept-reject step combines the
Cooper--Frye factor with the transverse density profile $\Omega(\tilde r)$ and
the local thermal distribution. In the numerical implementation, the
single-particle thermal Bose distribution is written as a finite sum,
\begin{equation}
\frac{1}{\exp(p_\mu u^\mu/T_{\mathrm{kin}})-1}
\simeq
\sum_{k=1}^{4}
\exp
\left[
-k\frac{p_\mu u^\mu}{T_{\mathrm{kin}}}
\right].
\end{equation}
Thus, apart from factors already included through the proposal distributions,
the unnormalized blast-wave weight used for a proposed phase-space point is
\begin{equation}
w_{\mathrm{BW}}(x,p)
\propto
m_T \cosh(\eta_s-Y)\,
\Omega(\tilde r)
\sum_{k=1}^{4}
\exp\left[
-k\frac{p_\mu u^\mu}{T_{\mathrm{kin}}}
\right].
\end{equation}
This weight is used as the target weight in the accept-reject sampling of the
proposed phase-space point $(x,p)$. Corrections associated with the auxiliary
momentum proposal distribution are described in the Monte Carlo implementation
section.

In the present implementation, the RLBW source is used to generate the
initial, undistorted pion and kaon samples. These particles are then propagated
through the residual Coulomb field described below. This separation is useful:
the blast-wave model defines the freeze-out baseline, while the subsequent
propagation isolates the additional momentum distortion caused by the residual
charged source.

\section{Residual Coulomb field model}

We now define the residual Coulomb field used to propagate the particles after
they are emitted from the blast-wave source. In this model, the field is treated
as an external, time-dependent electric field produced by the remaining charged
matter in the collision system. Each emitted particle is propagated through this
field independently, so the effect enters as a one-body post-emission momentum
distortion.

The residual source is assumed to carry an effective net positive charge,
$Z_{\mathrm{res}}$, which controls the strength of the residual Coulomb field.
This parameter should not be interpreted as a direct count of individual protons
remaining in the system at a particular time. The real post-freeze-out charge
distribution is spatially extended, evolves with time, and is not known event
by event. Therefore, $Z_{\mathrm{res}}$ is used as an effective charge scale for
the smooth residual field in the model.

In the default calculation we use $Z_{\mathrm{res}}=60$. This value is chosen as
a reasonable order-of-magnitude baseline for a central heavy-ion collision and
is later varied to test how sensitive the calculated correlation functions are
to the assumed strength of the residual Coulomb field. For a central Au+Au or
Pb+Pb collision, the participant charge can be estimated as
\begin{equation}
Z_{\mathrm{part}}
\approx
\frac{Z}{A}N_{\mathrm{part}} .
\end{equation}
For heavy nuclei, $Z/A \simeq 0.4$. Thus, a collision with
$N_{\mathrm{part}}\sim 300$--$400$ corresponds to a participant charge of order
$Z_{\mathrm{part}}\sim 120$--$160$~\cite{Miller:2007ri,ALICE:2013hur}. This estimate gives the overall
scale of the positive participant charge, but it should not be interpreted as
the charge that acts on an emitted particle as a compact static source.

In reality, the charge that contributes to the Coulomb field seen by an emitted
particle is distributed over a finite volume, spread over rapidity, and carried
by many particles that continue to move and interact. Some of the charge may
already be far from the particle of interest, and the effective field also
depends on the space-time evolution of the source. For this reason, the charge
entering the simplified residual-source model is expected to be smaller than
the full participant charge. The choice $Z_{\mathrm{res}}=60$ corresponds to a
moderate fraction of the estimated participant charge and is used as the
default baseline.

Experimentally, the effective charge $Z_{\mathrm{res}}$ cannot be determined
uniquely from a single observable. Charge-dependent single-particle spectra or
two-particle correlation functions may also contain contributions from other
physics effects. For example, the $\pi^-/\pi^+$ ratio can be influenced not
only by residual Coulomb fields, but also by isospin effects related to the
neutron-to-proton composition of the participant matter, possible neutron-skin
effects in the colliding nuclei, resonance production and decay, rescattering,
absorption, and other hadronic dynamics~\cite{Barz:1997su,SpiRIT:2021gtq,TMEP:2023ifw,Xu:2009fj}. Therefore, any experimental
constraint on $Z_{\mathrm{res}}$ is necessarily model dependent.
Our aim in the current study is not to determine this charge, but to make a reasonable assumption and estimate the expected magnitude of the effect on measurements.

The residual charge distribution used in the force calculation corresponds to a
spherically symmetric Gaussian source centered at the origin. This is a
simplified description of the charged matter left behind after particle
emission. It does not attempt to reproduce all details of the full fireball
geometry, but it provides a smooth charge distribution with a controllable size.

The corresponding charge density may be written as
\begin{equation}
\rho_C(r,t)
=
\mathcal{N}_C(t)
\exp
\left[
-\frac{r^2}{2\sigma_C^2(t)}
\right],
\end{equation}
where $r$ is the distance from the center of the source, $\sigma_C(t)$ is the
time-dependent width of the charge distribution, and $\mathcal{N}_C(t)$ is a
normalization constant. The normalization is fixed by requiring that the total
charge of the distribution is $Z_{\mathrm{res}}$,
\begin{equation}
\int \rho_C(\mathbf{r},t)\,d^3r
=
Z_{\mathrm{res}} .
\end{equation}
For a three-dimensional Gaussian,
\begin{equation}
\int
\exp
\left[
-\frac{r^2}{2\sigma_C^2(t)}
\right]
d^3r
=
(2\pi)^{3/2}\sigma_C^3(t).
\end{equation}
Therefore,
\begin{equation}
\mathcal{N}_C(t)
=
\frac{Z_{\mathrm{res}}}
{(2\pi)^{3/2}\sigma_C^3(t)} ,
\end{equation}
and the charge density becomes
\begin{equation}
\rho_C(r,t)
=
\frac{Z_{\mathrm{res}}}
{(2\pi)^{3/2}\sigma_C^3(t)}
\exp
\left[
-\frac{r^2}{2\sigma_C^2(t)}
\right].
\end{equation}

The numerical implementation does not use this density directly. Since the
residual source is spherically symmetric, the Coulomb force can instead be
written in terms of the fraction of charge enclosed inside the particle radius.
This enclosed-charge expression is obtained by integrating the Gaussian density
over a sphere of radius $r$.

The width of the residual charge distribution is allowed to increase with time,
\begin{equation}
\sigma_C(t)
=
\sigma_0 + v_{\mathrm{exp}} t .
\end{equation}
Here $\sigma_0$ is the initial width of the charge cloud and $v_{\mathrm{exp}}$
controls the expansion rate of the charge distribution. In the numerical
implementation, the relevant time is the absolute time of the particle during
propagation. If a particle is emitted at time $t_{\mathrm{emit}}$ and is then
propagated for a time $t_{\mathrm{prop}}$, the width used in the force
calculation is
\begin{equation}
\sigma_C
=
\sigma_0
+
v_{\mathrm{exp}}
\left(
t_{\mathrm{emit}} + t_{\mathrm{prop}}
\right).
\end{equation}
Thus, particles emitted later see a more expanded residual charge distribution
already at the beginning of their propagation.

The default values $\sigma_0=5.0~\mathrm{fm}$ and $v_{\mathrm{exp}}=0.3$ are
chosen as reasonable baseline parameters for an extended residual charge cloud
in a central heavy-ion collision. The value $\sigma_0=5.0~\mathrm{fm}$ does not
represent the full geometric radius of the fireball. Rather, it is the Gaussian
width of the effective charge distribution. For a three-dimensional Gaussian,
this corresponds to an rms radius
\begin{equation}
\sqrt{\langle r^2\rangle}
=
\sqrt{3}\,\sigma_0
\approx
8.7~\mathrm{fm},
\end{equation}
which is comparable to the freeze-out size scale used in the blast-wave source.
Thus, the residual charge is treated as an extended distribution, not as a
point-like charge at the center of the system.

The parameter $v_{\mathrm{exp}}$ controls how quickly this effective charge
distribution becomes diluted with time. The default value
$v_{\mathrm{exp}}=0.3$ corresponds to a moderate expansion velocity in units
where $c=1$. It allows the residual source to expand during the post-emission
propagation, while keeping the field sufficiently long lived to test its
possible influence on the final particle momenta. Both $\sigma_0$ and
$v_{\mathrm{exp}}$ are varied later to estimate how sensitive the calculated
correlation functions are to the assumed size and expansion rate of the
residual charged source.

For a spherically symmetric residual charge distribution, the electric field at
radius $r$ depends only on the charge enclosed inside that radius. The enclosed
charge fraction is
\begin{equation}
f(r,\sigma_C)
=
\frac{Q(<r)}{Z_{\mathrm{res}}}
=
\frac{1}{Z_{\mathrm{res}}}
\int_0^r
4\pi r'^2
\rho_C(r',\sigma_C)\,dr' .
\end{equation}
For the Gaussian density used here, this integral can be evaluated analytically:
\begin{equation}
f(r,\sigma_C)
=
\operatorname{erf}
\left(
\frac{r}{\sqrt{2}\sigma_C}
\right)
-
\sqrt{\frac{2}{\pi}}
\frac{r}{\sigma_C}
\exp
\left[
-\frac{r^2}{2\sigma_C^2}
\right].
\end{equation}
This is the expression used in the numerical force calculation. It has the
expected limiting behavior. Near the center of the source, the enclosed charge
goes to zero, so the Coulomb force does not have the singular behavior of a
point charge. At distances much larger than the Gaussian width, the enclosed
fraction approaches unity, and the field becomes equivalent to that of a point
charge with charge $Z_{\mathrm{res}}$.

For a particle with charge sign $q=\pm 1$ at position $\mathbf{r}$, the
momentum change due to the residual field is
\begin{equation}
\frac{d\mathbf{p}}{dt}
=
q Z_{\mathrm{res}} e^2
\,
\frac{f(r,\sigma_C)}{r^3}
\,
\mathbf{r},
\end{equation}
where $r=|\mathbf{r}|$ and $f(r,\sigma_C)$ is the fraction of the Gaussian
charge enclosed inside radius $r$. In the numerical implementation, this
equation is integrated in finite time steps,
\begin{equation}
\Delta \mathbf{p}
=
q Z_{\mathrm{res}} e^2
\,
\frac{f(r,\sigma_C)}{r^3}
\,
\mathbf{r}\,
\Delta t .
\end{equation}

In the numerical propagation, the Coulomb force is integrated up to a maximum
time $t_{\max}=30~\mathrm{fm}/c$. The base time step is
$\Delta t=0.01~\mathrm{fm}/c$. An adaptive step size is used: the step is kept
small, $\Delta t=0.01~\mathrm{fm}/c$, for the first $10~\mathrm{fm}/c$, when
the particle is still relatively close to the residual charge distribution and
the Coulomb field changes most rapidly. At later times the particle is
typically farther from the source and the force becomes weaker and more slowly
varying. The step size is therefore increased to $0.05~\mathrm{fm}/c$ for the
remaining evolution up to $30~\mathrm{fm}/c$.

This adaptive choice improves the numerical efficiency while keeping good
resolution during the early part of the trajectory, where most of the momentum
change is accumulated. The values of $\Delta t$ and $t_{\max}$ are used as the
default propagation settings and are varied later to check that the calculated
correlation functions are stable against the numerical details of the
propagation.

The sign of $q$ determines the direction of the effect. For a positive residual
source, positively charged particles are accelerated outward, while negatively
charged particles are accelerated inward. This charge dependence is the key
feature of the residual Coulomb effect studied here.

This model intentionally keeps the residual field simple. The charge cloud is
taken to be smooth, centered at the origin, and spherically symmetric, while the
blast-wave freeze-out source may have an elliptic transverse shape. This choice
allows us to isolate the effect of the post-emission Coulomb acceleration
without introducing additional assumptions about the detailed space-time
evolution of the remaining charged matter. The parameters
$Z_{\mathrm{res}}$, $\sigma_0$, and $v_{\mathrm{exp}}$ can then be varied to
study how the strength, size, and expansion of the residual source affect the
final pion and kaon correlation functions.

\section{Monte Carlo implementation}

The model described above is implemented as a Monte Carlo generator. Its purpose
is to produce charge-dependent final-state samples from a common freeze-out
baseline. Particles are first generated from the RLBW
source. The same generated particles are then propagated through the residual
Coulomb field twice: once with positive charge and once with negative charge.
This produces two distorted samples corresponding to positively and negatively
charged particles.

The main comparison in this study is between these two final charge states.
Because they start from the same freeze-out distribution, differences between
them can be associated with the opposite sign of the residual Coulomb force. The
undistorted sample is also stored as a reference, but the primary observable is
the charge-dependent difference between the final positive and negative
correlation functions.

\subsection{Particle sampling}

For each Monte Carlo sample, two independent pions and two independent kaons are
generated from the blast-wave source. Each particle is assigned an emission
point $x^\mu$ and an initial four-momentum $p^\mu$. The emission point is
sampled from the space-time source described in the previous section, while the
momentum is sampled from the local thermal distribution boosted by the
RLBW flow field.

The sampling is performed with an accept-reject procedure. Trial emission
points are proposed using the elliptical source coordinates, while trial
momenta are generated in terms of the transverse momentum $p_T$, the momentum
azimuth $\phi_p$, and the momentum rapidity $Y$. Here $p_T$ is the magnitude of
the particle momentum in the transverse plane, $\phi_p$ is the azimuthal angle
of this transverse momentum, and
\begin{equation}
Y =
\frac{1}{2}
\ln
\left(
\frac{E+p_z}{E-p_z}
\right)
\end{equation}
is the momentum rapidity.

In the numerical implementation, $\phi_p$ is sampled uniformly in the interval
$[0,2\pi)$, the momentum rapidity is proposed in the range
$|Y|<Y_{\max}$, and the transverse momentum is proposed from an exponential
distribution,
\begin{equation}
g(p_T)
\propto
\exp\left(-\frac{p_T}{p_0}\right),
\end{equation}
with the additional requirement $p_T<p_{T,\max}$. The proposal scale is
$p_0=0.3~\mathrm{GeV}/c$. This is not a physical parameter of the source; it is
used only to improve the efficiency of the Monte Carlo sampling. Since the
present analysis focuses on low-$k_T$ femtoscopic pairs, most accepted
particles come from the low-$p_T$ region. The exponential proposal therefore
places trial particles where the blast-wave emission weight is largest, while
still allowing a tail to higher transverse momentum.

The default proposal limits are $Y_{\max}=1.5$ and
$p_{T,\max}=4~\mathrm{GeV}/c$. These values are numerical sampling limits, not
detector acceptance cuts. The choice $Y_{\max}=1.5$ provides a broad
midrapidity momentum region for the low-$k_T$ femtoscopic analysis, while
avoiding inefficient sampling of very forward particles where the simplified
boost-invariant blast-wave description is not intended to be applied. The upper
limit $p_{T,\max}=4~\mathrm{GeV}/c$ is well above the momentum region that
dominates the low-$k_T$ correlation functions, but prevents the Monte Carlo
from spending time on very high-momentum particles with negligible thermal
weight.

The momentum-rapidity range should be distinguished from the source-rapidity
range. In the default setup, the source rapidity is sampled over
$|\eta_s|<\eta_{\max}$ with $\eta_{\max}=3.0$, while the particle momentum
rapidity is proposed over $|Y|<1.5$. This is natural because $\eta_s$ describes
the space-time rapidity of the emission point, whereas $Y$ describes the
rapidity of the emitted particle momentum. The larger value of $\eta_{\max}$
allows the source to have a finite but extended longitudinal size, while the
sampled particles remain focused on the central momentum-rapidity region.
Contributions from source elements with large $|Y-\eta_s|$ are naturally
suppressed by the thermal factor in the blast-wave weight. The values of
$Y_{\max}$ and $\eta_{\max}$ are varied later to verify that the calculated
correlation functions in the selected low-$k_T$ region are not driven by the
rapidity cutoffs.

The total sampling weight for a proposed particle is proportional to
\begin{equation}
w
\propto
m_T\cosh(\eta_s-Y)
\,
\Omega(\tilde r)
\,
\sum_{k=1}^{4}
\exp
\left[
-k\frac{p_\mu u^\mu}{T_{\mathrm{kin}}}
\right]
\,
w_{\mathrm{prop}} .
\end{equation}
The factor $w_{\mathrm{prop}}$ corrects for the fact that $p_T$ is proposed
from the auxiliary distribution $g(p_T)$. In an accept-reject procedure, the
weight must be proportional to the desired distribution divided by the proposal
distribution. Therefore,
\begin{equation}
w_{\mathrm{prop}}
\propto
\frac{1}{g(p_T)}
\propto
\exp\left(\frac{p_T}{p_0}\right).
\end{equation}
This correction removes the bias from the exponential proposal, so the accepted
particles follow the blast-wave emission weight rather than the proposal shape.
The variables $\phi_p$ and $Y$ are proposed uniformly, so their proposal
densities contribute only constant factors and do not require explicit
corrections in the acceptance weight.

The upper bound used in the accept-reject step is updated during the sampling
and is stored separately for pions and kaons, because the two particle species
have different masses and therefore different weight scales.

\subsection{Post-emission propagation}

After particles are generated from the blast-wave source, they are propagated
through the residual Coulomb field. The same freeze-out particles are used as
the starting point for both charge cases. Each particle is propagated once with
charge sign $q=+1$ and once with charge sign $q=-1$, corresponding to
positively and negatively charged pions or kaons. In this way, the positive and
negative samples differ only by the sign of the residual Coulomb force.

For each charge case, the two particles in the pair are propagated
independently through the same residual charge distribution. During the
propagation, the particle position and momentum are updated according to the
Coulomb force described above. After propagation, the final momenta are stored
and later used to construct the one-dimensional and three-dimensional
correlation functions.

This procedure produces three samples for each particle species:
\begin{enumerate}
\item the true sample, before residual Coulomb propagation;
\item the negatively charged sample, after propagation with $q=-1$;
\item the positively charged sample, after propagation with $q=+1$.
\end{enumerate}
The true sample is useful as a reference, but the main comparison in this study
is between the final positively and negatively charged samples. Since both
charge cases start from the same underlying freeze-out distribution,
differences between them can be directly associated, within this model, with the
opposite sign of the residual Coulomb force.

Unless stated otherwise, the calculations use the default parameter set listed
in Table~\ref{tab:default_parameters}. The parameters are grouped according to
their role in the simulation: the blast-wave source defines the initial
freeze-out distribution, the residual-charge parameters define the effective
Coulomb field, and the numerical propagation parameters control the time
evolution after emission.

\begin{table}[t]
\centering
\caption{Default parameters used in the residual Coulomb study. The blast-wave
parameters define the undistorted freeze-out source, while the residual-charge
and propagation parameters define the post-emission Coulomb evolution.}
\label{tab:default_parameters}
\begin{tabular}{lll}
\hline
Parameter & Meaning & Default value \\
\hline
\multicolumn{3}{l}{\textit{Blast-wave source}} \\
$T_{\mathrm{kin}}$ & kinetic freeze-out temperature & $0.110~\mathrm{GeV}$ \\
$\rho_0$ & average transverse flow rapidity & $0.8$ \\
$\rho_2$ & elliptic modulation of transverse flow & $0.01$ \\
$R_x$ & transverse radius in the $x$ direction & $11.5~\mathrm{fm}$ \\
$R_y$ & transverse radius in the $y$ direction & $10.0~\mathrm{fm}$ \\
$a_s$ & surface diffuseness & $0$ \\
$\tau_0$ & mean freeze-out proper time & $6.0~\mathrm{fm}/c$ \\
$\Delta\tau$ & emission duration & $2.0~\mathrm{fm}/c$ \\
$\eta_{\max}$ & maximum source rapidity & $3.0$ \\
$n$ & radial flow exponent & $1.0$ \\
\hline
\multicolumn{3}{l}{\textit{Residual charge model}} \\
$Z_{\mathrm{res}}$ & effective residual charge & $60$ \\
$\sigma_0$ & initial Gaussian charge width & $5.0~\mathrm{fm}$ \\
$v_{\mathrm{exp}}$ & expansion rate of charge width & $0.3$ \\
\hline
\multicolumn{3}{l}{\textit{Numerical propagation}} \\
$\Delta t$ & base propagation step & $0.01~\mathrm{fm}/c$ \\
$t_{\max}$ & maximum propagation time & $30~\mathrm{fm}/c$ \\
\hline
\end{tabular}
\end{table}

In the default setup, no additional transverse-momentum or pseudorapidity cuts
are applied to the distorted particles. The effect of detector-like acceptance
cuts on the calculated observables is studied later in this paper.

\section{Correlation function construction}

After the particle samples have been generated, and after the residual Coulomb
propagation has been applied when needed, the particles are used to construct
femtoscopic correlation functions. The same procedure is followed for pions and
kaons. For each particle species, pairs are formed from the generated emission
points and four-momenta.

For a pair with four-momenta $p_1^\mu$ and $p_2^\mu$, the relative
four-momentum is defined as
\begin{equation}
q^\mu = p_1^\mu - p_2^\mu .
\end{equation}
The invariant relative momentum is then calculated as
\begin{equation}
q_{\mathrm{inv}}
=
\sqrt{-q^\mu q_\mu}.
\end{equation}
In the numerical implementation, the argument of the square root is protected
against small numerical round-off effects by taking the positive part of
$-q^\mu q_\mu$.

The average pair momentum is defined as
\begin{equation}
K^\mu
=
\frac{1}{2}
\left(
p_1^\mu+p_2^\mu
\right),
\end{equation}
and the pair transverse momentum is
\begin{equation}
k_T = |\mathbf{K}_T| .
\end{equation}
The present study focuses on the low-$k_T$ region, where the residual Coulomb
effect is expected to be most important for femtoscopy. Slowly moving particles
spend more time near the residual source and are therefore more sensitive to the
post-emission Coulomb field. We therefore construct the correlation functions
for $0<k_T<0.4~\mathrm{GeV}/c$ using four intervals of width
$0.1~\mathrm{GeV}/c$.

The quantum-statistical part of the identical-particle correlation is included
through a plane-wave Bose--Einstein weight. For two identical bosons, the weight
is written as
\begin{equation}
w_{\mathrm{BE}}
=
1+
\cos
\left(
\frac{q_\mu \Delta x^\mu}{\hbar c}
\right),
\label{eq:w_BE}
\end{equation}
where
\begin{equation}
\Delta x^\mu = x_1^\mu-x_2^\mu
\end{equation}
is the four-vector separation between the two emission points. This term
represents the interference contribution from symmetrizing the two-particle
wave function. Here it is used as a simple femtoscopic weight.

The usual two-body Coulomb and strong final-state interactions within the pair
are not included. The reason is that the main observable is the difference
between identical positive and identical negative pairs, such as
$\pi^+\pi^+$ compared with $\pi^-\pi^-$, or $K^+K^+$ compared with
$K^-K^-$. For these same-charge identical pairs, the ordinary two-body Coulomb
interaction is repulsive in both cases, and the strong interaction is the same
up to charge symmetry. These two-body effects therefore do not generate the
charge-dependent difference that is the focus of this work. Accordingly, they
are omitted in order to isolate the three-body residual Coulomb effect and to
simplify the numerical simulation.

For each pair, two histograms are filled. The numerator is filled with the
Bose--Einstein weight $w_{\mathrm{BE}}$, while the denominator is filled with
unit weight. The one-dimensional correlation function is then constructed as
\begin{equation}
C(q_{\mathrm{inv}})
=
\frac{N(q_{\mathrm{inv}})}
     {D(q_{\mathrm{inv}})} ,
\end{equation}
where $N$ is the weighted numerator and $D$ is the unweighted denominator.

The same construction is used for the three-dimensional correlation function.
The relative momentum is decomposed into the Bertsch--Pratt components~\cite{Pratt:1986cc, Bertsch:1988db}
$q_{\mathrm{out}}$, $q_{\mathrm{side}}$, and $q_{\mathrm{long}}$ in the
longitudinally comoving system. To define this frame, the pair is first boosted
along the beam direction so that the total longitudinal momentum of the pair
vanishes. In this frame, the ``long'' direction is along the beam axis, the
``out'' direction is chosen along the transverse pair momentum, and the
``side'' direction is perpendicular to it in the transverse plane. The
three-dimensional correlation function is then constructed as
\begin{equation}
C(q_{\mathrm{out}},q_{\mathrm{side}},q_{\mathrm{long}})
=
\frac{
N(q_{\mathrm{out}},q_{\mathrm{side}},q_{\mathrm{long}})
}{
D(q_{\mathrm{out}},q_{\mathrm{side}},q_{\mathrm{long}})
}.
\end{equation}

The residual Coulomb field enters the correlation construction through the
final momenta of the particles. The undistorted reference sample is built
directly from the freeze-out momenta. For the distorted samples, the same
particles are first propagated through the residual Coulomb field with positive
or negative charge. The final propagated momenta are then used to recalculate
$q_{\mathrm{inv}}$, $k_T$, and the Bertsch--Pratt components.

The Bose--Einstein weight is evaluated from the original freeze-out emission
points and momenta. The residual Coulomb propagation changes the final relative
momentum of the pair, and therefore changes where the pair contributes in the
correlation histogram. In this way, the femtoscopic interference is set by the
emission configuration, while the residual Coulomb field modifies the measured
relative momentum.

\begin{figure}[h!]
    \centering
    \includegraphics[width=0.9\linewidth]{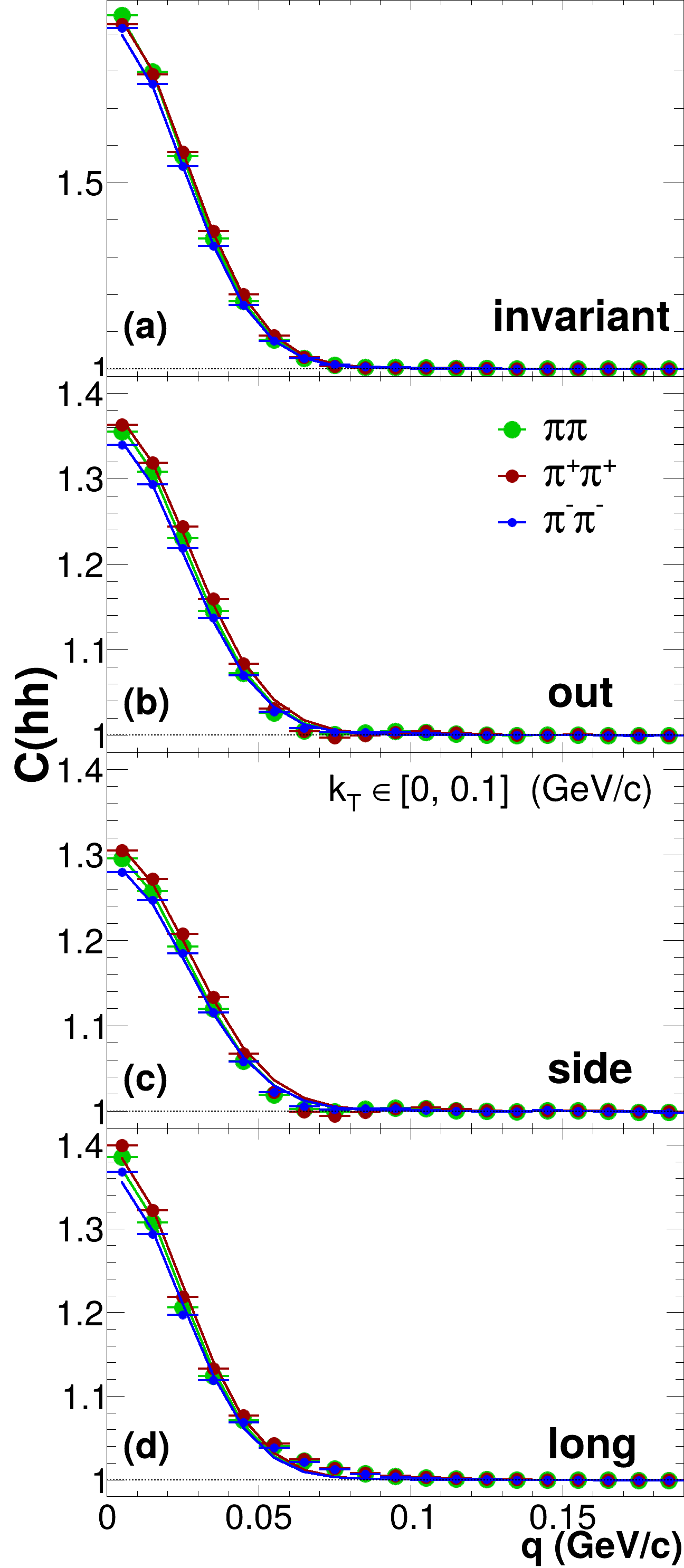}
\caption{
Example of the pion correlation function for the transverse-momentum interval
$0<k_T<0.1~\mathrm{GeV}/c$. Panel (a) shows the one-dimensional correlation
function as a function of $q_{\mathrm{inv}}$. Panels (b)--(d) show the
one-dimensional projections of the three-dimensional correlation function onto
$q_{\mathrm{out}}$, $q_{\mathrm{side}}$, and $q_{\mathrm{long}}$, respectively.
The undistorted reference $\pi\pi$ sample, generated before propagation through
the residual Coulomb field, is shown by green circles. The distorted
$\pi^+\pi^+$ and $\pi^-\pi^-$ samples, obtained after propagation, are shown by
red and blue circles, respectively.
}
    \label{fig:CF}
\end{figure}

\begin{figure}[h!]
    \centering
    \includegraphics[width=0.9\linewidth]{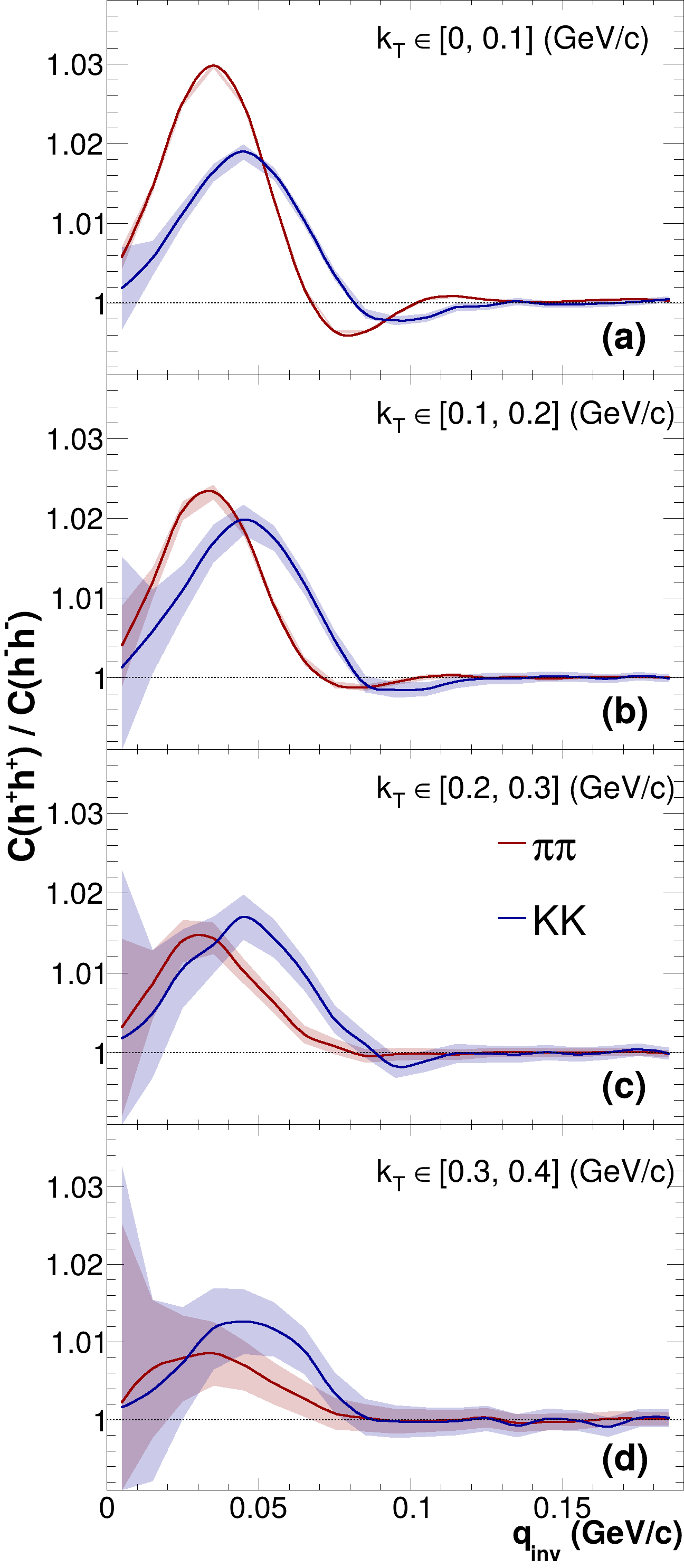}
\caption{
Ratio of like-sign positive to negative correlation functions,
$C(h^+h^+)/C(h^-h^-)$, as a function of the invariant relative momentum
$q_{\mathrm{inv}}$. Results are shown for pions, indicated by the red line,
and kaons, indicated by the blue line. The shaded bands around the lines
represent statistical uncertainties. The four panels correspond to
transverse-momentum intervals from $0$ to $0.4~\mathrm{GeV}/c$ in steps of
$0.1~\mathrm{GeV}/c$. The horizontal dashed line indicates unity.
}
    \label{fig:cf_rat_1d_def}
\end{figure}

\begin{figure}[h!]
    \centering
    \includegraphics[width=1.0\linewidth]{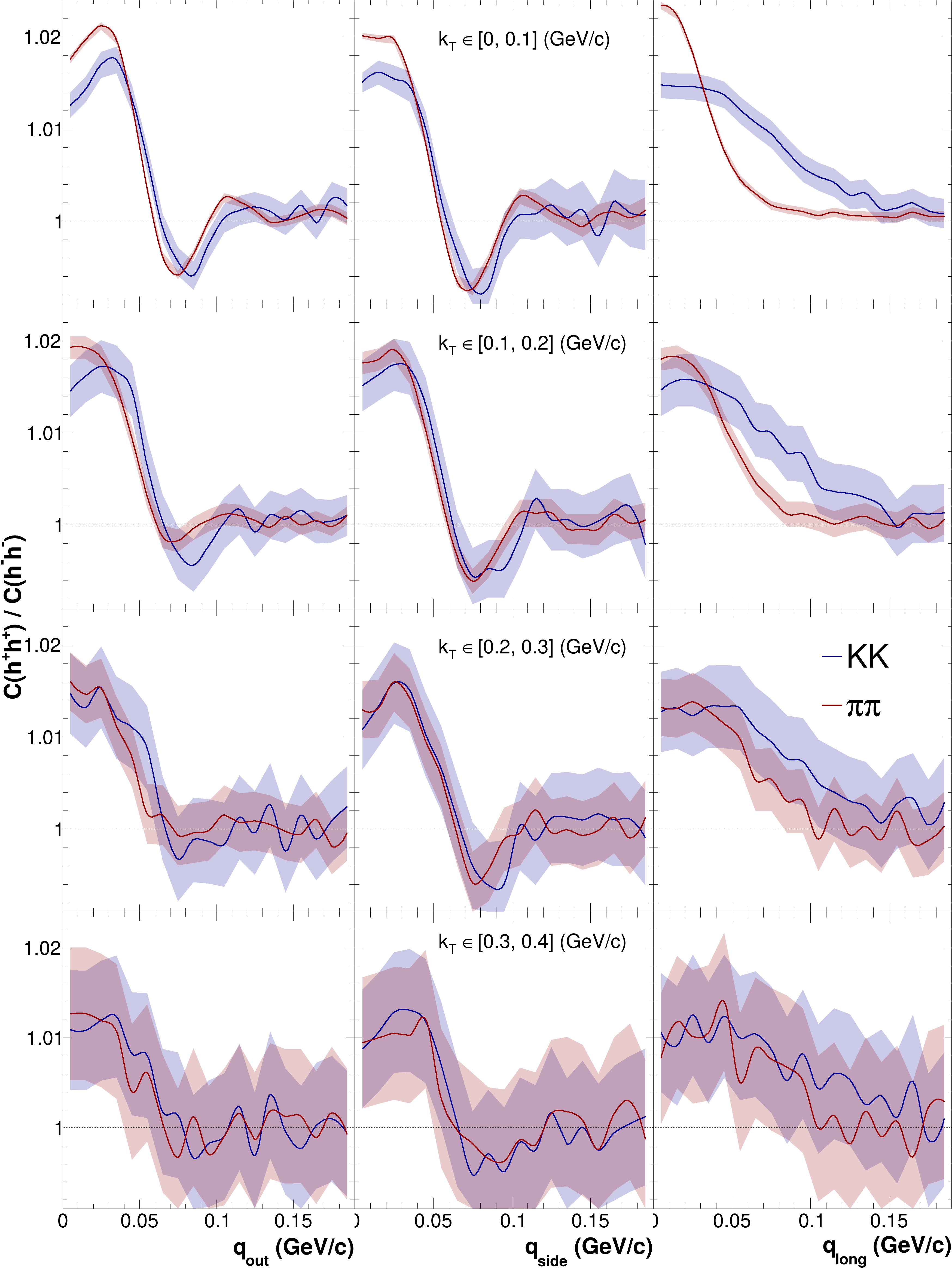}
\caption{
Example of the ratios of the one-dimensional projections of the
three-dimensional correlation functions for positive hadron pairs to those for
negative hadron pairs, $C(h^+h^+)/C(h^-h^-)$. Results are shown for pions,
indicated by the red line, and kaons, indicated by the blue line. The shaded
bands around the lines represent statistical uncertainties. The four rows
correspond to transverse-momentum intervals from $0$ to $0.4~\mathrm{GeV}/c$
in steps of $0.1~\mathrm{GeV}/c$. The three columns correspond to the
$q_{\mathrm{out}}$, $q_{\mathrm{side}}$, and $q_{\mathrm{long}}$ projections.
The horizontal dashed line indicates unity.
}
    \label{fig:cf_rat_3d_def}
\end{figure}

An example of the undistorted and distorted pion correlation functions is shown
in Fig.~\ref{fig:CF}. A small charge-dependent splitting is visible: the
correlation function for negative pion pairs lies slightly below that for
positive pion pairs.

This splitting comes from the opposite action of the positively charged
residual source on particles of different charge. Positive pions are repelled
from the source, while negative pions are attracted toward it. Since the two
particles in a pair are emitted from different positions and, in general, at
different times, they receive different momentum kicks. As a result, the
residual field changes the final relative momentum of the pair.

For negative pions, the attractive force pulls the particles back toward the
center of the source and modifies the original blast-wave position--momentum
correlations. This leads to a redistribution of pairs in final relative-momentum
space. Consequently, the Bose--Einstein enhancement is sampled slightly
differently after propagation, giving a correlation function that is lower than
the corresponding positive-pion case in the low-relative-momentum region. The
effect is small but systematic, and reflects the charge-dependent momentum
distortion induced by the residual Coulomb field.

\section{Fit of the correlation functions and extraction of femtoscopic parameters}

After the correlation functions are constructed, they are fitted in order to
summarize the effect of the residual Coulomb field in terms of standard
femtoscopic parameters. These fit parameters provide a compact way to compare
the true, positively charged, and negatively charged samples. However, they
should be interpreted as effective Gaussian parameters, because the residual
Coulomb propagation can also modify the non-Gaussian shape of the correlation
function.

For the one-dimensional correlation functions, the fit is performed as a
function of $q_{\mathrm{inv}}$. A Gaussian form is used, following the standard
practice in femtoscopic analyses:
\begin{equation}
C(q_{\mathrm{inv}})
=
N
\left[
1
+
\lambda
\exp
\left(
-
q_{\mathrm{inv}}^2 R_{\mathrm{inv}}^2
\right)
\right],
\end{equation}
where $N$ is an overall normalization, $\lambda$ is an effective
correlation-strength parameter, and $R_{\mathrm{inv}}$ is the one-dimensional
femtoscopic radius characterizing the homogeneity region of the source. In this
notation, the product $q_{\mathrm{inv}}R_{\mathrm{inv}}$ is understood to be
made dimensionless using the same $\hbar c$ convention as in the construction
of the Bose--Einstein weight.

The three-dimensional correlation functions are fitted in the Bertsch--Pratt
representation using
\begin{equation}
C(q_o,q_s,q_l)
=
N
\left[
1
+
\lambda
\exp
\left(
-
\sum_{i=o,s,l}
q_i^2 R_i^2
\right)
\right],
\end{equation}
where $o$, $s$, and $l$ denote the out, side, and long directions,
respectively. The corresponding radii are $R_{\mathrm{out}}$,
$R_{\mathrm{side}}$, and $R_{\mathrm{long}}$.

Cross terms, such as $R_{os}^2$, $R_{sl}^2$, and $R_{ol}^2$, are not included
in the fit. Since the analysis is performed at midrapidity and is integrated
over the azimuthal angle of the pair, the relevant reflection symmetries are
not broken, and these cross terms are expected to vanish within the precision
of the model. This simplified Gaussian form is sufficient for the present
study, where the goal is to compare the relative change of the extracted
parameters caused by the residual Coulomb field.

The same fitting procedure is applied to the true, positively charged, and
negatively charged samples for each particle species and each $k_T$ interval.
Using the same fit form and fit range for all samples ensures that differences
between the extracted parameters reflect changes in the correlation functions
rather than differences in the fitting procedure. The one-dimensional fits are
performed in the range $0<q_{\mathrm{inv}}<0.4~\mathrm{GeV}/c$. For the
three-dimensional analysis, the fits are performed in the range
$-0.4<q_i<0.4~\mathrm{GeV}/c$ for each Bertsch--Pratt component $i=o,s,l$.

\section{Residual Coulomb effect on correlation functions and fitted parameters}

Figures~\ref{fig:cf_rat_1d_def} and~\ref{fig:cf_rat_3d_def} show the ratios of
the correlation functions for positive hadron pairs to those for negative
hadron pairs. Figure~\ref{fig:cf_rat_1d_def} corresponds to the
one-dimensional $q_{\mathrm{inv}}$ correlation functions, while
Fig.~\ref{fig:cf_rat_3d_def} shows the ratios for the one-dimensional
projections of the three-dimensional correlation functions. In both figures,
the four rows correspond to different $k_T$ intervals. Deviations of the ratios
from unity show the charge-dependent splitting produced by the residual Coulomb
field.

The effect decreases with increasing $k_T$. This behavior is expected because
particles with larger transverse momentum move away from the residual source
more rapidly and therefore spend less time in the region where the Coulomb field
is strong. As a result, the accumulated momentum change from the residual field
becomes smaller at larger $k_T$, and the positive and negative correlation
functions become more similar.

A clear particle-species dependence is also visible. The residual Coulomb force
changes the particle momentum, but the size of the final distortion depends on
the particle mass through the particle velocity and trajectory. At fixed
$k_T$, kaons are slower than pions because of their larger mass. They can
therefore remain closer to the residual source for a longer time and accumulate
a different Coulomb kick. At the lowest $k_T$, the distortion is sizable for
both pions and kaons. At higher $k_T$, pions move away from the source more
quickly, while the heavier kaons remain more sensitive to the residual field.
This competition leads to a different $k_T$ dependence for pions and kaons.

The correlation-function ratios show that the residual Coulomb field modifies
both the width and the height of the correlation peak. Therefore, the effect can
influence both the extracted femtoscopic radii $R$ and the fitted
correlation-strength parameter $\lambda$. However, these parameters should be
interpreted as effective Gaussian fit parameters. Even without the residual
field, the simulated correlation function is not guaranteed to be perfectly
Gaussian, and the post-emission Coulomb propagation can make the shape even
less Gaussian. Therefore, a Gaussian fit provides a compact parametrization of
selected features of the correlation function, rather than a complete
description of the Coulomb distortion.

For each particle species and each $k_T$ interval, the fitted parameters are
compared between the positive and negative samples by forming
positive-to-negative ratios. This is done for the correlation-strength
parameter $\lambda$ and for the extracted femtoscopic radii. In the
one-dimensional analysis, the ratio is formed for $R_{\mathrm{inv}}$. In the
three-dimensional analysis, the same procedure is applied to
$R_{\mathrm{out}}$, $R_{\mathrm{side}}$, and $R_{\mathrm{long}}$. If the
residual Coulomb field produced no charge-dependent modification, these ratios
would be consistent with unity.

\begin{figure}
    \centering
    \includegraphics[width=1.0\linewidth]{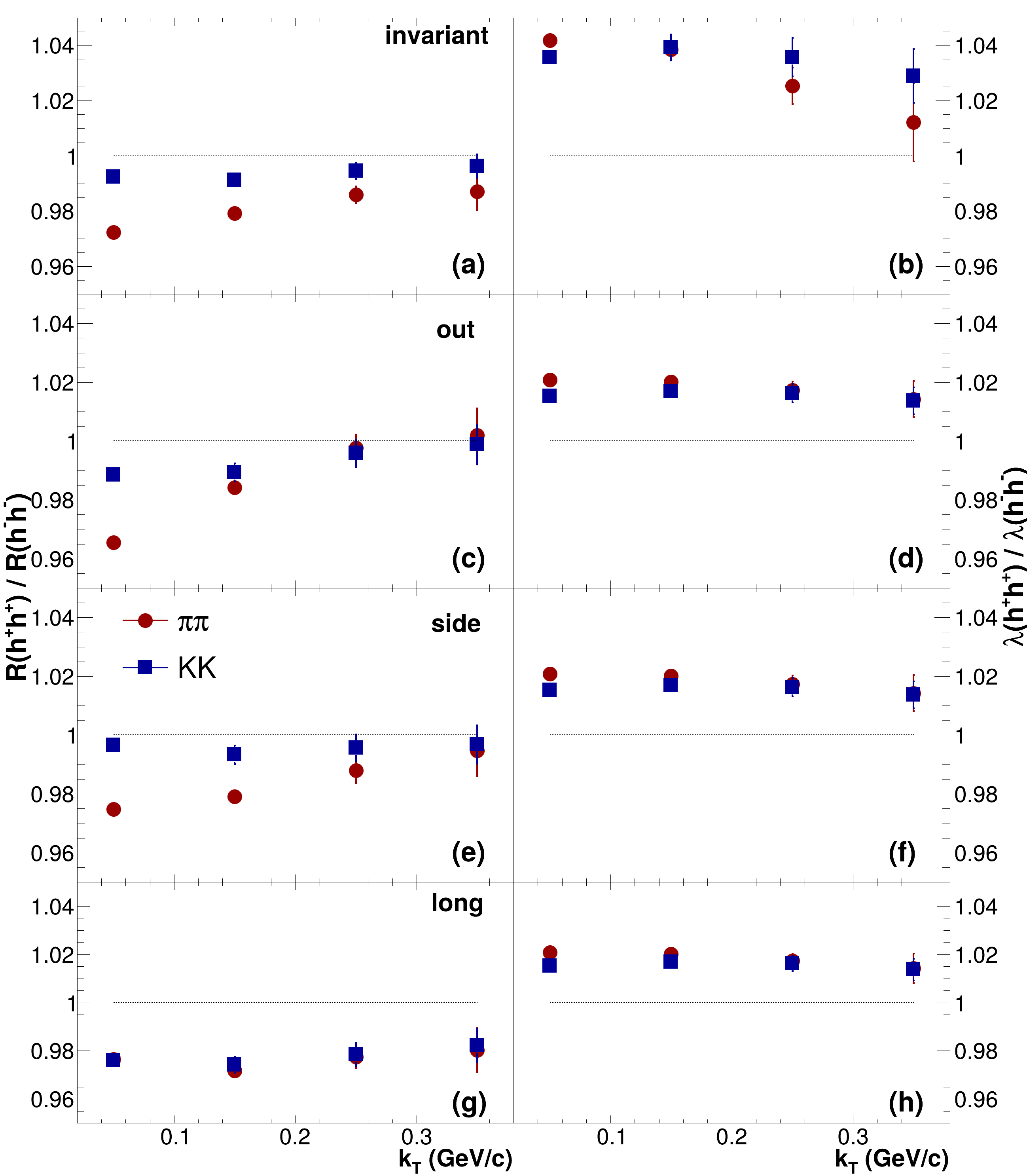}
\caption{
Ratios of the extracted femtoscopic parameters for positive hadron pairs to
the corresponding parameters for negative hadron pairs, shown as functions of
transverse momentum $k_T$. Results are shown for pions, indicated by red
circles, and kaons, indicated by blue circles. The left column shows the ratios
of the radii, $R(h^+h^+)/R(h^-h^-)$, from the one-dimensional and
three-dimensional analyses, while the right column shows the corresponding
ratios of the intercept parameters,
$\lambda(h^+h^+)/\lambda(h^-h^-)$. The horizontal line indicates unity.
}
    \label{fig:radii_rat_def}
\end{figure}

Figure~\ref{fig:radii_rat_def} shows the positive-to-negative ratios of the
fitted radii and $\lambda$ parameters. With the default parameter set listed in
Table~\ref{tab:default_parameters}, the deviations from unity are at the level
of a few percent, showing that both the fitted radii and the fitted
correlation-strength parameters are affected by the residual Coulomb
propagation.

The extracted radius ratios are generally below unity. This means that the
effective Gaussian radii obtained from positive-pair correlation functions are
smaller than those obtained from negative-pair correlation functions. The effect
is most pronounced at low $k_T$, consistent with the behavior observed directly
in the correlation-function ratios, and becomes weaker as the transverse
momentum increases.

In contrast, the ratios of the $\lambda$ parameters are mostly above unity,
showing that the residual Coulomb field also modifies the fitted correlation
strength. The radius ratios show a visible species dependence, while the
$\lambda$ ratios for pions and kaons are distorted in a more similar way. This
suggests that the radii are more sensitive to the mass-dependent change of the
pair relative momentum, whereas $\lambda$ mainly reflects the change in the
apparent height of the fitted Gaussian component.

The possible non-Gaussian distortion caused by the residual Coulomb field is
important when interpreting fitted femtoscopic parameters. In the existing
femtoscopy literature, deviations from a Gaussian shape are often associated
with resonance decays, which broaden the effective emission
profile~\cite{Pratt:1986cc,Csorgo:1995bi,Heinz:1999rw,Humanic:2006ve}. Such
contributions can naturally lead to non-Gaussian, or L\'evy-like, emission
patterns, which may be described within a L\'evy-walk picture related to
anomalous diffusion and resonance-decay
kinematics~\cite{Csorgo:2003uv,Csanad:2007fr,Kincses:2024lnv}. The present
study suggests that the residual Coulomb field can provide another possible
source of non-Gaussian distortion. In this case, the non-Gaussianity does not
come from the initial emission profile itself, but from the charge-dependent
post-emission modification of the particle momenta.

\section{Systematic variations of model parameters}

The results discussed above were obtained using the default parameter set listed
in Table~\ref{tab:default_parameters}. This set defines the baseline
configuration for the blast-wave freeze-out source and the residual Coulomb
field. Several quantities in the model are effective parameters and are not
fixed uniquely by a single experimental observable. It is therefore important to
test how the calculated charge-dependent splitting changes when these
assumptions are varied.

The purpose of this systematic study is twofold. First, it checks whether the
observed residual Coulomb effect is stable against reasonable changes of the
model inputs. Second, it helps identify which physical ingredients control the
size of the distortion. In the following subsections, positive-to-negative
ratios of the correlation functions and fitted femtoscopic parameters are
compared for several variations around the default setup. The same analysis
procedure is used for all parameter sets, so that changes in the ratios can be
associated with the corresponding parameter variation.

\subsection{Variation of blast-wave source parameters}

The first group of variations concerns the blast-wave freeze-out source. These
parameters determine the space-time and momentum distribution of particles
before the residual Coulomb propagation is applied. They control the size and
shape of the source, the emission time, the freeze-out temperature, the
collective flow field, and the strength of position--momentum correlations.

Varying these parameters allows us to scan over different freeze-out
conditions. In a qualitative sense, such changes can be associated with moving
between different collision centralities and beam energies. For example, a
larger source or a later freeze-out time corresponds to a system that has
expanded for a longer time, while changing the temperature or transverse flow
modifies the initial particle velocities. Since the residual Coulomb field acts
only after emission, all of these changes can influence how long the particles
remain close to the residual charged source and how much momentum kick they
accumulate.

The varied blast-wave, numerical sampling, and propagation parameters are
summarized in Table~\ref{tab:bw_variations}. Only one parameter is changed at a
time, while all other parameters are kept at their default values from
Table~\ref{tab:default_parameters}. The chosen values are not meant to
represent one specific experimental system. Instead, they bracket the default
configuration and test the response of the residual Coulomb effect to smaller
and larger sources, weaker and stronger flow, earlier and later emission, and
narrower or wider numerical sampling ranges.

\begin{table}[h!]
\centering
\caption{Blast-wave source, numerical sampling, and propagation parameters
varied in the systematic study. For each row, only the listed parameter is
changed, while the remaining parameters are kept at their default values from
Table~\ref{tab:default_parameters}.}
\label{tab:bw_variations}
\begin{tabular}{lll}
\hline
Parameter & Default value & Varied values \\
\hline
\multicolumn{3}{l}{\textit{Blast-wave source}} \\
$T_{\mathrm{kin}}$ & $0.110~\mathrm{GeV}$ & $0.100,\;0.135~\mathrm{GeV}$ \\
$\rho_0$           & $0.8$                & $0.35,\;0.95$ \\
$\rho_2$           & $0.01$               & $0,\;0.06$ \\
$(R_x,R_y)$        & $(11.5,\;10)~\mathrm{fm}$ 
                   & $(10,\;10),\;(14,\;13),\;(5,\;5.5)~\mathrm{fm}$ \\
$a_s$              & $0$                  & $0.1,\;0.3$ \\
$\tau_0$           & $6.0~\mathrm{fm}/c$  & $3.0,\;10.0~\mathrm{fm}/c$ \\
$\Delta\tau$       & $2.0~\mathrm{fm}/c$  & $1.0,\;3.0~\mathrm{fm}/c$ \\
$n$                & $1$                  & $0,\;4$ \\
$\eta_{\max}$      & $3.0$                & $5.0$ \\
\hline
\multicolumn{3}{l}{\textit{Numerical sampling and propagation}} \\
$t_{\max}$         & $30~\mathrm{fm}/c$   & $60,\;200~\mathrm{fm}/c$ \\
$p_{T,\max}$       & $4~\mathrm{GeV}/c$   & $2,\;7~\mathrm{GeV}/c$ \\
$Y_{\max}$         & $1.5$                & $1.0,\;3.0$ \\
\hline
\end{tabular}
\end{table}

\begin{figure}[h!]
    \centering
    \includegraphics[width=1.0\linewidth]{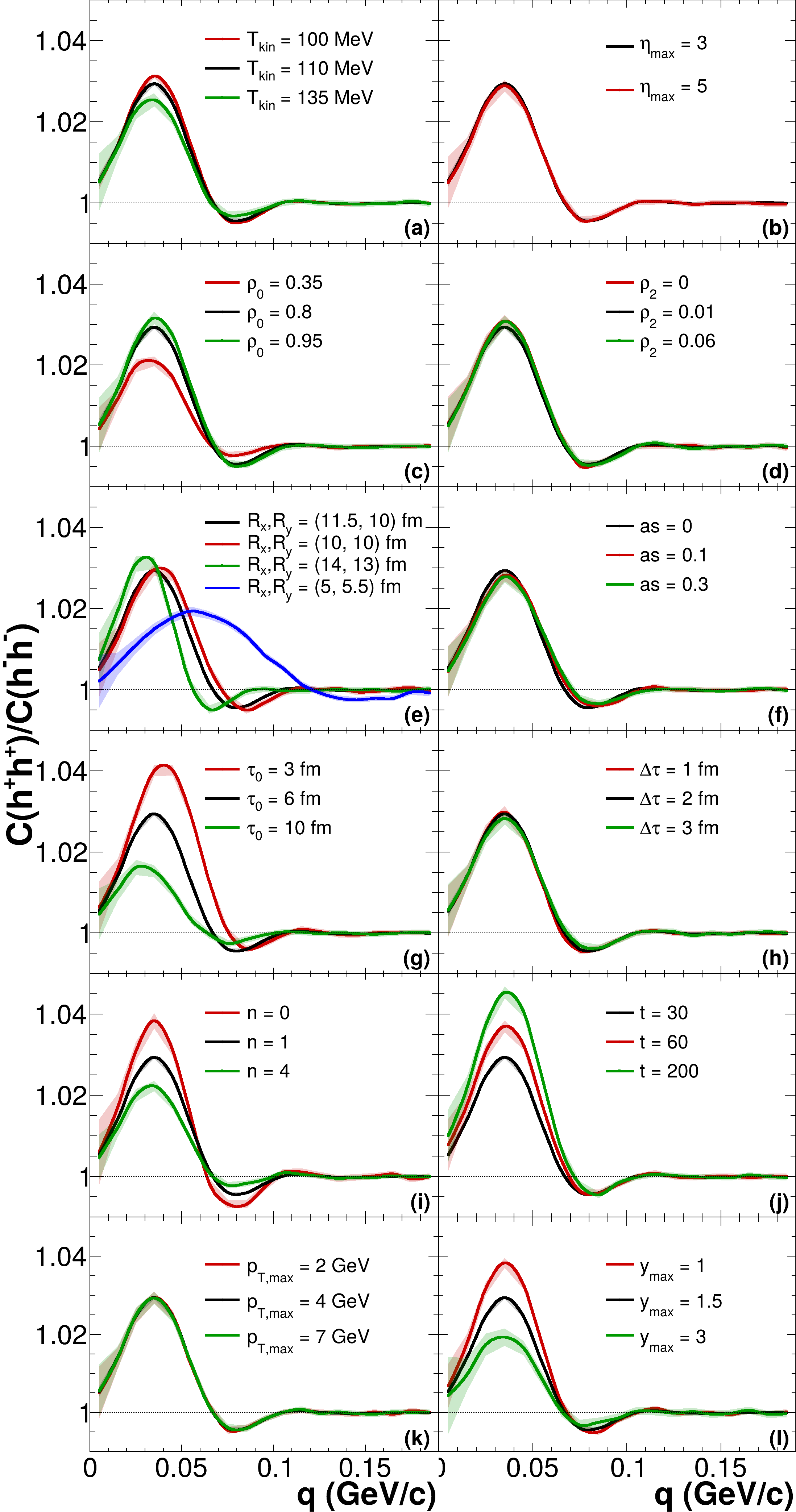}
\caption{
Systematic variations of the one-dimensional positive-to-negative pion
correlation-function ratio,
$C(\pi^+\pi^+)/C(\pi^-\pi^-)$, shown as a function of the invariant relative
momentum $q_{\mathrm{inv}}$. In each panel, one analysis parameter is varied
while all other parameters are kept at their default values. The black line
shows the default calculation, while the colored lines show the corresponding
variations. The horizontal line indicates unity. The shaded bands represent
statistical uncertainties.
}
    \label{fig:bw_ratios_pions}
\end{figure}

\begin{figure}[h!]
    \centering
    \includegraphics[width=1.0\linewidth]{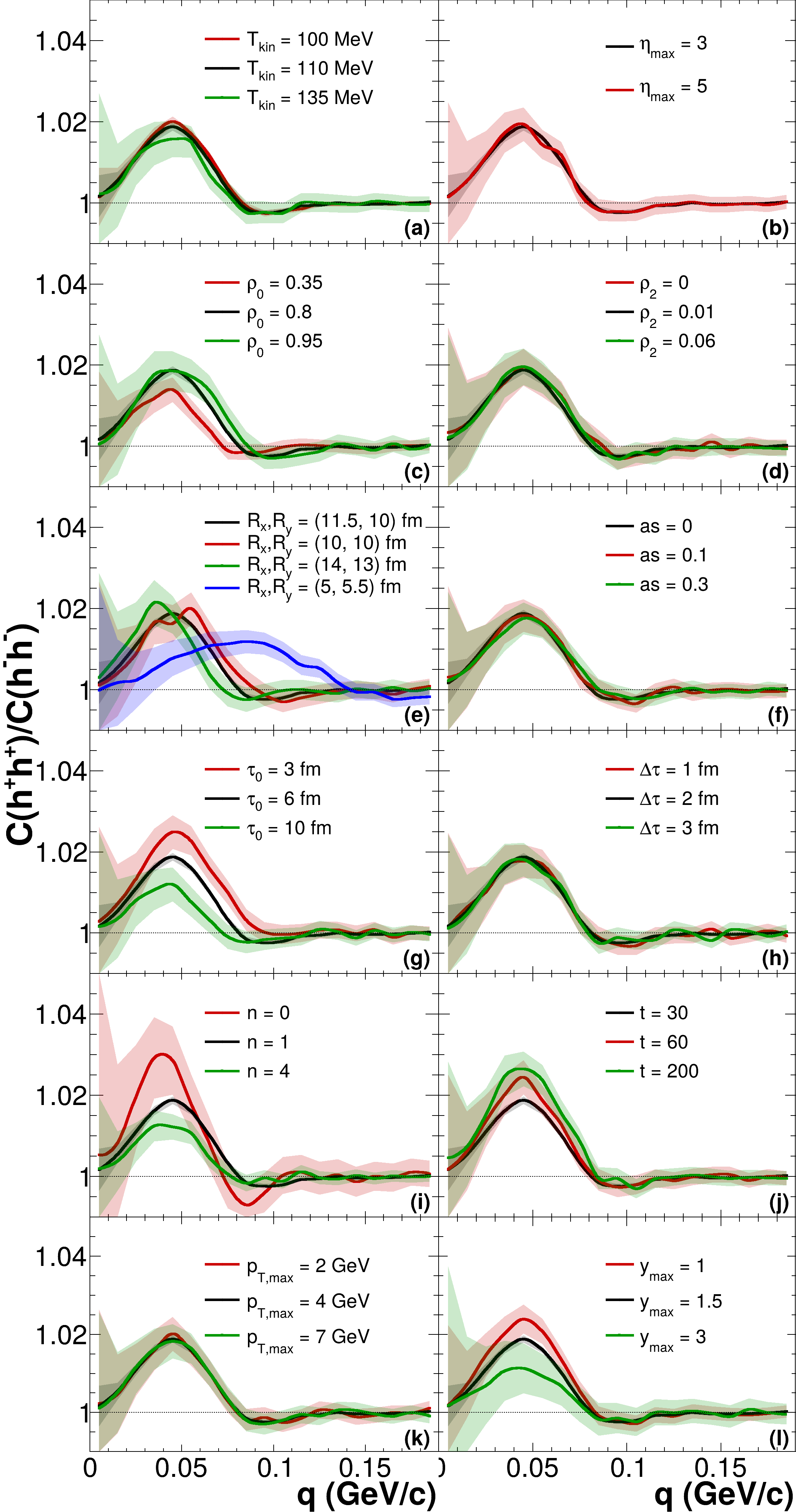}
\caption{
Same as Fig.~\ref{fig:bw_ratios_pions}, but for the kaon ratio
$C(K^+K^+)/C(K^-K^-)$.
}
    \label{fig:bw_ratios_kaons}
\end{figure}

Figures~\ref{fig:bw_ratios_pions} and~\ref{fig:bw_ratios_kaons} show how the
positive-to-negative correlation-function ratios respond to these variations
for pions and kaons, respectively. The discussion below focuses on the pion
case, where the trends are easier to see, while the corresponding kaon results
show the same variations for the heavier particle species.

The temperature variations test how the residual Coulomb effect changes when
particles decouple from a cooler or hotter source. A lower value of
$T_{\mathrm{kin}}$ can be associated with a cooler kinetic freeze-out, while a
higher value corresponds to stronger thermal smearing. The effect of varying
$T_{\mathrm{kin}}$ is shown in Fig.~\ref{fig:bw_ratios_pions}(a). The change is
visible but relatively small. The residual Coulomb distortion becomes slightly
stronger when the temperature is decreased. This is expected because a cooler
source produces particles with smaller thermal velocities on average. These
particles spend more time near the residual charge distribution and therefore
accumulate a larger Coulomb momentum kick.

The parameter $\eta_{\max}$ controls the longitudinal extent of the source in
space-time rapidity. The result of varying this parameter is shown in
Fig.~\ref{fig:bw_ratios_pions}(b). This variation tests whether the calculated
correlation-function ratio is sensitive to the finite longitudinal cutoff of
the blast-wave source. Since Eq.~\ref{eq:coord} relates $\eta_s$ to the
longitudinal coordinate $z$, increasing $\eta_{\max}$ extends the source in the
beam direction.

The comparison between the default value $\eta_{\max}=3$ and the larger value
$\eta_{\max}=5$ shows almost no change in the positive-to-negative ratio.
This indicates that the default value is already close to the extended-source
regime for the present setup, and that the calculated low-$k_T$ correlation
ratios are not strongly driven by the finite source-rapidity cutoff.

The variations of $\rho_0$ test the role of the average transverse collective
expansion. Physically, the radial flow strength reflects how efficiently
pressure gradients accelerate the system before kinetic freeze-out. A smaller
value of $\rho_0$ represents a source with weaker collective expansion, while a
larger value represents a source with stronger radial flow, as may occur for a
larger or longer-lived system. The effect of varying $\rho_0$ is shown in
Fig.~\ref{fig:bw_ratios_pions}(c). In the present setup, the residual Coulomb
distortion becomes smaller when $\rho_0$ is decreased. This indicates that the
charge-dependent splitting is sensitive not only to the time particles spend
near the residual source, but also to the initial position--momentum
correlations. Stronger radial flow creates a stronger connection between the
emission point and the particle momentum, so the residual Coulomb field can
modify the final relative-momentum distribution more visibly.

The parameter $\rho_2$ controls the azimuthal modulation of the transverse flow.
This modulation is connected to the conversion of the initial spatial
anisotropy of the collision zone into anisotropic collective expansion. The
case $\rho_2=0$ removes this modulation, while larger values test a source with
stronger elliptic-flow modulation. The effect of varying $\rho_2$ is shown in
Fig.~\ref{fig:bw_ratios_pions}(d). No sizable change is observed. This is
expected because the present analysis is integrated over the pair azimuthal
angle and the residual Coulomb field is taken to be spherically symmetric. As a
result, the second-harmonic modulation of the blast-wave flow is largely
averaged out. The weak dependence on $\rho_2$ therefore provides a useful
consistency check: it indicates that the observed positive-to-negative
splitting is controlled mainly by the radial expansion, emission time, and
residual Coulomb field, rather than by the elliptic modulation of the source.

The geometric variations of $(R_x,R_y)$ test three source-size scenarios in
addition to the default configuration: an approximately circular source, a
larger source, and a smaller source. These variations are useful because the
residual Coulomb kick depends on where the particles are emitted relative to
the residual charge distribution. They also change the femtoscopic width of the
correlation function, since a larger homogeneity region produces a narrower
correlation in relative momentum, while a smaller source produces a broader
correlation.

The effect of varying $(R_x,R_y)$ is shown in
Fig.~\ref{fig:bw_ratios_pions}(e). The approximately circular source, shown by
the red line, does not strongly change the amplitude of the positive-to-negative
ratio compared with the default case. This is expected because the overall
source size remains similar to the default configuration. The main visible
change is a shift of the structure toward slightly larger $q_{\mathrm{inv}}$,
which follows from the somewhat smaller effective source size and therefore a
broader correlation function.

The larger source, shown by the green line, produces a stronger deviation from
unity and shifts the structure toward smaller $q_{\mathrm{inv}}$. The shift is
consistent with the inverse relation between the source size and the width of
the femtoscopic correlation function: a larger emitting region gives a narrower
correlation in relative momentum. The increased amplitude can be understood
from the larger spread of emission points. Particles emitted from more
separated regions of the source experience different Coulomb kicks, which leads
to a stronger modification of the final relative-momentum distribution.

The strongest qualitative change is observed for the smaller source, shown by
the blue line. In this case, the effect is reduced in amplitude and shifted
toward larger $q_{\mathrm{inv}}$. The shift again follows from the smaller
geometric scale, which broadens the correlation function. The reduced amplitude
is consistent with the fact that, in a compact source, the two particles in a
pair are more likely to receive similar Coulomb kicks. A nearly common momentum
shift affects the single-particle momenta, but it has a smaller impact on the
relative momentum of the pair, which is the quantity entering the correlation
function.

The surface diffuseness parameter $a_s$ is varied to check how sensitive the
effect is to replacing a hard-edged source by a smoother transverse density
profile. The result is shown in Fig.~\ref{fig:bw_ratios_pions}(f). The
dependence on $a_s$ is small. As the surface becomes more diffuse, the
structure in the ratio is slightly smeared and shifted toward larger
$q_{\mathrm{inv}}$.

This behavior is expected because increasing $a_s$ softens the sharp transverse
edge of the source. Instead of emitting particles from a source with a sharply
defined boundary, the emission probability decreases more gradually around the
nominal radius. This changes the effective homogeneity region and slightly
shifts the structure in relative momentum. The small size of the observed
change indicates that the residual Coulomb effect is not strongly controlled by
the details of the source surface profile for the parameter range studied here.

The time parameters $\tau_0$ and $\Delta\tau$ control when particles are
emitted and over what duration. The parameter $\tau_0$ is the mean proper time
of kinetic freeze-out. Varying it therefore tests different freeze-out
scenarios: a smaller $\tau_0$ corresponds to earlier particle emission, while a
larger $\tau_0$ corresponds to later emission from a more evolved source. This
is important for the residual Coulomb effect because the charge distribution is
allowed to expand with time. Particles emitted earlier begin their propagation
when the residual charge cloud is still more compact and the Coulomb field is
stronger. Particles emitted later start from a configuration in which the
residual charge distribution is already more diluted.

The effect of varying $\tau_0$ is shown in
Fig.~\ref{fig:bw_ratios_pions}(g). The residual Coulomb distortion is strongest
for the smallest value of $\tau_0$ and becomes weaker as $\tau_0$ increases.
This behavior is consistent with the picture described above: earlier-emitted
particles spend more time in a stronger residual field and therefore accumulate
a larger momentum kick.

The parameter $\Delta\tau$ controls the duration of emission around the mean
freeze-out time. A smaller $\Delta\tau$ corresponds to a more sudden
freeze-out, while a larger $\Delta\tau$ describes a broader emission period.
The effect of varying $\Delta\tau$ is shown in
Fig.~\ref{fig:bw_ratios_pions}(h). In contrast to the strong dependence on
$\tau_0$, the dependence on $\Delta\tau$ is small for the range studied here.
This indicates that the average emission time is more important for the
residual Coulomb distortion than the width of the emission-time distribution,
at least for the present default setup.

The radial flow exponent $n$ controls how quickly the transverse flow increases
with the dimensionless radius $\tilde r$. The default value $n=1$ corresponds
to a linear radial dependence. The variation $n=0$ gives a nearly flat flow
profile, where the transverse flow strength is much less dependent on radius,
while $n=4$ produces a much steeper profile, with most of the flow concentrated
near the outer part of the source. This variation tests how the radial structure
of the position--momentum correlations affects the residual Coulomb distortion.

The result is shown in Fig.~\ref{fig:bw_ratios_pions}(i). The effect is largest
for the flatter flow profile and becomes smaller for the steeper profile. This
can be understood from the way the flow field organizes the initial particle
momenta. When the flow profile is flatter, particles emitted from different
radii have more similar collective velocities, so the residual Coulomb kick can
produce a larger relative change in the final momentum distribution. For a
steeper profile, the strong radial dependence of the initial flow already
separates particles according to their emission radius, making the additional
Coulomb-induced modification less pronounced.

The numerical sampling and propagation variations are used as stability checks.
Increasing $t_{\max}$ tests whether the Coulomb momentum kick has already
saturated within the default propagation time. The result is shown in
Fig.~\ref{fig:bw_ratios_pions}(j). The change from
$t_{\max}=30~\mathrm{fm}/c$ to larger values increases the size of the effect,
which indicates that some additional Coulomb acceleration is still accumulated
after the default propagation time. However, the shape of the ratio remains
similar, suggesting that the default value already captures the main qualitative
behavior of the distortion.

The parameters $p_{T,\max}$ and $Y_{\max}$ define the phase-space region used to
propose particles in the Monte Carlo sampling. Their variations are shown in
Figs.~\ref{fig:bw_ratios_pions}(k) and~\ref{fig:bw_ratios_pions}(l),
respectively. The ratio is essentially unchanged when $p_{T,\max}$ is varied,
showing that the low-$k_T$ correlation functions are not driven by the upper
limit of the transverse-momentum proposal. This is expected because the
analysis focuses on low-$k_T$ pairs, while the varied $p_{T,\max}$ values remain
well above the momentum region that dominates the accepted pairs.

The dependence on $Y_{\max}$ is more visible. A smaller rapidity proposal range
selects particles closer to midrapidity, while a larger value allows a broader
range of longitudinal momenta. Since the residual Coulomb distortion depends on
the particle trajectory and on the time spent near the residual source, changing
the sampled rapidity range can modify the positive-to-negative ratio. This
variation should therefore be interpreted as a sampling acceptance check: it
tests how sensitive the calculated low-$k_T$ ratios are to the range of
particle rapidities included in the Monte Carlo sample.

\subsection{Variation of residual Coulomb-field parameters}

The second group of variations concerns the residual Coulomb field itself.
These parameters control the charge distribution through which the emitted
particles propagate after freeze-out. The initial Gaussian width
$\sigma_0$ determines how concentrated or diffuse the residual charge
distribution is at the beginning of the propagation, the expansion velocity
$v_{\mathrm{exp}}$ controls how quickly the charge cloud becomes diluted with
time, and the effective residual charge $Z_{\mathrm{res}}$ sets the overall
strength of the field.

The residual-field parameters varied in this study are summarized in
Table~\ref{tab:coulomb_variations}. As in the blast-wave variations, only one
parameter is changed at a time, while the remaining parameters are kept at their
default values unless stated otherwise.

\begin{table}[h!]
\centering
\caption{Residual Coulomb-field parameters varied in the systematic study. For
each row, only the listed parameter is changed, while the remaining parameters
are kept at their default values from Table~\ref{tab:default_parameters}. The
expansion parameter $v_{\mathrm{exp}}$ is given in units of $c$.}
\label{tab:coulomb_variations}
\begin{tabular}{lll}
\hline
Parameter & Default value & Varied values \\
\hline
$Z_{\mathrm{res}}$ & $60$                 & $10,\;40,\;80$ \\
$\sigma_0$         & $5.0~\mathrm{fm}$    & $1,\;3,\;10~\mathrm{fm}$ \\
$v_{\mathrm{exp}}$ & $0.3$                & $0.01,\;0.1,\;0.2,\;0.5$ \\
\hline
\end{tabular}
\end{table}

\begin{figure}[h!]
    \centering
    \includegraphics[width=1.0\linewidth]{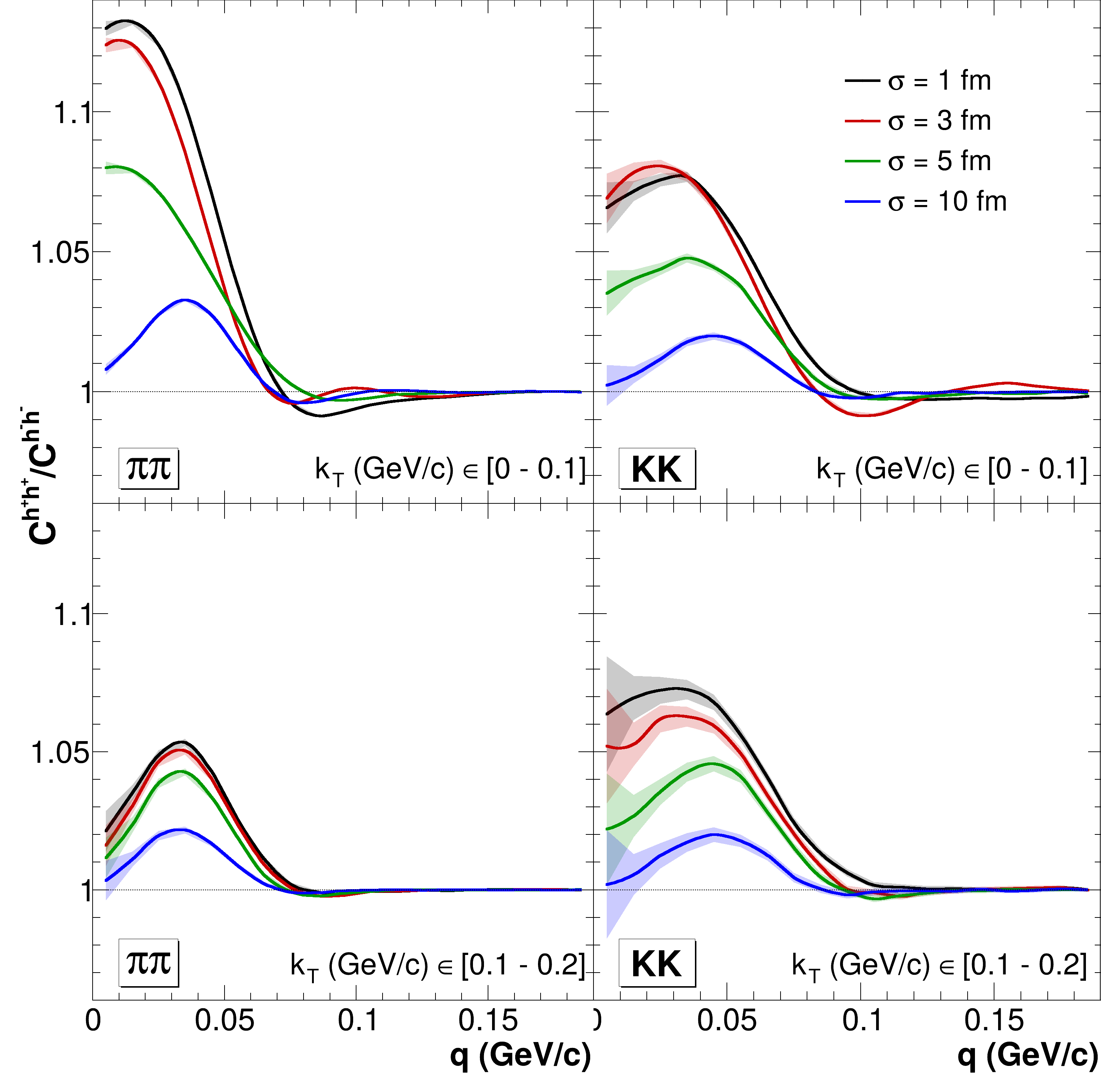}
    \caption{
Effect of varying the initial width $\sigma_0$ of the residual charge
distribution on the positive-to-negative correlation-function ratio,
$C(h^+h^+)/C(h^-h^-)$, shown as a function of the invariant relative momentum
$q_{\mathrm{inv}}$. Results are shown for pions in the left column and kaons
in the right column. The top row corresponds to
$0<k_T<0.1~\mathrm{GeV}/c$, while the bottom row corresponds to
$0.1<k_T<0.2~\mathrm{GeV}/c$. The different colored lines show calculations
with different values of $\sigma_0$. The shaded bands represent statistical
uncertainties.
}
    \label{fig:cf_rat_charge_sigma}
\end{figure}

\begin{figure}[h!]
    \centering
    \includegraphics[width=1.0\linewidth]{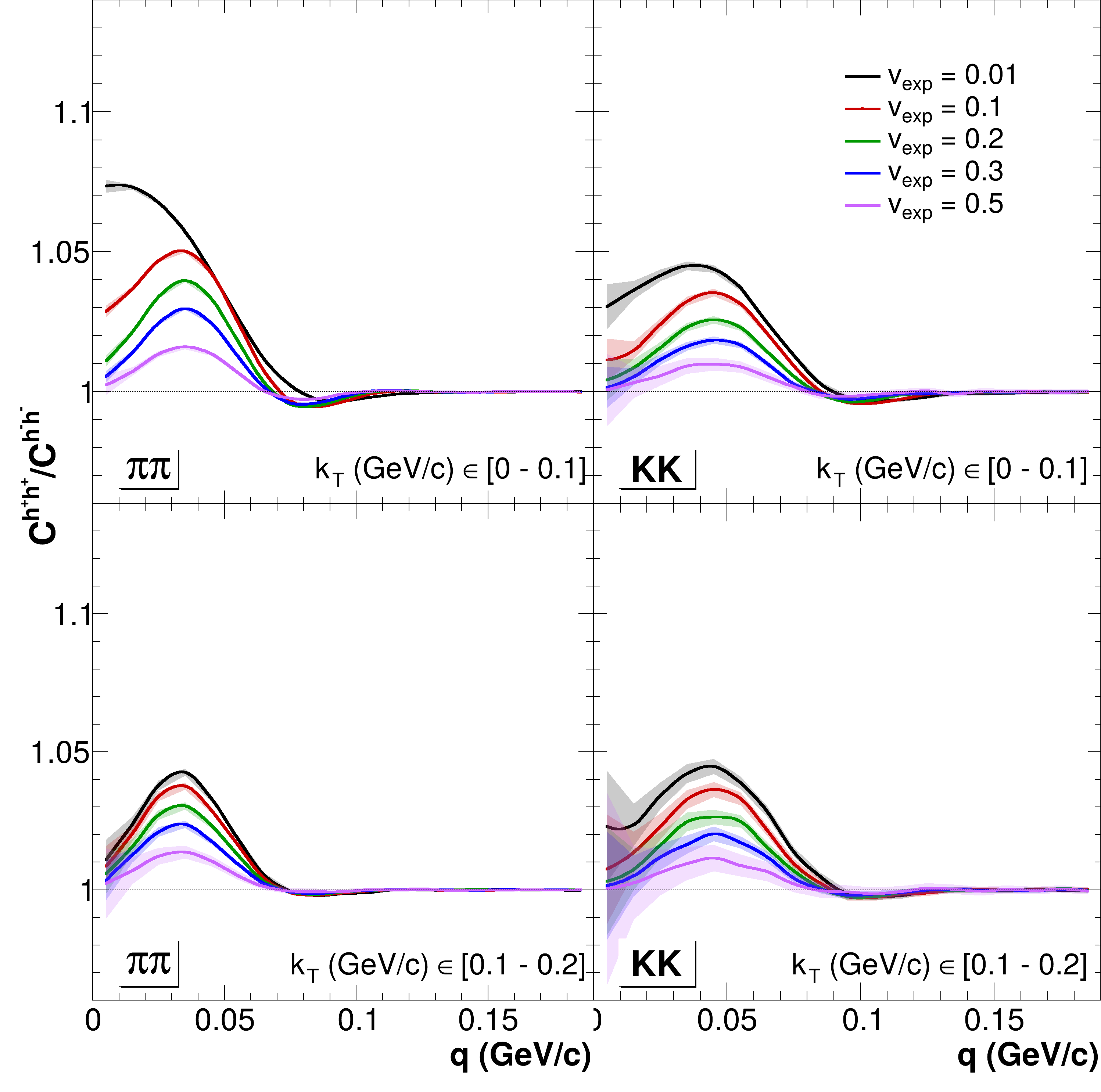}
    \caption{
Same as Fig.~\ref{fig:cf_rat_charge_sigma}, but for different expansion
velocities of the residual charge distribution, $v_{\mathrm{exp}}$.
}
    \label{fig:cf_rat_charge_velocity}
\end{figure}

\begin{figure}[h!]
    \centering
    \includegraphics[width=1.0\linewidth]{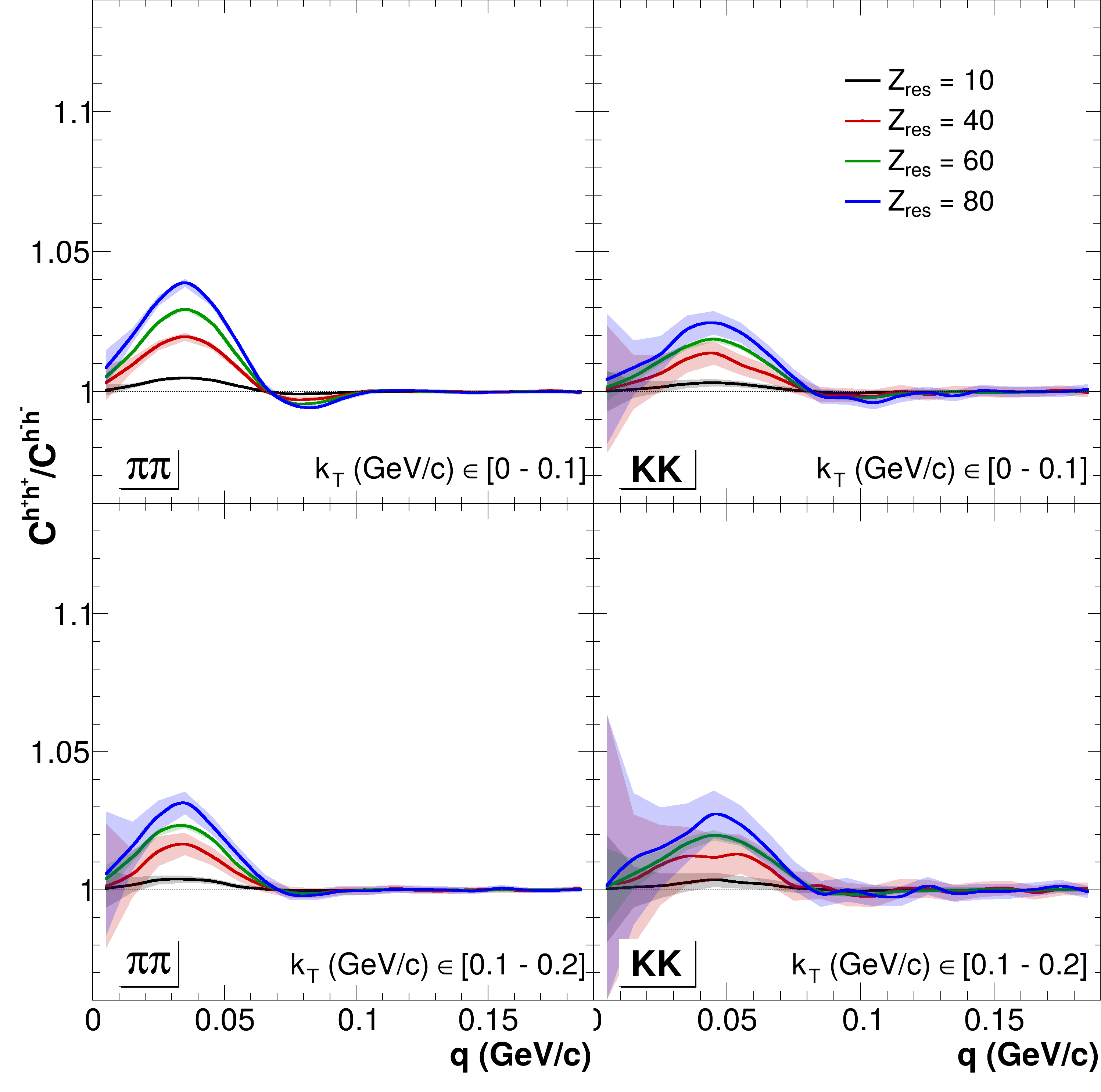}
    \caption{
Same as Figs.~\ref{fig:cf_rat_charge_sigma} and
\ref{fig:cf_rat_charge_velocity}, but for different effective residual charges,
$Z_{\mathrm{res}}$.
}
    \label{fig:cf_rat_charge_z}
\end{figure}

Figures~\ref{fig:cf_rat_charge_sigma}--\ref{fig:cf_rat_charge_z} show the
ratios of the one-dimensional correlation functions for positively charged
pairs to those for negatively charged pairs. In each figure, the results are
shown for the first two pair-transverse-momentum intervals, in order to
illustrate how quickly the effect changes with $k_T$. The left panels
correspond to pions, while the right panels correspond to kaons. Different
colors and bands represent different values of the varied residual-field
parameter.

Among the variations considered here, the strongest shape change is observed
when the initial width of the residual charge distribution is varied, as shown
in Fig.~\ref{fig:cf_rat_charge_sigma}. This is expected because $\sigma_0$
controls how concentrated the residual charge is around the emission region. A
smaller $\sigma_0$ corresponds to a more compact charge distribution and
therefore to a stronger Coulomb field near the center of the source. A larger
$\sigma_0$ spreads the same effective charge over a wider volume, making the
field more diffuse and reducing the momentum kick received by the emitted
particles.

Physically, different values of $\sigma_0$ may correspond to different
space-time configurations of the residual charged source. The charge
distribution can depend on the collision centrality, collision energy, amount
of longitudinal stopping, freeze-out time, and the degree of expansion before
particle emission. More central or longer-lived systems may produce a more
extended residual charge cloud, while a more compact or earlier-emitting system
may correspond to a smaller effective width. Thus, varying $\sigma_0$ tests how
sensitive the charge-dependent correlation splitting is to the assumed spatial
size of the residual field.

The $k_T$ dependence is also different for pions and kaons. At the same $k_T$,
pions have a larger velocity because of their smaller mass. As $k_T$ increases,
they leave the region of strong Coulomb field more quickly, and the accumulated
momentum kick decreases rapidly. Kaons are heavier and move more slowly at the
same $k_T$, so they remain near the residual source for a longer time and show a
weaker $k_T$ dependence.

The variations of $Z_{\mathrm{res}}$ and $v_{\mathrm{exp}}$ follow the same
general physical trend. Increasing $Z_{\mathrm{res}}$ strengthens the Coulomb
field and increases the positive-to-negative splitting of the correlation
functions, while decreasing $Z_{\mathrm{res}}$ weakens the effect. The expansion
velocity controls how quickly the residual charge distribution becomes diluted:
a slower expansion keeps the charge cloud more compact for a longer time,
allowing particles to accumulate a larger momentum kick, while a faster
expansion reduces the field more quickly and weakens the charge-dependent
distortion.

Overall, the correlation-function splitting becomes larger when the residual
Coulomb field is stronger, more compact, or longer lived. These variations show
that the observed charge-dependent effect is controlled not only by the total
effective charge, but also by the spatial size and time evolution of the
residual charged source.

\begin{figure}[h!]
    \centering
    \includegraphics[width=1.0\linewidth]{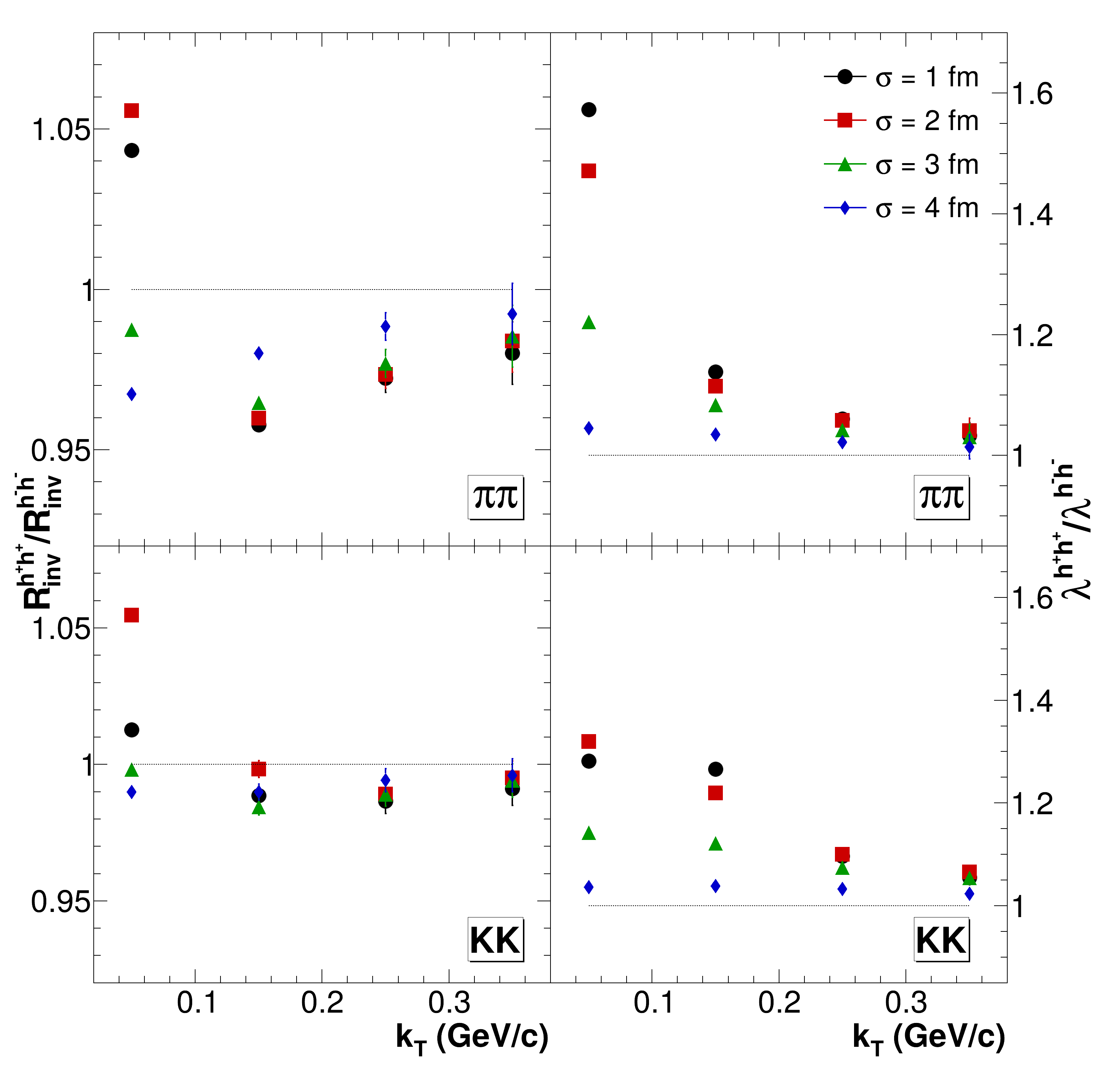}
    \caption{
Effect of varying the initial width $\sigma_0$ of the residual charge
distribution on the extracted one-dimensional femtoscopic-parameter ratios.
The left column shows
$R_{\mathrm{inv}}^{h^+h^+}/R_{\mathrm{inv}}^{h^-h^-}$, while the right column
shows $\lambda^{h^+h^+}/\lambda^{h^-h^-}$. The top row corresponds to pions
and the bottom row to kaons. Different marker colors correspond to different
values of $\sigma_0$. The horizontal line indicates unity.
}
    \label{fig:radii_rat_charge_sigma}
\end{figure}

Figure~\ref{fig:radii_rat_charge_sigma} shows an example of how the extracted
one-dimensional femtoscopic parameters, $R_{\mathrm{inv}}$ and $\lambda$, are
affected by varying the initial width of the residual charge distribution. The
results are shown for both pions and kaons.

For some values of $\sigma_0$, the radius ratios do not follow the simple trend
suggested by the direct correlation-function ratios. One might expect a
correlation-function ratio above unity to correspond to a radius ratio below
unity, because a narrower positive-pair correlation function would generally
lead to a smaller extracted radius. However, in some cases the fitted
$R_{\mathrm{inv}}$ ratio becomes larger than unity.

This behavior indicates that the residual Coulomb distortion changes more than
just the Gaussian width of the correlation function. It also modifies the height
of the correlation peak, as reflected by the change in the fitted $\lambda$
parameter. More generally, varying the residual-source parameters can make the
correlation function less Gaussian. In such cases, the Gaussian fit parameters
should be interpreted with caution: they still provide a useful compact
comparison between positive and negative pairs, but they may not fully describe
the shape change caused by the residual Coulomb field.


\section{Isospin effect and charge-dependent dynamics in UrQMD}

The previous sections isolated the effect of the residual Coulomb field using a
controlled blast-wave calculation. In that study, the positive and negative
samples were generated from the same freeze-out distribution, and the only
difference between them was the sign of the post-emission Coulomb force. The
initial space-time and momentum distributions of positive and negative
particles were therefore identical by construction.

In this section we ask a different question: can a positive-to-negative
difference appear even when no residual Coulomb propagation is applied? In a
microscopic transport model, positive and negative particles are produced
dynamically and are not forced to have identical emission distributions. Their
space-time and momentum distributions can differ because of the isospin content
of the initial state, resonance production and decay, absorption, and hadronic
rescattering. Such differences can then lead to different identical-particle
correlation functions even without an explicit residual Coulomb field.

To test this, an additional study was performed with UrQMD 3.4 in cascade
mode~\cite{Bass:1998ca,Bleicher:1999xi}. UrQMD provides a microscopic
transport description of the collision and includes hadron production,
resonance decays, and hadronic rescattering.

The baseline UrQMD results shown in this section correspond to Au+Au collisions
at $\sqrt{s_{NN}}=7.7~\mathrm{GeV}$. This collision energy was chosen because it
is a region where several charge-dependent effects may be relevant at the same
time, including isospin effects, baryon stopping, resonance contributions, and
hadronic rescattering. It is also experimentally accessible, which makes it a
useful case for future comparison with data. To make the comparison more
realistic, a simple detector-like pseudorapidity acceptance cut,
$|\eta|<1$, is applied.

In this UrQMD study, no post-emission residual Coulomb propagation is applied.
The usual two-body Coulomb and strong femtoscopic final-state interactions are
also not included. The numerator of the correlation function is filled only with
the plane-wave Bose--Einstein symmetrization weight defined in
Eq.~\ref{eq:w_BE}, while the denominator is filled with unit weight. The same
weighting procedure is used for the three-dimensional correlation functions in
the Bertsch--Pratt variables.

Therefore, any observed difference between $\pi^+\pi^+$ and $\pi^-\pi^-$, or
between $K^+K^+$ and $K^-K^-$, must originate from the particle production and
hadronic dynamics already present in UrQMD, rather than from an explicit pair
final-state interaction or from the residual three-body Coulomb field considered
above.

\begin{figure}[h!]
    \centering
    \includegraphics[width=1\linewidth]{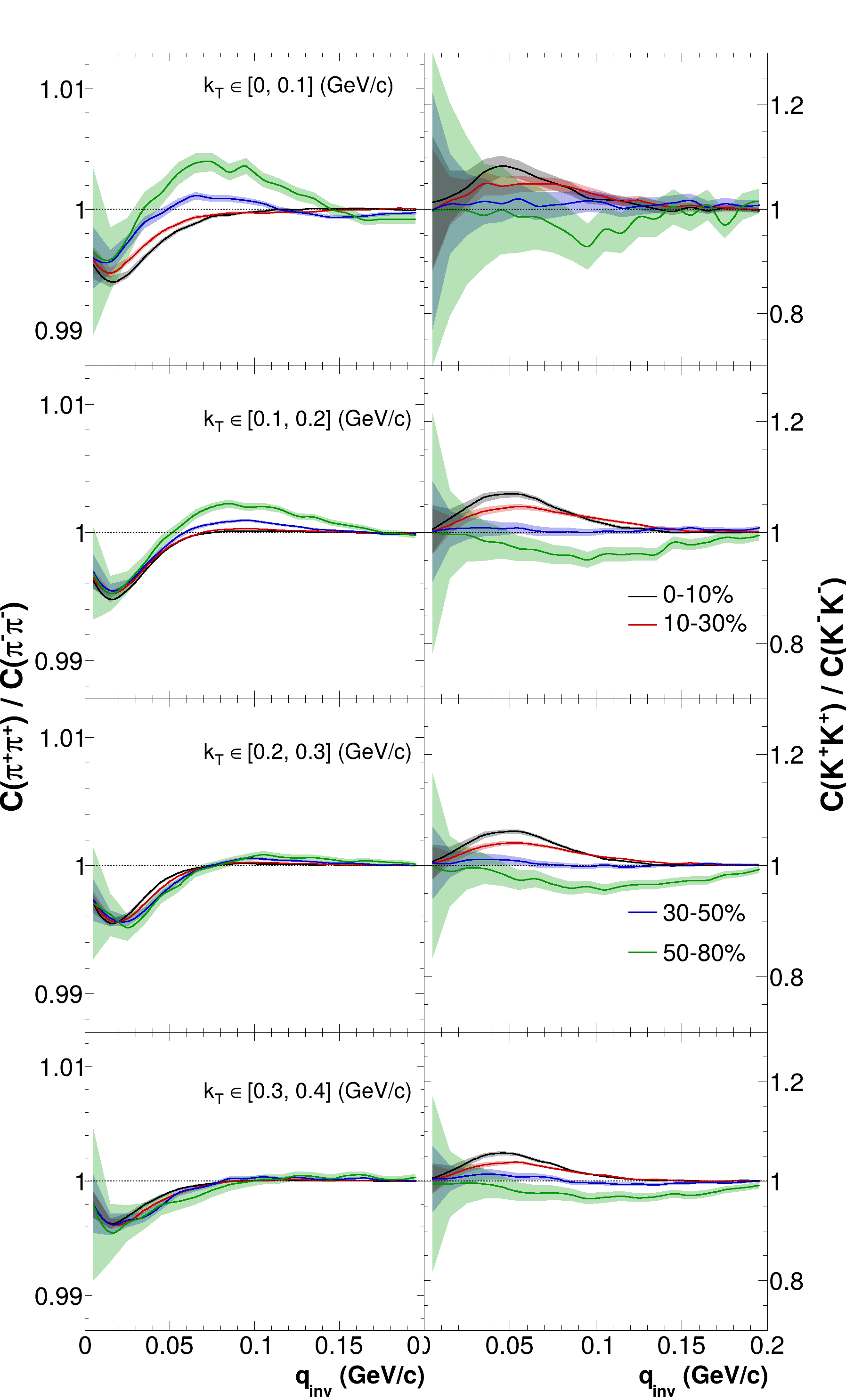}
    \caption{
Positive-to-negative correlation-function ratios from UrQMD 3.4 in cascade
mode, shown as functions of the invariant relative momentum $q_{\mathrm{inv}}$
for different centrality classes and transverse-momentum intervals. The left
column shows the pion ratio $C(\pi^+\pi^+)/C(\pi^-\pi^-)$, while the right
column shows the kaon ratio $C(K^+K^+)/C(K^-K^-)$. From top to bottom, the rows
correspond to $k_T \in [0,0.1]$, $[0.1,0.2]$, $[0.2,0.3]$, and
$[0.3,0.4]~\mathrm{GeV}/c$. Different colored lines indicate different
centrality intervals: 0--10\%, 10--30\%, 30--50\%, and 50--80\%. The shaded
bands represent statistical uncertainties, and the horizontal dashed line
indicates unity.
}
    \label{fig:urqmd_kt_cent}
\end{figure}

Figure~\ref{fig:urqmd_kt_cent} shows the result of this check for both particle
species: pions in the left column and kaons in the right column. A
positive-to-negative difference is visible even though no additional residual
Coulomb propagation is applied. This is an important observation, because
charge-dependent differences in such correlation functions are often interpreted
or assumed to originate mainly from the Coulomb field of the residual charged
source~\cite{HADES:2019lek}. The UrQMD result shows that this interpretation is not unique:
a difference between positive and negative pairs can also be generated by the
microscopic particle production and hadronic evolution present in the transport
model.

Interestingly, the direction of the effect is different for pions and kaons.
For pions, the positive-to-negative difference goes in the opposite direction to
the one obtained in the blast-wave calculation with residual Coulomb
propagation. For kaons, the effect is mostly in the same direction as in the
residual-Coulomb study. This suggests that the observed charge dependence is
not controlled by a single mechanism. Instead, different contributions, such as
isospin, resonance production and decay, hadronic rescattering, and possible
Coulomb effects, may compete with each other and can affect pions and kaons in
different ways.

The different colors in Fig.~\ref{fig:urqmd_kt_cent} correspond to different
collision centralities. In this UrQMD study, the centrality classes are defined
using impact-parameter intervals, as listed in
Table~\ref{tab:urqmd_centrality}. The mapping between centrality and impact
parameter was performed using a Glauber nuclear-overlap calculation, following
the standard geometric interpretation of centrality as a percentile of the total
inelastic nucleus--nucleus cross section~\cite{Miller:2007ri}.

\begin{table}[h!]
\centering
\caption{Centrality classes used in the UrQMD study, defined by impact
parameter intervals.}
\label{tab:urqmd_centrality}
\begin{tabular}{cc}
\hline
Centrality class & Impact parameter interval \\
\hline
$0$--$10\%$  & $0.0 < b < 4.7~\mathrm{fm}$ \\
$10$--$30\%$ & $4.7 < b < 8.1~\mathrm{fm}$ \\
$30$--$50\%$ & $8.1 < b < 10.4~\mathrm{fm}$ \\
$50$--$80\%$ & $10.4 < b < 13.2~\mathrm{fm}$ \\
\hline
\end{tabular}
\end{table}

This definition is sufficient for the present model study, where the goal is to
demonstrate that a positive-to-negative difference can arise from the hadronic
dynamics in UrQMD itself. For a direct comparison with experimental data, a
closer matching of the centrality definition would be needed. In particular,
experiments usually define centrality using a reference multiplicity or a
related event-activity measure rather than the true impact parameter. Therefore,
a future data comparison should determine the UrQMD centrality classes using the
same reference-multiplicity procedure as in the experiment.

The next question is where this effect comes from. One natural possibility is
that it is related to isospin. In a neutron-rich system, such as Au+Au, the
participant matter contains more neutrons than protons. Since the strong
interaction approximately conserves isospin, this initial neutron-to-proton
asymmetry can influence the relative population of different hadronic charge
channels during the collision. In particular, pion production in the hadronic
stage is strongly connected to baryon-resonance dynamics, especially through
intermediate $\Delta$ states. The charge composition of the participant
matter can therefore affect the relative production of $\Delta^{-}$,
$\Delta^{0}$, $\Delta^{+}$, and $\Delta^{++}$ resonances, which subsequently
feed different charged-pion channels. As a result, the initial isospin
imbalance can survive through the hadronic evolution and appear in the final
$\pi^+$ and $\pi^-$ yields, spectra, and space-time emission distributions.

This mechanism is well known in studies of charged-pion production in
neutron-rich heavy-ion collisions, where the $\pi^-/\pi^+$ ratio can be used
as an isospin-sensitive observable. In the present work, we are not
using the pion yield ratio itself as the observable. Instead, we ask whether the
same isospin-driven differences in the underlying $\pi^+$ and $\pi^-$ emission
distributions can also lead to a difference between the identical-pion
correlation functions, $\pi^+\pi^+$ and $\pi^-\pi^-$.

\begin{figure}[h!]
    \centering
    \includegraphics[width=1.0\linewidth]{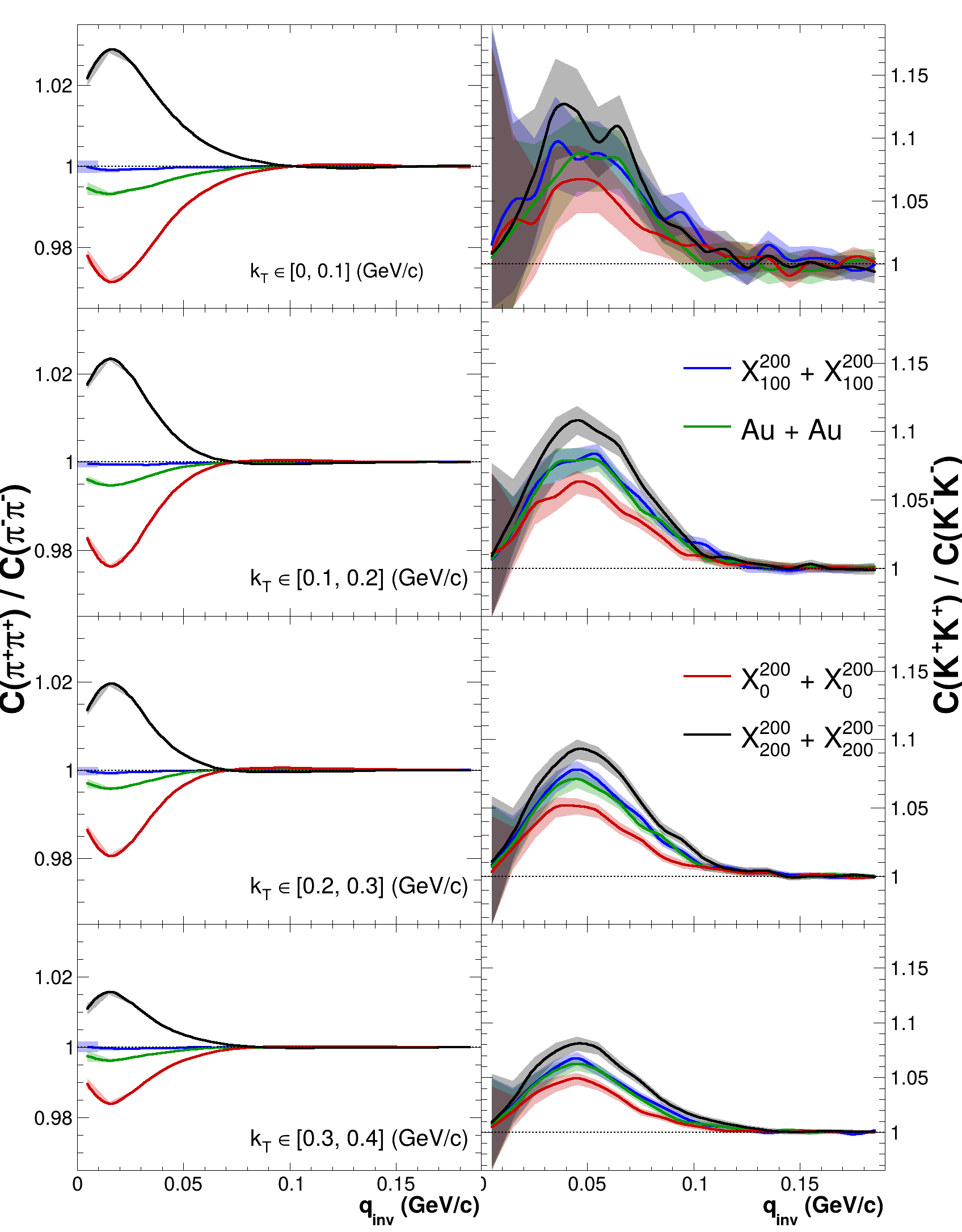}
    \caption{
Ratios of positive-pair to negative-pair one-dimensional correlation functions,
shown as functions of $q_{\mathrm{inv}}$ for several collision systems at
$\sqrt{s_{NN}}=7.7~\mathrm{GeV}$. The pion ratios
$C(\pi^+\pi^+)/C(\pi^-\pi^-)$ are shown in the left column, while the kaon
ratios $C(K^+K^+)/C(K^-K^-)$ are shown in the right column. Green triangles
denote ${}^{197}_{79}\mathrm{Au}+{}^{197}_{79}\mathrm{Au}$ collisions, black
circles denote ${}^{200}_{200}X+{}^{200}_{200}X$, red squares denote
${}^{200}_{0}X+{}^{200}_{0}X$, and blue diamonds denote
${}^{200}_{100}X+{}^{200}_{100}X$. From top to bottom, the panels correspond to
different pair-transverse-momentum intervals.
}
    \label{fig:1D_diffsystems}
\end{figure}

\begin{figure}[h!]
    \centering
    \includegraphics[width=1.0\linewidth]{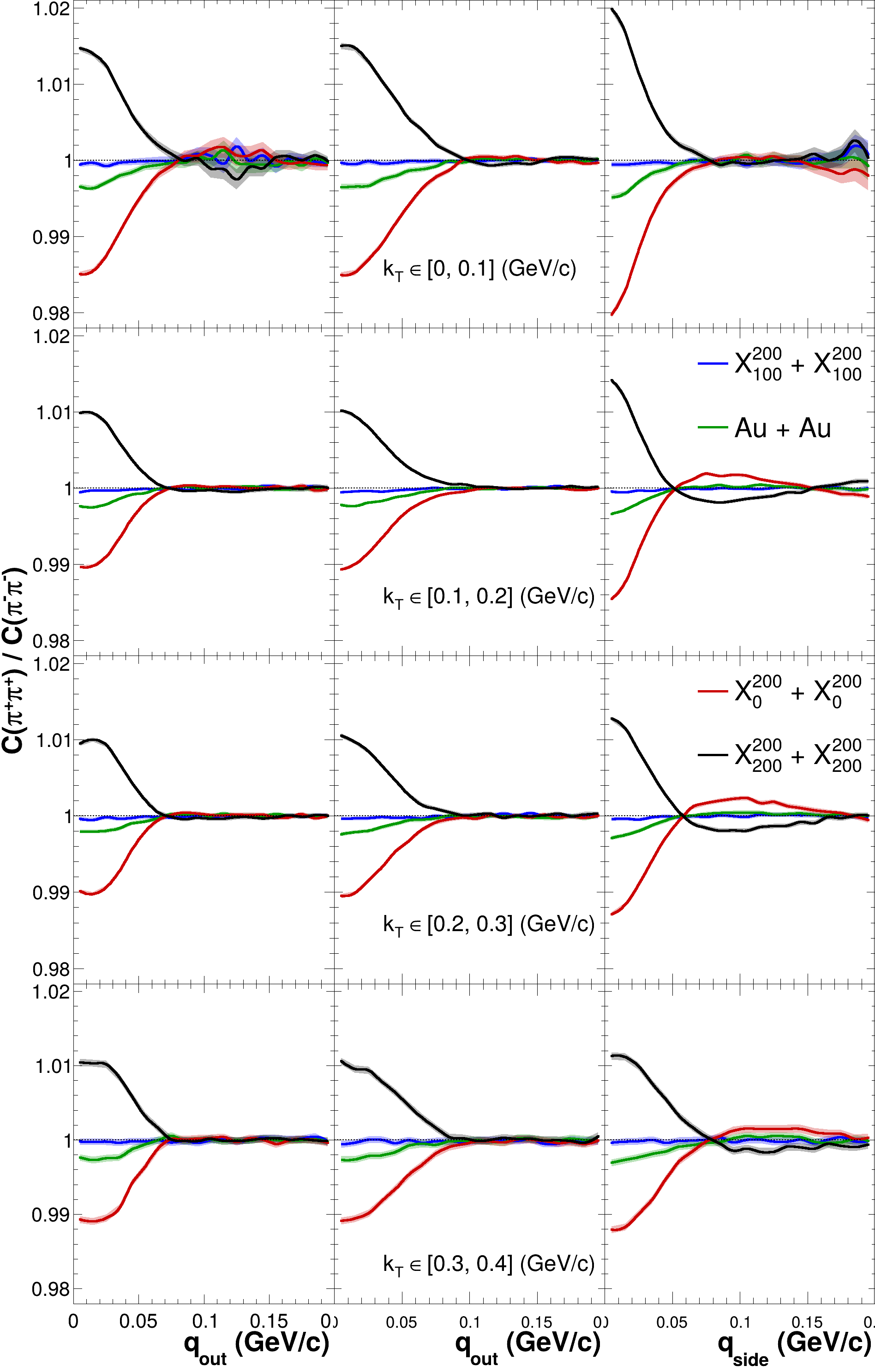}
    \caption{
Ratios of the like-sign two-pion correlation functions,
$C(\pi^+\pi^+)/C(\pi^-\pi^-)$, for several collision systems at
$\sqrt{s_{NN}}=7.7~\mathrm{GeV}$, projected onto the $q_{\mathrm{out}}$,
$q_{\mathrm{side}}$, and $q_{\mathrm{long}}$ directions. Green triangles denote
${}^{197}_{79}\mathrm{Au}+{}^{197}_{79}\mathrm{Au}$ collisions, black circles
denote ${}^{200}_{200}X+{}^{200}_{200}X$, red squares denote
${}^{200}_{0}X+{}^{200}_{0}X$, and blue diamonds denote
${}^{200}_{100}X+{}^{200}_{100}X$. From top to bottom, the panels correspond to
different pair-transverse-momentum intervals; from left to right, they show the
$q_{\mathrm{out}}$, $q_{\mathrm{side}}$, and $q_{\mathrm{long}}$ projections.
}
    \label{fig:3D_diffsystems}
\end{figure}

To isolate this contribution, four collision systems were generated. In addition
to realistic Au+Au collisions, artificial nuclei with the same mass number but
different charge content were used:
\begin{itemize}
\item $X^{200}_{100}+X^{200}_{100}$, an isospin-symmetric system with
$Z=N=100$;
\item Au+Au, a realistic neutron-rich heavy-ion system;
\item $X^{200}_{0}+X^{200}_{0}$, an extreme neutron-only system;
\item $X^{200}_{200}+X^{200}_{200}$, an extreme proton-only system.
\end{itemize}
Here the superscript denotes the mass number and the subscript denotes the
charge of the artificial nucleus. These artificial systems are not intended to
represent realistic nuclei in all details. Instead, they provide controlled
limits that help test how the charge and isospin composition of the initial
state can influence the final pion correlation functions.

Figures~\ref{fig:1D_diffsystems} and~\ref{fig:3D_diffsystems} show the ratios
of positive-pair to negative-pair correlation functions obtained from the UrQMD
samples with different initial isospin compositions. Figure~\ref{fig:1D_diffsystems}
shows the one-dimensional $q_{\mathrm{inv}}$ ratios for pions and kaons, while
Fig.~\ref{fig:3D_diffsystems} shows the corresponding three-dimensional
projections for pions.

The artificial systems show a clear dependence on the charge composition of the
colliding nuclei. The neutron-only and proton-only systems produce the largest
deviations from unity, with opposite signs. The isospin-symmetric system,
$X^{200}_{100}+X^{200}_{100}$, stays much closer to unity, as expected when the
initial proton and neutron numbers are balanced. The realistic Au+Au result
lies between these limiting cases and shows a smaller but still visible
deviation from unity. This indicates that, within UrQMD, the positive-to-negative
difference for pions is driven mainly by the isospin content of the initial
state.

\begin{figure}[h!]
    \centering
    \includegraphics[width=1.0\linewidth]{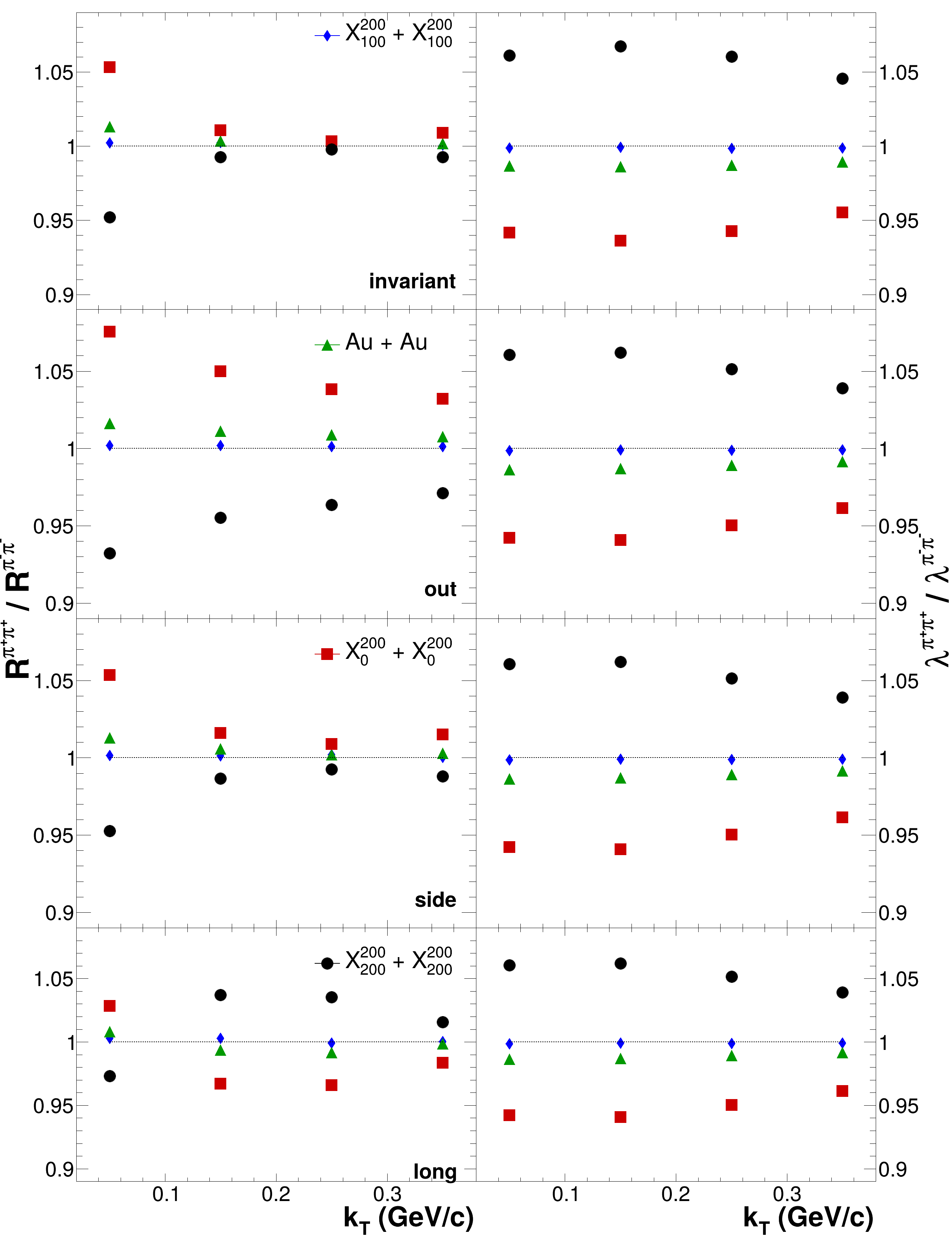}
    \caption{
Ratios of the extracted pion femtoscopic parameters for positive pion pairs to
the corresponding parameters for negative pion pairs, shown as functions of
transverse momentum $k_T$ for different colliding systems. The left column
shows the radius ratios,
$R^{\pi^+\pi^+}/R^{\pi^-\pi^-}$, while the right column shows the
corresponding intercept-parameter ratios,
$\lambda^{\pi^+\pi^+}/\lambda^{\pi^-\pi^-}$. From top to bottom, the rows
correspond to the invariant, out, side, and long components. The different
markers indicate different collision systems, including Au+Au and the
corresponding artificial systems used to separate charge and isospin effects.
The horizontal line indicates unity.
}
    \label{fig:radii_pions_ur}
\end{figure}

\begin{figure}[h!]
    \centering
    \includegraphics[width=1.0\linewidth]{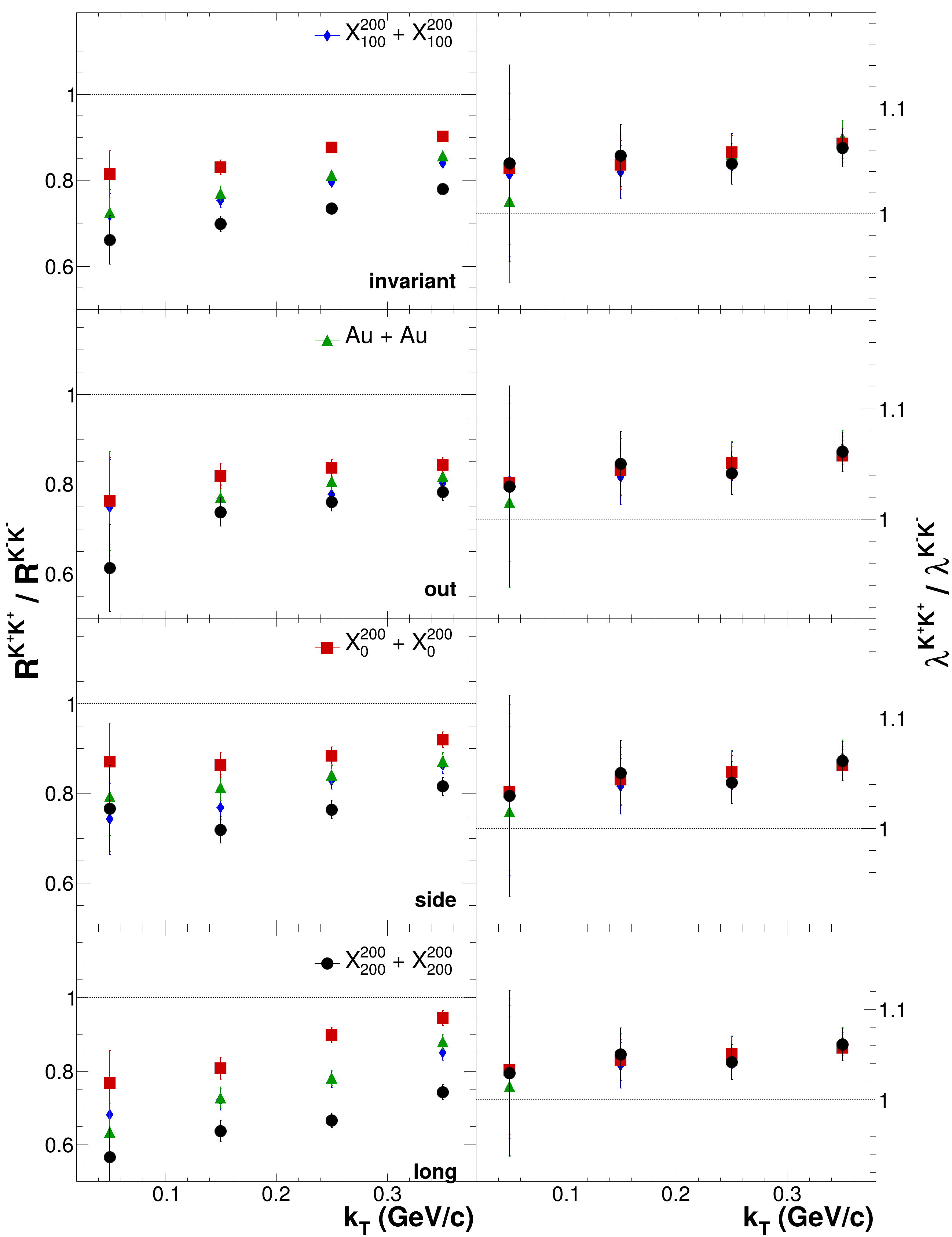}
    \caption{
Ratios of the extracted kaon femtoscopic parameters for positive kaon pairs to
the corresponding parameters for negative kaon pairs, shown as functions of
transverse momentum $k_T$ for different colliding systems. The left column
shows the radius ratios,
$R^{K^+K^+}/R^{K^-K^-}$, while the right column shows the corresponding
intercept-parameter ratios,
$\lambda^{K^+K^+}/\lambda^{K^-K^-}$. From top to bottom, the rows correspond
to the invariant, out, side, and long components. The different markers
indicate Au+Au and the artificial systems used to separate charge and isospin
effects. The horizontal line indicates unity.
}
    \label{fig:radii_kaons_ur}
\end{figure}

The situation for kaons is more complicated. The right column of
Fig.~\ref{fig:1D_diffsystems} shows that changing the initial isospin content
also affects the kaon correlation ratios, but this is not the only source of the
$K^+K^+$ to $K^-K^-$ difference. Even for the isospin-symmetric
$X^{200}_{100}+X^{200}_{100}$ system, the ratio is not unity. This indicates
that the kaon charge splitting is not driven solely by the initial
neutron-to-proton imbalance.

For charged kaons, a large part of the effect comes from the fact that $K^+$
and $K^-$ are produced and transported through the hadronic medium in different
ways. At these collision energies, $K^+$ mesons are produced largely through
associated strangeness-production channels, for example together with a hyperon,
schematically
\begin{equation}
N N \rightarrow N Y K^+ ,
\end{equation}
where $Y$ denotes a hyperon such as $\Lambda$ or $\Sigma$~\cite{Hartnack:2011cn, Fuchs:2005zg}. The strange
quark is carried by the hyperon and the antistrange quark by the $K^+$, so
strangeness is conserved. Once produced, $K^+$ mesons have a relatively small
absorption probability in baryon-rich matter.

Negative kaons are different. A significant part of the $K^-$ population can be
produced through secondary strangeness-exchange reactions, such as
\begin{equation}
\pi Y \leftrightarrow K^- N .
\end{equation}
In addition, $K^-$ mesons interact more strongly with the baryon-rich medium
and can be absorbed or regenerated during the hadronic evolution~\cite{Hartnack:2011cn, Fuchs:2005zg}.
Therefore, the space-time emission distribution of $K^-$ can differ
substantially from that of $K^+$. The two species can freeze out at different
times, from different regions of the source, and with different momentum
distributions.

As a result, the $K^+K^+$ and $K^-K^-$ correlation functions can differ in ways
that are not directly caused by the initial isospin imbalance alone. The isospin
scan shows that the initial charge composition contributes to the kaon ratios,
but the larger charge-dependent difference for kaons is driven mainly by their
different production, absorption, and rescattering histories in the hadronic
medium.

Finally, Figs.~\ref{fig:radii_pions_ur} and~\ref{fig:radii_kaons_ur} show the
$k_T$ dependence of the positive-to-negative ratios of the extracted
femtoscopic parameters for pions and kaons, respectively. The ratios are shown
for both the one-dimensional and three-dimensional fits.

For pions, Fig.~\ref{fig:radii_pions_ur} shows that the effect on the extracted
homogeneity radii decreases rather quickly as the pair transverse momentum
increases. It is especially useful to compare the one-dimensional and
three-dimensional results. In the one-dimensional case, the ratio of
$R_{\mathrm{inv}}$ becomes almost indistinguishable from unity for
$k_T>0.1~\mathrm{GeV}/c$. However, this does not mean that the
charge-dependent effect is absent.

In the three-dimensional analysis, distortions of comparable size remain visible
in the different Bertsch--Pratt directions, but they do not all go in the same
direction. The out and side radii, as well as the invariant radius, are above
unity, while the long radius is below unity. When the three-dimensional
structure is projected into a single one-dimensional radius, these opposite
distortions can partially cancel. As a result, the one-dimensional fit can
underestimate the size of the charge-dependent modification of the source.

This provides another example of why a three-dimensional femtoscopic analysis is
important. A one-dimensional analysis is useful as a compact summary, but it can
average over direction-dependent effects and hide distortions that are visible
only when the out, side, and long components are studied separately.

The same figure also shows that the isospin-driven difference affects the
extracted $\lambda$ parameter. In a Gaussian femtoscopic fit, $\lambda$ is often
interpreted as an effective correlation strength, or intercept parameter. In
experimental analyses it can be influenced by several effects, such as particle
purity, long-lived resonance contributions, partial coherence, and deviations
from a purely Gaussian source shape. In the present UrQMD study, the change in
$\lambda$ does not come from detector purity or an added final-state
interaction, but from the fact that the positive and negative particles have
different space-time and momentum emission distributions. These differences
change the height and shape of the weighted correlation peak and therefore
modify the effective $\lambda$ extracted from a Gaussian fit.

This means that $\lambda$ should not be interpreted only as a coherence
parameter. In realistic correlation functions, and even in controlled model
studies such as this one, $\lambda$ can absorb several physical and
analysis-related effects. Therefore, changes in $\lambda$ should be interpreted
together with the full correlation shape and the extracted radii, rather than as
a standalone measure of source coherence.

This point is especially important when comparing $\lambda$ values between
different collision systems, beam energies, experiments, or analysis methods.
Such comparisons can be meaningful only if the possible contributions from
particle purity, resonance feed-down, non-Gaussian source shape, final-state
interactions, acceptance effects, and charge-dependent emission dynamics are
understood and treated consistently.

\section*{Discussion and Summary}

In this work, we studied charge-dependent modifications of identical-pion and
identical-kaon femtoscopic correlation functions. Two sources of such
modifications were considered: the residual Coulomb field of the charged matter
remaining after particle emission, and isospin-related effects in the hadronic
evolution.

The residual Coulomb effect was studied using a modified Retière--Lisa
blast-wave source. The blast-wave model was used to generate a controlled
freeze-out distribution with finite geometry, collective flow, emission
duration, and position--momentum correlations. The same generated particles
were then propagated through an effective residual Coulomb field with positive
and negative charge signs. This allowed the positive and negative correlation
functions to be compared starting from the same underlying freeze-out source.

The residual Coulomb field produces a small but systematic splitting between
positive and negative correlation functions. With the default parameter set, the
effect appears at the level of a few percent in the extracted femtoscopic
parameters. The splitting is largest at low pair transverse momentum and
decreases with increasing $k_T$, because faster particles spend less time near
the residual charged source and accumulate a smaller Coulomb momentum kick. A
particle-species dependence is also observed: because kaons are heavier than
pions, their velocity and trajectory at the same $k_T$ are different, which
changes their sensitivity to the residual field.

The residual Coulomb propagation affects not only the apparent width of the
correlation function, but also its height and shape. Therefore, both the fitted
radii and the fitted $\lambda$ parameters can be modified. The extracted
Gaussian parameters provide a useful compact comparison between positive and
negative samples, but they do not fully describe the residual Coulomb effect
when the correlation function becomes non-Gaussian.

Systematic parameter variations show that the residual Coulomb effect is
strongly model dependent. It depends not only on the effective residual charge
$Z_{\mathrm{res}}$, but also on the initial width of the charge distribution
$\sigma_0$, its expansion velocity $v_{\mathrm{exp}}$, and the freeze-out
configuration of the emitted particles. The splitting becomes larger when the
residual field is stronger, more compact, or longer lived. Therefore, any
quantitative estimate of the residual Coulomb contribution requires constraints
on several model ingredients, not only on the total effective charge.

An additional study was performed with UrQMD 3.4 in cascade mode. In this case,
no post-emission residual Coulomb propagation and no explicit two-body Coulomb
or strong femtoscopic final-state interactions were applied. The numerator was
filled only with the plane-wave Bose--Einstein symmetrization weight. Even in
this setup, a positive-to-negative difference was observed. This demonstrates
that charge-dependent femtoscopic splittings do not have to originate only from
a residual Coulomb field; they can also arise from the microscopic production
and hadronic evolution in the transport model.

For pions, the UrQMD study shows that the positive-to-negative difference is
mainly driven by the isospin content of the initial state. Artificial systems
with different neutron-to-proton compositions produce different pion
correlation-function ratios: neutron-only and proton-only systems give the
largest deviations with opposite signs, while the isospin-symmetric system
stays much closer to unity. For kaons, the situation is more complex. The
initial isospin content contributes, but the dominant charge dependence is also
connected to the different production, absorption, and rescattering histories of
$K^+$ and $K^-$ in the baryon-rich hadronic medium.

The UrQMD results also show that one-dimensional femtoscopic fits can hide
direction-dependent effects. In some cases, the one-dimensional radius ratio is
close to unity, while the three-dimensional out, side, and long radii still show
visible charge-dependent distortions. This happens because distortions in
different directions can partially cancel when projected into a single
one-dimensional radius.

The main conclusion is that both residual Coulomb effects and isospin-related
effects are strongly model dependent and can compete with each other. Without
constraining the residual Coulomb field, including its effective charge, initial
spatial spread, and expansion velocity, one cannot uniquely assign a measured
positive-to-negative splitting to isospin effects. Conversely, one also cannot
exclude an isospin contribution without controlling the possible residual
Coulomb distortion. A reliable interpretation of charge-dependent femtoscopic
correlations therefore requires treating residual-source Coulomb effects and
charge-dependent hadronic dynamics within the same constrained model framework.

\begin{acknowledgments}
This work supported by The U.S. Department of Energy Grant DE-SC0020651.

\end{acknowledgments}

\bibliography{apssamp}

\begin{thebibliography}{58}%
\makeatletter
\providecommand \@ifxundefined [1]{%
 \@ifx{#1\undefined}
}%
\providecommand \@ifnum [1]{%
 \ifnum #1\expandafter \@firstoftwo
 \else \expandafter \@secondoftwo
 \fi
}%
\providecommand \@ifx [1]{%
 \ifx #1\expandafter \@firstoftwo
 \else \expandafter \@secondoftwo
 \fi
}%
\providecommand \natexlab [1]{#1}%
\providecommand \enquote  [1]{``#1''}%
\providecommand \bibnamefont  [1]{#1}%
\providecommand \bibfnamefont [1]{#1}%
\providecommand \citenamefont [1]{#1}%
\providecommand \href@noop [0]{\@secondoftwo}%
\providecommand \href [0]{\begingroup \@sanitize@url \@href}%
\providecommand \@href[1]{\@@startlink{#1}\@@href}%
\providecommand \@@href[1]{\endgroup#1\@@endlink}%
\providecommand \@sanitize@url [0]{\catcode `\\12\catcode `\$12\catcode
  `\&12\catcode `\#12\catcode `\^12\catcode `\_12\catcode `\%12\relax}%
\providecommand \@@startlink[1]{}%
\providecommand \@@endlink[0]{}%
\providecommand \url  [0]{\begingroup\@sanitize@url \@url }%
\providecommand \@url [1]{\endgroup\@href {#1}{\urlprefix }}%
\providecommand \urlprefix  [0]{URL }%
\providecommand \Eprint [0]{\href }%
\providecommand \doibase [0]{https://doi.org/}%
\providecommand \selectlanguage [0]{\@gobble}%
\providecommand \bibinfo  [0]{\@secondoftwo}%
\providecommand \bibfield  [0]{\@secondoftwo}%
\providecommand \translation [1]{[#1]}%
\providecommand \BibitemOpen [0]{}%
\providecommand \bibitemStop [0]{}%
\providecommand \bibitemNoStop [0]{.\EOS\space}%
\providecommand \EOS [0]{\spacefactor3000\relax}%
\providecommand \BibitemShut  [1]{\csname bibitem#1\endcsname}%
\let\auto@bib@innerbib\@empty
\bibitem [{\citenamefont {Lisa}\ \emph {et~al.}(2005)\citenamefont {Lisa},
  \citenamefont {Pratt}, \citenamefont {Soltz},\ and\ \citenamefont
  {Wiedemann}}]{Lisa:2005dd}%
  \BibitemOpen
  \bibfield  {author} {\bibinfo {author} {\bibfnamefont {M.~A.}\ \bibnamefont
  {Lisa}}, \bibinfo {author} {\bibfnamefont {S.}~\bibnamefont {Pratt}},
  \bibinfo {author} {\bibfnamefont {R.}~\bibnamefont {Soltz}},\ and\ \bibinfo
  {author} {\bibfnamefont {U.}~\bibnamefont {Wiedemann}},\ }\bibfield  {title}
  {\bibinfo {title} {{Femtoscopy in relativistic heavy ion collisions}},\
  }\href {https://doi.org/10.1146/annurev.nucl.55.090704.151533} {\bibfield
  {journal} {\bibinfo  {journal} {Ann. Rev. Nucl. Part. Sci.}\ }\textbf
  {\bibinfo {volume} {55}},\ \bibinfo {pages} {357} (\bibinfo {year} {2005})},\
  \Eprint {https://arxiv.org/abs/nucl-ex/0505014} {arXiv:nucl-ex/0505014}
  \BibitemShut {NoStop}%
\bibitem [{\citenamefont {Heinz}\ and\ \citenamefont
  {Jacak}(1999)}]{Heinz:1999rw}%
  \BibitemOpen
  \bibfield  {author} {\bibinfo {author} {\bibfnamefont {U.~W.}\ \bibnamefont
  {Heinz}}\ and\ \bibinfo {author} {\bibfnamefont {B.~V.}\ \bibnamefont
  {Jacak}},\ }\bibfield  {title} {\bibinfo {title} {{Two particle correlations
  in relativistic heavy ion collisions}},\ }\href
  {https://doi.org/10.1146/annurev.nucl.49.1.529} {\bibfield  {journal}
  {\bibinfo  {journal} {Ann. Rev. Nucl. Part. Sci.}\ }\textbf {\bibinfo
  {volume} {49}},\ \bibinfo {pages} {529} (\bibinfo {year} {1999})},\ \Eprint
  {https://arxiv.org/abs/nucl-th/9902020} {arXiv:nucl-th/9902020} \BibitemShut
  {NoStop}%
\bibitem [{\citenamefont {Kopylov}\ and\ \citenamefont
  {Podgoretsky}(1972)}]{Kopylov:1972qw}%
  \BibitemOpen
  \bibfield  {author} {\bibinfo {author} {\bibfnamefont {G.~I.}\ \bibnamefont
  {Kopylov}}\ and\ \bibinfo {author} {\bibfnamefont {M.~I.}\ \bibnamefont
  {Podgoretsky}},\ }\bibfield  {title} {\bibinfo {title} {{Correlations of
  identical particles emitted by highly excited nuclei}},\ }\href@noop {}
  {\bibfield  {journal} {\bibinfo  {journal} {Sov. J. Nucl. Phys.}\ }\textbf
  {\bibinfo {volume} {15}},\ \bibinfo {pages} {219} (\bibinfo {year}
  {1972})}\BibitemShut {NoStop}%
\bibitem [{\citenamefont {Pratt}(1986)}]{Pratt:1986cc}%
  \BibitemOpen
  \bibfield  {author} {\bibinfo {author} {\bibfnamefont {S.}~\bibnamefont
  {Pratt}},\ }\bibfield  {title} {\bibinfo {title} {{Pion Interferometry of
  Quark-Gluon Plasma}},\ }\href {https://doi.org/10.1103/PhysRevD.33.1314}
  {\bibfield  {journal} {\bibinfo  {journal} {Phys. Rev. D}\ }\textbf {\bibinfo
  {volume} {33}},\ \bibinfo {pages} {1314} (\bibinfo {year}
  {1986})}\BibitemShut {NoStop}%
\bibitem [{\citenamefont {Bertsch}\ \emph {et~al.}(1988)\citenamefont
  {Bertsch}, \citenamefont {Gong},\ and\ \citenamefont
  {Tohyama}}]{Bertsch:1988db}%
  \BibitemOpen
  \bibfield  {author} {\bibinfo {author} {\bibfnamefont {G.}~\bibnamefont
  {Bertsch}}, \bibinfo {author} {\bibfnamefont {M.}~\bibnamefont {Gong}},\ and\
  \bibinfo {author} {\bibfnamefont {M.}~\bibnamefont {Tohyama}},\ }\bibfield
  {title} {\bibinfo {title} {{Pion Interferometry in Ultrarelativistic Heavy
  Ion Collisions}},\ }\href {https://doi.org/10.1103/PhysRevC.37.1896}
  {\bibfield  {journal} {\bibinfo  {journal} {Phys. Rev. C}\ }\textbf {\bibinfo
  {volume} {37}},\ \bibinfo {pages} {1896} (\bibinfo {year}
  {1988})}\BibitemShut {NoStop}%
\bibitem [{\citenamefont {Akkelin}\ and\ \citenamefont
  {Sinyukov}(1995)}]{Akkelin:1995gh}%
  \BibitemOpen
  \bibfield  {author} {\bibinfo {author} {\bibfnamefont {S.~V.}\ \bibnamefont
  {Akkelin}}\ and\ \bibinfo {author} {\bibfnamefont {Y.~M.}\ \bibnamefont
  {Sinyukov}},\ }\bibfield  {title} {\bibinfo {title} {{The HBT interferometry
  of expanding sources}},\ }\href
  {https://doi.org/10.1016/0370-2693(95)00765-D} {\bibfield  {journal}
  {\bibinfo  {journal} {Phys. Lett. B}\ }\textbf {\bibinfo {volume} {356}},\
  \bibinfo {pages} {525} (\bibinfo {year} {1995})}\BibitemShut {NoStop}%
\bibitem [{\citenamefont {Retiere}\ and\ \citenamefont
  {Lisa}(2004)}]{Retiere:2003kf}%
  \BibitemOpen
  \bibfield  {author} {\bibinfo {author} {\bibfnamefont {F.}~\bibnamefont
  {Retiere}}\ and\ \bibinfo {author} {\bibfnamefont {M.~A.}\ \bibnamefont
  {Lisa}},\ }\bibfield  {title} {\bibinfo {title} {{Observable implications of
  geometrical and dynamical aspects of freeze out in heavy ion collisions}},\
  }\href {https://doi.org/10.1103/PhysRevC.70.044907} {\bibfield  {journal}
  {\bibinfo  {journal} {Phys. Rev. C}\ }\textbf {\bibinfo {volume} {70}},\
  \bibinfo {pages} {044907} (\bibinfo {year} {2004})},\ \Eprint
  {https://arxiv.org/abs/nucl-th/0312024} {arXiv:nucl-th/0312024} \BibitemShut
  {NoStop}%
\bibitem [{\citenamefont {Lednicky}\ and\ \citenamefont
  {Lyuboshits}(1981)}]{Lednicky:1981su}%
  \BibitemOpen
  \bibfield  {author} {\bibinfo {author} {\bibfnamefont {R.}~\bibnamefont
  {Lednicky}}\ and\ \bibinfo {author} {\bibfnamefont {V.~L.}\ \bibnamefont
  {Lyuboshits}},\ }\bibfield  {title} {\bibinfo {title} {{Final State
  Interaction Effect on Pairing Correlations Between Particles with Small
  Relative Momenta}},\ }\href@noop {} {\bibfield  {journal} {\bibinfo
  {journal} {Yad. Fiz.}\ }\textbf {\bibinfo {volume} {35}},\ \bibinfo {pages}
  {1316} (\bibinfo {year} {1981})}\BibitemShut {NoStop}%
\bibitem [{\citenamefont {Lednicky}(2009)}]{Lednicky:2005tb}%
  \BibitemOpen
  \bibfield  {author} {\bibinfo {author} {\bibfnamefont {R.}~\bibnamefont
  {Lednicky}},\ }\bibfield  {title} {\bibinfo {title} {{Finite-size effects on
  two-particle production in continuous and discrete spectrum}},\ }\href
  {https://doi.org/10.1134/S1063779609030034} {\bibfield  {journal} {\bibinfo
  {journal} {Phys. Part. Nucl.}\ }\textbf {\bibinfo {volume} {40}},\ \bibinfo
  {pages} {307} (\bibinfo {year} {2009})},\ \Eprint
  {https://arxiv.org/abs/nucl-th/0501065} {arXiv:nucl-th/0501065} \BibitemShut
  {NoStop}%
\bibitem [{\citenamefont {Voloshin}\ \emph {et~al.}(1997)\citenamefont
  {Voloshin}, \citenamefont {Lednicky}, \citenamefont {Panitkin},\ and\
  \citenamefont {Xu}}]{Voloshin:1997jh}%
  \BibitemOpen
  \bibfield  {author} {\bibinfo {author} {\bibfnamefont {S.}~\bibnamefont
  {Voloshin}}, \bibinfo {author} {\bibfnamefont {R.}~\bibnamefont {Lednicky}},
  \bibinfo {author} {\bibfnamefont {S.}~\bibnamefont {Panitkin}},\ and\
  \bibinfo {author} {\bibfnamefont {N.}~\bibnamefont {Xu}},\ }\bibfield
  {title} {\bibinfo {title} {{Relative space-time asymmetries in pion and
  nucleon production in noncentral nucleus-nucleus collisions at
  high-energies}},\ }\href {https://doi.org/10.1103/PhysRevLett.79.4766}
  {\bibfield  {journal} {\bibinfo  {journal} {Phys. Rev. Lett.}\ }\textbf
  {\bibinfo {volume} {79}},\ \bibinfo {pages} {4766} (\bibinfo {year}
  {1997})},\ \Eprint {https://arxiv.org/abs/nucl-th/9708044}
  {arXiv:nucl-th/9708044} \BibitemShut {NoStop}%
\bibitem [{\citenamefont {Lednicky}\ \emph {et~al.}(1996)\citenamefont
  {Lednicky}, \citenamefont {Lyuboshits}, \citenamefont {Erazmus},\ and\
  \citenamefont {Nouais}}]{Lednicky:1995vk}%
  \BibitemOpen
  \bibfield  {author} {\bibinfo {author} {\bibfnamefont {R.}~\bibnamefont
  {Lednicky}}, \bibinfo {author} {\bibfnamefont {V.~L.}\ \bibnamefont
  {Lyuboshits}}, \bibinfo {author} {\bibfnamefont {B.}~\bibnamefont
  {Erazmus}},\ and\ \bibinfo {author} {\bibfnamefont {D.}~\bibnamefont
  {Nouais}},\ }\bibfield  {title} {\bibinfo {title} {{How to measure which sort
  of particles was emitted earlier and which later}},\ }\href
  {https://doi.org/10.1016/0370-2693(96)00124-4} {\bibfield  {journal}
  {\bibinfo  {journal} {Phys. Lett. B}\ }\textbf {\bibinfo {volume} {373}},\
  \bibinfo {pages} {30} (\bibinfo {year} {1996})}\BibitemShut {NoStop}%
\bibitem [{\citenamefont {Adams}\ \emph {et~al.}(2003)\citenamefont {Adams}
  \emph {et~al.}}]{STAR:2003cqe}%
  \BibitemOpen
  \bibfield  {author} {\bibinfo {author} {\bibfnamefont {J.}~\bibnamefont
  {Adams}} \emph {et~al.} (\bibinfo {collaboration} {STAR}),\ }\bibfield
  {title} {\bibinfo {title} {{Pion kaon correlations in Au+Au collisions at
  s(NN)**1/2 = 130-GeV}},\ }\href
  {https://doi.org/10.1103/PhysRevLett.91.262302} {\bibfield  {journal}
  {\bibinfo  {journal} {Phys. Rev. Lett.}\ }\textbf {\bibinfo {volume} {91}},\
  \bibinfo {pages} {262302} (\bibinfo {year} {2003})},\ \Eprint
  {https://arxiv.org/abs/nucl-ex/0307025} {arXiv:nucl-ex/0307025} \BibitemShut
  {NoStop}%
\bibitem [{\citenamefont {Koonin}(1977)}]{Koonin:1977fh}%
  \BibitemOpen
  \bibfield  {author} {\bibinfo {author} {\bibfnamefont {S.~E.}\ \bibnamefont
  {Koonin}},\ }\bibfield  {title} {\bibinfo {title} {{Proton Pictures of
  High-Energy Nuclear Collisions}},\ }\href
  {https://doi.org/10.1016/0370-2693(77)90340-9} {\bibfield  {journal}
  {\bibinfo  {journal} {Phys. Lett. B}\ }\textbf {\bibinfo {volume} {70}},\
  \bibinfo {pages} {43} (\bibinfo {year} {1977})}\BibitemShut {NoStop}%
\bibitem [{\citenamefont {Collaboration}\ \emph {et~al.}(2020)\citenamefont
  {Collaboration} \emph {et~al.}}]{ALICE:2020mfd}%
  \BibitemOpen
  \bibfield  {author} {\bibinfo {author} {\bibfnamefont {A.}~\bibnamefont
  {Collaboration}} \emph {et~al.} (\bibinfo {collaboration} {ALICE}),\
  }\bibfield  {title} {\bibinfo {title} {{Unveiling the strong interaction
  among hadrons at the LHC}},\ }\href
  {https://doi.org/10.1038/s41586-020-3001-6} {\bibfield  {journal} {\bibinfo
  {journal} {Nature}\ }\textbf {\bibinfo {volume} {588}},\ \bibinfo {pages}
  {232} (\bibinfo {year} {2020})},\ \bibinfo {note} {[Erratum: Nature 590, E13
  (2021)]},\ \Eprint {https://arxiv.org/abs/2005.11495} {arXiv:2005.11495
  [nucl-ex]} \BibitemShut {NoStop}%
\bibitem [{\citenamefont {Fabbietti}\ \emph {et~al.}(2021)\citenamefont
  {Fabbietti}, \citenamefont {Mantovani~Sarti},\ and\ \citenamefont
  {Vazquez~Doce}}]{Fabbietti:2020bfg}%
  \BibitemOpen
  \bibfield  {author} {\bibinfo {author} {\bibfnamefont {L.}~\bibnamefont
  {Fabbietti}}, \bibinfo {author} {\bibfnamefont {V.}~\bibnamefont
  {Mantovani~Sarti}},\ and\ \bibinfo {author} {\bibfnamefont {O.}~\bibnamefont
  {Vazquez~Doce}},\ }\bibfield  {title} {\bibinfo {title} {{Study of the Strong
  Interaction Among Hadrons with Correlations at the LHC}},\ }\href
  {https://doi.org/10.1146/annurev-nucl-102419-034438} {\bibfield  {journal}
  {\bibinfo  {journal} {Ann. Rev. Nucl. Part. Sci.}\ }\textbf {\bibinfo
  {volume} {71}},\ \bibinfo {pages} {377} (\bibinfo {year} {2021})},\ \Eprint
  {https://arxiv.org/abs/2012.09806} {arXiv:2012.09806 [nucl-ex]} \BibitemShut
  {NoStop}%
\bibitem [{\citenamefont {Adamczyk}\ \emph
  {et~al.}(2015{\natexlab{a}})\citenamefont {Adamczyk} \emph
  {et~al.}}]{STAR:2015kha}%
  \BibitemOpen
  \bibfield  {author} {\bibinfo {author} {\bibfnamefont {L.}~\bibnamefont
  {Adamczyk}} \emph {et~al.} (\bibinfo {collaboration} {STAR}),\ }\bibfield
  {title} {\bibinfo {title} {{Measurement of Interaction between
  Antiprotons}},\ }\href {https://doi.org/10.1038/nature15724} {\bibfield
  {journal} {\bibinfo  {journal} {Nature}\ }\textbf {\bibinfo {volume} {527}},\
  \bibinfo {pages} {345} (\bibinfo {year} {2015}{\natexlab{a}})},\ \Eprint
  {https://arxiv.org/abs/1507.07158} {arXiv:1507.07158 [nucl-ex]} \BibitemShut
  {NoStop}%
\bibitem [{\citenamefont {Kamiya}\ \emph {et~al.}(2020)\citenamefont {Kamiya},
  \citenamefont {Hyodo}, \citenamefont {Morita}, \citenamefont {Ohnishi},\ and\
  \citenamefont {Weise}}]{Kamiya:2019uiw}%
  \BibitemOpen
  \bibfield  {author} {\bibinfo {author} {\bibfnamefont {Y.}~\bibnamefont
  {Kamiya}}, \bibinfo {author} {\bibfnamefont {T.}~\bibnamefont {Hyodo}},
  \bibinfo {author} {\bibfnamefont {K.}~\bibnamefont {Morita}}, \bibinfo
  {author} {\bibfnamefont {A.}~\bibnamefont {Ohnishi}},\ and\ \bibinfo {author}
  {\bibfnamefont {W.}~\bibnamefont {Weise}},\ }\bibfield  {title} {\bibinfo
  {title} {{$K^-p$ Correlation Function from High-Energy Nuclear Collisions and
  Chiral SU(3) Dynamics}},\ }\href
  {https://doi.org/10.1103/PhysRevLett.124.132501} {\bibfield  {journal}
  {\bibinfo  {journal} {Phys. Rev. Lett.}\ }\textbf {\bibinfo {volume} {124}},\
  \bibinfo {pages} {132501} (\bibinfo {year} {2020})},\ \Eprint
  {https://arxiv.org/abs/1911.01041} {arXiv:1911.01041 [nucl-th]} \BibitemShut
  {NoStop}%
\bibitem [{\citenamefont {Adamczyk}\ \emph
  {et~al.}(2015{\natexlab{b}})\citenamefont {Adamczyk} \emph
  {et~al.}}]{STAR:2014dcy}%
  \BibitemOpen
  \bibfield  {author} {\bibinfo {author} {\bibfnamefont {L.}~\bibnamefont
  {Adamczyk}} \emph {et~al.} (\bibinfo {collaboration} {STAR}),\ }\bibfield
  {title} {\bibinfo {title} {{$\Lambda\Lambda$ Correlation Function in Au+Au
  collisions at $\sqrt{s_{NN}}=$ 200 GeV}},\ }\href
  {https://doi.org/10.1103/PhysRevLett.114.022301} {\bibfield  {journal}
  {\bibinfo  {journal} {Phys. Rev. Lett.}\ }\textbf {\bibinfo {volume} {114}},\
  \bibinfo {pages} {022301} (\bibinfo {year} {2015}{\natexlab{b}})},\ \Eprint
  {https://arxiv.org/abs/1408.4360} {arXiv:1408.4360 [nucl-ex]} \BibitemShut
  {NoStop}%
\bibitem [{\citenamefont {Adam}\ \emph {et~al.}(2019)\citenamefont {Adam} \emph
  {et~al.}}]{STAR:2018uho}%
  \BibitemOpen
  \bibfield  {author} {\bibinfo {author} {\bibfnamefont {J.}~\bibnamefont
  {Adam}} \emph {et~al.} (\bibinfo {collaboration} {STAR}),\ }\bibfield
  {title} {\bibinfo {title} {{The Proton-$\Omega$ correlation function in Au+Au
  collisions at $\sqrt{s_{NN}}$=200 GeV}},\ }\href
  {https://doi.org/10.1016/j.physletb.2019.01.055} {\bibfield  {journal}
  {\bibinfo  {journal} {Phys. Lett. B}\ }\textbf {\bibinfo {volume} {790}},\
  \bibinfo {pages} {490} (\bibinfo {year} {2019})},\ \Eprint
  {https://arxiv.org/abs/1808.02511} {arXiv:1808.02511 [hep-ex]} \BibitemShut
  {NoStop}%
\bibitem [{\citenamefont {Acharya}\ \emph
  {et~al.}(2019{\natexlab{a}})\citenamefont {Acharya} \emph
  {et~al.}}]{ALICE:2019eol}%
  \BibitemOpen
  \bibfield  {author} {\bibinfo {author} {\bibfnamefont {S.}~\bibnamefont
  {Acharya}} \emph {et~al.} (\bibinfo {collaboration} {ALICE}),\ }\bibfield
  {title} {\bibinfo {title} {{Study of the $\Lambda$-$\Lambda$ interaction with
  femtoscopy correlations in pp and p-Pb collisions at the LHC}},\ }\href
  {https://doi.org/10.1016/j.physletb.2019.134822} {\bibfield  {journal}
  {\bibinfo  {journal} {Phys. Lett. B}\ }\textbf {\bibinfo {volume} {797}},\
  \bibinfo {pages} {134822} (\bibinfo {year} {2019}{\natexlab{a}})},\ \Eprint
  {https://arxiv.org/abs/1905.07209} {arXiv:1905.07209 [nucl-ex]} \BibitemShut
  {NoStop}%
\bibitem [{\citenamefont {Acharya}\ \emph {et~al.}(2021)\citenamefont {Acharya}
  \emph {et~al.}}]{ALICE:2021cpv}%
  \BibitemOpen
  \bibfield  {author} {\bibinfo {author} {\bibfnamefont {S.}~\bibnamefont
  {Acharya}} \emph {et~al.} (\bibinfo {collaboration} {ALICE}),\ }\bibfield
  {title} {\bibinfo {title} {{Experimental Evidence for an Attractive p-$\phi$
  Interaction}},\ }\href {https://doi.org/10.1103/PhysRevLett.127.172301}
  {\bibfield  {journal} {\bibinfo  {journal} {Phys. Rev. Lett.}\ }\textbf
  {\bibinfo {volume} {127}},\ \bibinfo {pages} {172301} (\bibinfo {year}
  {2021})},\ \Eprint {https://arxiv.org/abs/2105.05578} {arXiv:2105.05578
  [nucl-ex]} \BibitemShut {NoStop}%
\bibitem [{\citenamefont {Abelev}\ \emph {et~al.}(2006)\citenamefont {Abelev}
  \emph {et~al.}}]{STAR:2006ykx}%
  \BibitemOpen
  \bibfield  {author} {\bibinfo {author} {\bibfnamefont {B.~I.}\ \bibnamefont
  {Abelev}} \emph {et~al.} (\bibinfo {collaboration} {STAR}),\ }\bibfield
  {title} {\bibinfo {title} {{Neutral kaon interferometry in Au+Au collisions
  at s(NN)**(1/2) = 200-GeV}},\ }\href
  {https://doi.org/10.1103/PhysRevC.74.054902} {\bibfield  {journal} {\bibinfo
  {journal} {Phys. Rev. C}\ }\textbf {\bibinfo {volume} {74}},\ \bibinfo
  {pages} {054902} (\bibinfo {year} {2006})},\ \Eprint
  {https://arxiv.org/abs/nucl-ex/0608012} {arXiv:nucl-ex/0608012} \BibitemShut
  {NoStop}%
\bibitem [{\citenamefont {Mihaylov}\ \emph {et~al.}(2018)\citenamefont
  {Mihaylov}, \citenamefont {Mantovani~Sarti}, \citenamefont {Arnold},
  \citenamefont {Fabbietti}, \citenamefont {Hohlweger},\ and\ \citenamefont
  {Mathis}}]{Mihaylov:2018rva}%
  \BibitemOpen
  \bibfield  {author} {\bibinfo {author} {\bibfnamefont {D.~L.}\ \bibnamefont
  {Mihaylov}}, \bibinfo {author} {\bibfnamefont {V.}~\bibnamefont
  {Mantovani~Sarti}}, \bibinfo {author} {\bibfnamefont {O.~W.}\ \bibnamefont
  {Arnold}}, \bibinfo {author} {\bibfnamefont {L.}~\bibnamefont {Fabbietti}},
  \bibinfo {author} {\bibfnamefont {B.}~\bibnamefont {Hohlweger}},\ and\
  \bibinfo {author} {\bibfnamefont {A.~M.}\ \bibnamefont {Mathis}},\ }\bibfield
   {title} {\bibinfo {title} {{A femtoscopic Correlation Analysis Tool using
  the Schr{\"o}dinger equation (CATS)}},\ }\href
  {https://doi.org/10.1140/epjc/s10052-018-5859-0} {\bibfield  {journal}
  {\bibinfo  {journal} {Eur. Phys. J. C}\ }\textbf {\bibinfo {volume} {78}},\
  \bibinfo {pages} {394} (\bibinfo {year} {2018})},\ \Eprint
  {https://arxiv.org/abs/1802.08481} {arXiv:1802.08481 [hep-ph]} \BibitemShut
  {NoStop}%
\bibitem [{\citenamefont {Kisiel}\ \emph {et~al.}(2014)\citenamefont {Kisiel},
  \citenamefont {Zbroszczyk},\ and\ \citenamefont
  {Szyma{\'n}ski}}]{Kisiel:2014mma}%
  \BibitemOpen
  \bibfield  {author} {\bibinfo {author} {\bibfnamefont {A.}~\bibnamefont
  {Kisiel}}, \bibinfo {author} {\bibfnamefont {H.}~\bibnamefont {Zbroszczyk}},\
  and\ \bibinfo {author} {\bibfnamefont {M.}~\bibnamefont {Szyma{\'n}ski}},\
  }\bibfield  {title} {\bibinfo {title} {{Extracting baryon-antibaryon strong
  interaction potentials from p$\bar{\Lambda}$ femtoscopic correlation
  functions}},\ }\href {https://doi.org/10.1103/PhysRevC.89.054916} {\bibfield
  {journal} {\bibinfo  {journal} {Phys. Rev. C}\ }\textbf {\bibinfo {volume}
  {89}},\ \bibinfo {pages} {054916} (\bibinfo {year} {2014})},\ \Eprint
  {https://arxiv.org/abs/1403.0433} {arXiv:1403.0433 [nucl-th]} \BibitemShut
  {NoStop}%
\bibitem [{\citenamefont {Acharya}\ \emph
  {et~al.}(2019{\natexlab{b}})\citenamefont {Acharya} \emph
  {et~al.}}]{ALICE:2018ysd}%
  \BibitemOpen
  \bibfield  {author} {\bibinfo {author} {\bibfnamefont {S.}~\bibnamefont
  {Acharya}} \emph {et~al.} (\bibinfo {collaboration} {ALICE}),\ }\bibfield
  {title} {\bibinfo {title} {{p-p, p-$\Lambda$ and $\Lambda$-$\Lambda$
  correlations studied via femtoscopy in pp reactions at $\sqrt{s}$ = 7 TeV}},\
  }\href {https://doi.org/10.1103/PhysRevC.99.024001} {\bibfield  {journal}
  {\bibinfo  {journal} {Phys. Rev. C}\ }\textbf {\bibinfo {volume} {99}},\
  \bibinfo {pages} {024001} (\bibinfo {year} {2019}{\natexlab{b}})},\ \Eprint
  {https://arxiv.org/abs/1805.12455} {arXiv:1805.12455 [nucl-ex]} \BibitemShut
  {NoStop}%
\bibitem [{\citenamefont {Maj}\ and\ \citenamefont
  {Mrowczynski}(2009)}]{Maj:2009ue}%
  \BibitemOpen
  \bibfield  {author} {\bibinfo {author} {\bibfnamefont {R.}~\bibnamefont
  {Maj}}\ and\ \bibinfo {author} {\bibfnamefont {S.}~\bibnamefont
  {Mrowczynski}},\ }\bibfield  {title} {\bibinfo {title} {{Coulomb Effects in
  Femtoscopy}},\ }\href {https://doi.org/10.1103/PhysRevC.80.034907} {\bibfield
   {journal} {\bibinfo  {journal} {Phys. Rev. C}\ }\textbf {\bibinfo {volume}
  {80}},\ \bibinfo {pages} {034907} (\bibinfo {year} {2009})},\ \Eprint
  {https://arxiv.org/abs/0903.0111} {arXiv:0903.0111 [nucl-th]} \BibitemShut
  {NoStop}%
\bibitem [{\citenamefont {Bowler}(1991)}]{Bowler:1991vx}%
  \BibitemOpen
  \bibfield  {author} {\bibinfo {author} {\bibfnamefont {M.~G.}\ \bibnamefont
  {Bowler}},\ }\bibfield  {title} {\bibinfo {title} {{Coulomb corrections to
  Bose-Einstein correlations have been greatly exaggerated}},\ }\href
  {https://doi.org/10.1016/0370-2693(91)91541-3} {\bibfield  {journal}
  {\bibinfo  {journal} {Phys. Lett. B}\ }\textbf {\bibinfo {volume} {270}},\
  \bibinfo {pages} {69} (\bibinfo {year} {1991})}\BibitemShut {NoStop}%
\bibitem [{\citenamefont {Sinyukov}\ \emph {et~al.}(1998)\citenamefont
  {Sinyukov}, \citenamefont {Lednicky}, \citenamefont {Akkelin}, \citenamefont
  {Pluta},\ and\ \citenamefont {Erazmus}}]{Sinyukov:1998fc}%
  \BibitemOpen
  \bibfield  {author} {\bibinfo {author} {\bibfnamefont {Y.}~\bibnamefont
  {Sinyukov}}, \bibinfo {author} {\bibfnamefont {R.}~\bibnamefont {Lednicky}},
  \bibinfo {author} {\bibfnamefont {S.~V.}\ \bibnamefont {Akkelin}}, \bibinfo
  {author} {\bibfnamefont {J.}~\bibnamefont {Pluta}},\ and\ \bibinfo {author}
  {\bibfnamefont {B.}~\bibnamefont {Erazmus}},\ }\bibfield  {title} {\bibinfo
  {title} {{Coulomb corrections for interferometry analysis of expanding hadron
  systems}},\ }\href {https://doi.org/10.1016/S0370-2693(98)00653-4} {\bibfield
   {journal} {\bibinfo  {journal} {Phys. Lett. B}\ }\textbf {\bibinfo {volume}
  {432}},\ \bibinfo {pages} {248} (\bibinfo {year} {1998})}\BibitemShut
  {NoStop}%
\bibitem [{\citenamefont {Barz}(1996)}]{Barz:1996gr}%
  \BibitemOpen
  \bibfield  {author} {\bibinfo {author} {\bibfnamefont {H.~W.}\ \bibnamefont
  {Barz}},\ }\bibfield  {title} {\bibinfo {title} {{Effects of nuclear Coulomb
  field on two meson correlations}},\ }\href
  {https://doi.org/10.1103/PhysRevC.53.2536} {\bibfield  {journal} {\bibinfo
  {journal} {Phys. Rev. C}\ }\textbf {\bibinfo {volume} {53}},\ \bibinfo
  {pages} {2536} (\bibinfo {year} {1996})}\BibitemShut {NoStop}%
\bibitem [{\citenamefont {Barz}\ \emph {et~al.}(1997)\citenamefont {Barz},
  \citenamefont {Bondorf}, \citenamefont {Gaardhoje},\ and\ \citenamefont
  {Heiselberg}}]{Barz:1997es}%
  \BibitemOpen
  \bibfield  {author} {\bibinfo {author} {\bibfnamefont {H.~W.}\ \bibnamefont
  {Barz}}, \bibinfo {author} {\bibfnamefont {J.~P.}\ \bibnamefont {Bondorf}},
  \bibinfo {author} {\bibfnamefont {J.~J.}\ \bibnamefont {Gaardhoje}},\ and\
  \bibinfo {author} {\bibfnamefont {H.}~\bibnamefont {Heiselberg}},\ }\bibfield
   {title} {\bibinfo {title} {{Freezeout time in ultrarelativistic heavy ion
  collisions from Coulomb effects in transverse pion spectra}},\ }\href
  {https://doi.org/10.1103/PhysRevC.56.1553} {\bibfield  {journal} {\bibinfo
  {journal} {Phys. Rev. C}\ }\textbf {\bibinfo {volume} {56}},\ \bibinfo
  {pages} {1553} (\bibinfo {year} {1997})},\ \Eprint
  {https://arxiv.org/abs/nucl-th/9704045} {arXiv:nucl-th/9704045} \BibitemShut
  {NoStop}%
\bibitem [{\citenamefont {Shoppa}\ \emph {et~al.}(2000)\citenamefont {Shoppa},
  \citenamefont {Koonin},\ and\ \citenamefont {Seki}}]{Shoppa:1998sw}%
  \BibitemOpen
  \bibfield  {author} {\bibinfo {author} {\bibfnamefont {T.~D.}\ \bibnamefont
  {Shoppa}}, \bibinfo {author} {\bibfnamefont {S.~E.}\ \bibnamefont {Koonin}},\
  and\ \bibinfo {author} {\bibfnamefont {R.}~\bibnamefont {Seki}},\ }\bibfield
  {title} {\bibinfo {title} {{Effect of the source charge on charged boson
  interferometry}},\ }\href {https://doi.org/10.1103/PhysRevC.61.054902}
  {\bibfield  {journal} {\bibinfo  {journal} {Phys. Rev. C}\ }\textbf {\bibinfo
  {volume} {61}},\ \bibinfo {pages} {054902} (\bibinfo {year} {2000})},\
  \Eprint {https://arxiv.org/abs/nucl-th/9811075} {arXiv:nucl-th/9811075}
  \BibitemShut {NoStop}%
\bibitem [{\citenamefont {Adamczewski-Musch}\ \emph {et~al.}(2020)\citenamefont
  {Adamczewski-Musch} \emph {et~al.}}]{HADES:2019lek}%
  \BibitemOpen
  \bibfield  {author} {\bibinfo {author} {\bibfnamefont {J.}~\bibnamefont
  {Adamczewski-Musch}} \emph {et~al.} (\bibinfo {collaboration} {HADES}),\
  }\bibfield  {title} {\bibinfo {title} {{Identical pion intensity
  interferometry at $\sqrt{s_{\mathrm{NN}}}=2.4~\hbox {GeV}$: HADES
  collaboration}},\ }\href {https://doi.org/10.1140/epja/s10050-020-00116-w}
  {\bibfield  {journal} {\bibinfo  {journal} {Eur. Phys. J. A}\ }\textbf
  {\bibinfo {volume} {56}},\ \bibinfo {pages} {140} (\bibinfo {year} {2020})},\
  \Eprint {https://arxiv.org/abs/1910.07885} {arXiv:1910.07885 [nucl-ex]}
  \BibitemShut {NoStop}%
\bibitem [{\citenamefont {Reisdorf}\ \emph {et~al.}(2007)\citenamefont
  {Reisdorf} \emph {et~al.}}]{FOPI:2006ifg}%
  \BibitemOpen
  \bibfield  {author} {\bibinfo {author} {\bibfnamefont {W.}~\bibnamefont
  {Reisdorf}} \emph {et~al.} (\bibinfo {collaboration} {FOPI}),\ }\bibfield
  {title} {\bibinfo {title} {{Systematics of pion emission in heavy ion
  collisions in the 1A- GeV regime}},\ }\href
  {https://doi.org/10.1016/j.nuclphysa.2006.10.085} {\bibfield  {journal}
  {\bibinfo  {journal} {Nucl. Phys. A}\ }\textbf {\bibinfo {volume} {781}},\
  \bibinfo {pages} {459} (\bibinfo {year} {2007})},\ \Eprint
  {https://arxiv.org/abs/nucl-ex/0610025} {arXiv:nucl-ex/0610025} \BibitemShut
  {NoStop}%
\bibitem [{\citenamefont {Estee}\ \emph {et~al.}(2021)\citenamefont {Estee}
  \emph {et~al.}}]{SpiRIT:2021gtq}%
  \BibitemOpen
  \bibfield  {author} {\bibinfo {author} {\bibfnamefont {J.}~\bibnamefont
  {Estee}} \emph {et~al.} (\bibinfo {collaboration} {SpiRIT}),\ }\bibfield
  {title} {\bibinfo {title} {{Probing the Symmetry Energy with the Spectral
  Pion Ratio}},\ }\href {https://doi.org/10.1103/PhysRevLett.126.162701}
  {\bibfield  {journal} {\bibinfo  {journal} {Phys. Rev. Lett.}\ }\textbf
  {\bibinfo {volume} {126}},\ \bibinfo {pages} {162701} (\bibinfo {year}
  {2021})},\ \Eprint {https://arxiv.org/abs/2103.06861} {arXiv:2103.06861
  [nucl-ex]} \BibitemShut {NoStop}%
\bibitem [{\citenamefont {Klay}\ \emph {et~al.}(2003)\citenamefont {Klay} \emph
  {et~al.}}]{E-0895:2003oas}%
  \BibitemOpen
  \bibfield  {author} {\bibinfo {author} {\bibfnamefont {J.~L.}\ \bibnamefont
  {Klay}} \emph {et~al.} (\bibinfo {collaboration} {E-0895}),\ }\bibfield
  {title} {\bibinfo {title} {{Charged pion production in 2 to 8 agev central
  au+au collisions}},\ }\href {https://doi.org/10.1103/PhysRevC.68.054905}
  {\bibfield  {journal} {\bibinfo  {journal} {Phys. Rev. C}\ }\textbf {\bibinfo
  {volume} {68}},\ \bibinfo {pages} {054905} (\bibinfo {year} {2003})},\
  \Eprint {https://arxiv.org/abs/nucl-ex/0306033} {arXiv:nucl-ex/0306033}
  \BibitemShut {NoStop}%
\bibitem [{\citenamefont {Li}\ \emph {et~al.}(2008)\citenamefont {Li},
  \citenamefont {Chen},\ and\ \citenamefont {Ko}}]{Li:2008gp}%
  \BibitemOpen
  \bibfield  {author} {\bibinfo {author} {\bibfnamefont {B.-A.}\ \bibnamefont
  {Li}}, \bibinfo {author} {\bibfnamefont {L.-W.}\ \bibnamefont {Chen}},\ and\
  \bibinfo {author} {\bibfnamefont {C.~M.}\ \bibnamefont {Ko}},\ }\bibfield
  {title} {\bibinfo {title} {{Recent Progress and New Challenges in Isospin
  Physics with Heavy-Ion Reactions}},\ }\href
  {https://doi.org/10.1016/j.physrep.2008.04.005} {\bibfield  {journal}
  {\bibinfo  {journal} {Phys. Rept.}\ }\textbf {\bibinfo {volume} {464}},\
  \bibinfo {pages} {113} (\bibinfo {year} {2008})},\ \Eprint
  {https://arxiv.org/abs/0804.3580} {arXiv:0804.3580 [nucl-th]} \BibitemShut
  {NoStop}%
\bibitem [{\citenamefont {Schnedermann}\ \emph {et~al.}(1993)\citenamefont
  {Schnedermann}, \citenamefont {Sollfrank},\ and\ \citenamefont
  {Heinz}}]{Schnedermann:1993ws}%
  \BibitemOpen
  \bibfield  {author} {\bibinfo {author} {\bibfnamefont {E.}~\bibnamefont
  {Schnedermann}}, \bibinfo {author} {\bibfnamefont {J.}~\bibnamefont
  {Sollfrank}},\ and\ \bibinfo {author} {\bibfnamefont {U.~W.}\ \bibnamefont
  {Heinz}},\ }\bibfield  {title} {\bibinfo {title} {{Thermal phenomenology of
  hadrons from 200-A/GeV S+S collisions}},\ }\href
  {https://doi.org/10.1103/PhysRevC.48.2462} {\bibfield  {journal} {\bibinfo
  {journal} {Phys. Rev. C}\ }\textbf {\bibinfo {volume} {48}},\ \bibinfo
  {pages} {2462} (\bibinfo {year} {1993})},\ \Eprint
  {https://arxiv.org/abs/nucl-th/9307020} {arXiv:nucl-th/9307020} \BibitemShut
  {NoStop}%
\bibitem [{\citenamefont {Ollitrault}(1992)}]{Ollitrault:1992bk}%
  \BibitemOpen
  \bibfield  {author} {\bibinfo {author} {\bibfnamefont {J.-Y.}\ \bibnamefont
  {Ollitrault}},\ }\bibfield  {title} {\bibinfo {title} {{Anisotropy as a
  signature of transverse collective flow}},\ }\href
  {https://doi.org/10.1103/PhysRevD.46.229} {\bibfield  {journal} {\bibinfo
  {journal} {Phys. Rev. D}\ }\textbf {\bibinfo {volume} {46}},\ \bibinfo
  {pages} {229} (\bibinfo {year} {1992})}\BibitemShut {NoStop}%
\bibitem [{\citenamefont {Voloshin}\ \emph {et~al.}(2010)\citenamefont
  {Voloshin}, \citenamefont {Poskanzer},\ and\ \citenamefont
  {Snellings}}]{Voloshin:2008dg}%
  \BibitemOpen
  \bibfield  {author} {\bibinfo {author} {\bibfnamefont {S.~A.}\ \bibnamefont
  {Voloshin}}, \bibinfo {author} {\bibfnamefont {A.~M.}\ \bibnamefont
  {Poskanzer}},\ and\ \bibinfo {author} {\bibfnamefont {R.}~\bibnamefont
  {Snellings}},\ }\bibfield  {title} {\bibinfo {title} {{Collective phenomena
  in non-central nuclear collisions}},\ }\href
  {https://doi.org/10.1007/978-3-642-01539-7_10} {\bibfield  {journal}
  {\bibinfo  {journal} {Landolt-Bornstein}\ }\textbf {\bibinfo {volume} {23}},\
  \bibinfo {pages} {293} (\bibinfo {year} {2010})},\ \Eprint
  {https://arxiv.org/abs/0809.2949} {arXiv:0809.2949 [nucl-ex]} \BibitemShut
  {NoStop}%
\bibitem [{\citenamefont {De~Vries}\ \emph {et~al.}(1987)\citenamefont
  {De~Vries}, \citenamefont {De~Jager},\ and\ \citenamefont
  {De~Vries}}]{DeVries:1987atn}%
  \BibitemOpen
  \bibfield  {author} {\bibinfo {author} {\bibfnamefont {H.}~\bibnamefont
  {De~Vries}}, \bibinfo {author} {\bibfnamefont {C.~W.}\ \bibnamefont
  {De~Jager}},\ and\ \bibinfo {author} {\bibfnamefont {C.}~\bibnamefont
  {De~Vries}},\ }\bibfield  {title} {\bibinfo {title} {{Nuclear charge and
  magnetization density distribution parameters from elastic electron
  scattering}},\ }\href {https://doi.org/10.1016/0092-640X(87)90013-1}
  {\bibfield  {journal} {\bibinfo  {journal} {Atom. Data Nucl. Data Tabl.}\
  }\textbf {\bibinfo {volume} {36}},\ \bibinfo {pages} {495} (\bibinfo {year}
  {1987})}\BibitemShut {NoStop}%
\bibitem [{\citenamefont {Hofstadter}(1956)}]{Hofstadter:1956qs}%
  \BibitemOpen
  \bibfield  {author} {\bibinfo {author} {\bibfnamefont {R.}~\bibnamefont
  {Hofstadter}},\ }\bibfield  {title} {\bibinfo {title} {{Electron scattering
  and nuclear structure}},\ }\href {https://doi.org/10.1103/RevModPhys.28.214}
  {\bibfield  {journal} {\bibinfo  {journal} {Rev. Mod. Phys.}\ }\textbf
  {\bibinfo {volume} {28}},\ \bibinfo {pages} {214} (\bibinfo {year}
  {1956})}\BibitemShut {NoStop}%
\bibitem [{\citenamefont {Batyuk}\ \emph {et~al.}(2017)\citenamefont {Batyuk},
  \citenamefont {Karpenko}, \citenamefont {Lednicky}, \citenamefont {Malinina},
  \citenamefont {Mikhaylov}, \citenamefont {Rogachevsky},\ and\ \citenamefont
  {Wielanek}}]{Batyuk:2017smw}%
  \BibitemOpen
  \bibfield  {author} {\bibinfo {author} {\bibfnamefont {P.}~\bibnamefont
  {Batyuk}}, \bibinfo {author} {\bibfnamefont {I.}~\bibnamefont {Karpenko}},
  \bibinfo {author} {\bibfnamefont {R.}~\bibnamefont {Lednicky}}, \bibinfo
  {author} {\bibfnamefont {L.}~\bibnamefont {Malinina}}, \bibinfo {author}
  {\bibfnamefont {K.}~\bibnamefont {Mikhaylov}}, \bibinfo {author}
  {\bibfnamefont {O.}~\bibnamefont {Rogachevsky}},\ and\ \bibinfo {author}
  {\bibfnamefont {D.}~\bibnamefont {Wielanek}},\ }\bibfield  {title} {\bibinfo
  {title} {{Correlation femtoscopy study at energies available at the JINR
  Nuclotron-based Ion Collider fAcility and the BNL Relativistic Heavy Ion
  Collider within a viscous hydrodynamic plus cascade model}},\ }\href
  {https://doi.org/10.1103/PhysRevC.96.024911} {\bibfield  {journal} {\bibinfo
  {journal} {Phys. Rev. C}\ }\textbf {\bibinfo {volume} {96}},\ \bibinfo
  {pages} {024911} (\bibinfo {year} {2017})},\ \Eprint
  {https://arxiv.org/abs/1703.09628} {arXiv:1703.09628 [nucl-th]} \BibitemShut
  {NoStop}%
\bibitem [{\citenamefont {Abdulhamid}\ \emph {et~al.}(2026)\citenamefont
  {Abdulhamid} \emph {et~al.}}]{STAR:2026skb}%
  \BibitemOpen
  \bibfield  {author} {\bibinfo {author} {\bibfnamefont {M.~I.}\ \bibnamefont
  {Abdulhamid}} \emph {et~al.} (\bibinfo {collaboration} {STAR}),\ }\bibfield
  {title} {\bibinfo {title} {{Tilted geometry of the pion emission source in
  Au+Au collisions in the RHIC Beam Energy Scan}},\ }\href@noop {} {\bibfield
  {journal} {\bibinfo  {journal} {arXiv e-prints}\ ,\ \bibinfo {pages}
  {arXiv:2605.15013}} (\bibinfo {year} {2026})},\ \Eprint
  {https://arxiv.org/abs/2605.15013} {arXiv:2605.15013 [nucl-ex]} \BibitemShut
  {NoStop}%
\bibitem [{\citenamefont {Adamczyk}\ \emph
  {et~al.}(2015{\natexlab{c}})\citenamefont {Adamczyk} \emph
  {et~al.}}]{STAR:2014shf}%
  \BibitemOpen
  \bibfield  {author} {\bibinfo {author} {\bibfnamefont {L.}~\bibnamefont
  {Adamczyk}} \emph {et~al.} (\bibinfo {collaboration} {STAR}),\ }\bibfield
  {title} {\bibinfo {title} {{Beam-energy-dependent two-pion interferometry and
  the freeze-out eccentricity of pions measured in heavy ion collisions at the
  STAR detector}},\ }\href {https://doi.org/10.1103/PhysRevC.92.014904}
  {\bibfield  {journal} {\bibinfo  {journal} {Phys. Rev. C}\ }\textbf {\bibinfo
  {volume} {92}},\ \bibinfo {pages} {014904} (\bibinfo {year}
  {2015}{\natexlab{c}})},\ \Eprint {https://arxiv.org/abs/1403.4972}
  {arXiv:1403.4972 [nucl-ex]} \BibitemShut {NoStop}%
\bibitem [{\citenamefont {Miller}\ \emph {et~al.}(2007)\citenamefont {Miller},
  \citenamefont {Reygers}, \citenamefont {Sanders},\ and\ \citenamefont
  {Steinberg}}]{Miller:2007ri}%
  \BibitemOpen
  \bibfield  {author} {\bibinfo {author} {\bibfnamefont {M.~L.}\ \bibnamefont
  {Miller}}, \bibinfo {author} {\bibfnamefont {K.}~\bibnamefont {Reygers}},
  \bibinfo {author} {\bibfnamefont {S.~J.}\ \bibnamefont {Sanders}},\ and\
  \bibinfo {author} {\bibfnamefont {P.}~\bibnamefont {Steinberg}},\ }\bibfield
  {title} {\bibinfo {title} {{Glauber modeling in high energy nuclear
  collisions}},\ }\href {https://doi.org/10.1146/annurev.nucl.57.090506.123020}
  {\bibfield  {journal} {\bibinfo  {journal} {Ann. Rev. Nucl. Part. Sci.}\
  }\textbf {\bibinfo {volume} {57}},\ \bibinfo {pages} {205} (\bibinfo {year}
  {2007})},\ \Eprint {https://arxiv.org/abs/nucl-ex/0701025}
  {arXiv:nucl-ex/0701025} \BibitemShut {NoStop}%
\bibitem [{\citenamefont {Abelev}\ \emph {et~al.}(2013)\citenamefont {Abelev}
  \emph {et~al.}}]{ALICE:2013hur}%
  \BibitemOpen
  \bibfield  {author} {\bibinfo {author} {\bibfnamefont {B.}~\bibnamefont
  {Abelev}} \emph {et~al.} (\bibinfo {collaboration} {ALICE}),\ }\bibfield
  {title} {\bibinfo {title} {{Centrality determination of Pb-Pb collisions at
  $\sqrt{s_{NN}}$ = 2.76 TeV with ALICE}},\ }\href
  {https://doi.org/10.1103/PhysRevC.88.044909} {\bibfield  {journal} {\bibinfo
  {journal} {Phys. Rev. C}\ }\textbf {\bibinfo {volume} {88}},\ \bibinfo
  {pages} {044909} (\bibinfo {year} {2013})},\ \Eprint
  {https://arxiv.org/abs/1301.4361} {arXiv:1301.4361 [nucl-ex]} \BibitemShut
  {NoStop}%
\bibitem [{\citenamefont {Barz}\ \emph {et~al.}(1998)\citenamefont {Barz},
  \citenamefont {Bondorf}, \citenamefont {Gaardhoje},\ and\ \citenamefont
  {Heiselberg}}]{Barz:1997su}%
  \BibitemOpen
  \bibfield  {author} {\bibinfo {author} {\bibfnamefont {H.~W.}\ \bibnamefont
  {Barz}}, \bibinfo {author} {\bibfnamefont {J.~P.}\ \bibnamefont {Bondorf}},
  \bibinfo {author} {\bibfnamefont {J.~J.}\ \bibnamefont {Gaardhoje}},\ and\
  \bibinfo {author} {\bibfnamefont {H.}~\bibnamefont {Heiselberg}},\ }\bibfield
   {title} {\bibinfo {title} {{Coulomb effects on particle spectra in
  relativistic nuclear collisions}},\ }\href
  {https://doi.org/10.1103/PhysRevC.57.2536} {\bibfield  {journal} {\bibinfo
  {journal} {Phys. Rev. C}\ }\textbf {\bibinfo {volume} {57}},\ \bibinfo
  {pages} {2536} (\bibinfo {year} {1998})},\ \Eprint
  {https://arxiv.org/abs/nucl-th/9711064} {arXiv:nucl-th/9711064} \BibitemShut
  {NoStop}%
\bibitem [{\citenamefont {Xu}\ \emph {et~al.}(2024)\citenamefont {Xu} \emph
  {et~al.}}]{TMEP:2023ifw}%
  \BibitemOpen
  \bibfield  {author} {\bibinfo {author} {\bibfnamefont {J.}~\bibnamefont {Xu}}
  \emph {et~al.} (\bibinfo {collaboration} {TMEP}),\ }\bibfield  {title}
  {\bibinfo {title} {{Comparing pion production in transport simulations of
  heavy-ion collisions at 270AMeV under controlled conditions}},\ }\href
  {https://doi.org/10.1103/PhysRevC.109.044609} {\bibfield  {journal} {\bibinfo
   {journal} {Phys. Rev. C}\ }\textbf {\bibinfo {volume} {109}},\ \bibinfo
  {pages} {044609} (\bibinfo {year} {2024})},\ \Eprint
  {https://arxiv.org/abs/2308.05347} {arXiv:2308.05347 [nucl-th]} \BibitemShut
  {NoStop}%
\bibitem [{\citenamefont {Xu}\ \emph {et~al.}(2010)\citenamefont {Xu},
  \citenamefont {Ko},\ and\ \citenamefont {Oh}}]{Xu:2009fj}%
  \BibitemOpen
  \bibfield  {author} {\bibinfo {author} {\bibfnamefont {J.}~\bibnamefont
  {Xu}}, \bibinfo {author} {\bibfnamefont {C.~M.}\ \bibnamefont {Ko}},\ and\
  \bibinfo {author} {\bibfnamefont {Y.}~\bibnamefont {Oh}},\ }\bibfield
  {title} {\bibinfo {title} {{Isospin-dependent pion in-medium effects on
  charged pion ratio in heavy ion collisions}},\ }\href
  {https://doi.org/10.1103/PhysRevC.81.024910} {\bibfield  {journal} {\bibinfo
  {journal} {Phys. Rev. C}\ }\textbf {\bibinfo {volume} {81}},\ \bibinfo
  {pages} {024910} (\bibinfo {year} {2010})},\ \Eprint
  {https://arxiv.org/abs/0906.1602} {arXiv:0906.1602 [nucl-th]} \BibitemShut
  {NoStop}%
\bibitem [{\citenamefont {Csorgo}\ and\ \citenamefont
  {Lorstad}(1996)}]{Csorgo:1995bi}%
  \BibitemOpen
  \bibfield  {author} {\bibinfo {author} {\bibfnamefont {T.}~\bibnamefont
  {Csorgo}}\ and\ \bibinfo {author} {\bibfnamefont {B.}~\bibnamefont
  {Lorstad}},\ }\bibfield  {title} {\bibinfo {title} {{Bose-Einstein
  correlations for three-dimensionally expanding, cylindrically symmetric,
  finite systems}},\ }\href {https://doi.org/10.1103/PhysRevC.54.1390}
  {\bibfield  {journal} {\bibinfo  {journal} {Phys. Rev. C}\ }\textbf {\bibinfo
  {volume} {54}},\ \bibinfo {pages} {1390} (\bibinfo {year} {1996})},\ \Eprint
  {https://arxiv.org/abs/hep-ph/9509213} {arXiv:hep-ph/9509213} \BibitemShut
  {NoStop}%
\bibitem [{\citenamefont {Humanic}(2006)}]{Humanic:2006ve}%
  \BibitemOpen
  \bibfield  {author} {\bibinfo {author} {\bibfnamefont {T.~J.}\ \bibnamefont
  {Humanic}},\ }\bibfield  {title} {\bibinfo {title} {{Effects of non-causal
  artifacts in a hadronic rescattering model for RHIC collisions}},\ }\href
  {https://doi.org/10.1103/PhysRevC.73.054902} {\bibfield  {journal} {\bibinfo
  {journal} {Phys. Rev. C}\ }\textbf {\bibinfo {volume} {73}},\ \bibinfo
  {pages} {054902} (\bibinfo {year} {2006})},\ \Eprint
  {https://arxiv.org/abs/nucl-th/0602027} {arXiv:nucl-th/0602027} \BibitemShut
  {NoStop}%
\bibitem [{\citenamefont {Csorgo}\ \emph {et~al.}(2004)\citenamefont {Csorgo},
  \citenamefont {Hegyi},\ and\ \citenamefont {Zajc}}]{Csorgo:2003uv}%
  \BibitemOpen
  \bibfield  {author} {\bibinfo {author} {\bibfnamefont {T.}~\bibnamefont
  {Csorgo}}, \bibinfo {author} {\bibfnamefont {S.}~\bibnamefont {Hegyi}},\ and\
  \bibinfo {author} {\bibfnamefont {W.~A.}\ \bibnamefont {Zajc}},\ }\bibfield
  {title} {\bibinfo {title} {{Bose-Einstein correlations for Levy stable source
  distributions}},\ }\href {https://doi.org/10.1140/epjc/s2004-01870-9}
  {\bibfield  {journal} {\bibinfo  {journal} {Eur. Phys. J. C}\ }\textbf
  {\bibinfo {volume} {36}},\ \bibinfo {pages} {67} (\bibinfo {year} {2004})},\
  \Eprint {https://arxiv.org/abs/nucl-th/0310042} {arXiv:nucl-th/0310042}
  \BibitemShut {NoStop}%
\bibitem [{\citenamefont {Csanad}\ \emph {et~al.}(2007)\citenamefont {Csanad},
  \citenamefont {Csorgo},\ and\ \citenamefont {Nagy}}]{Csanad:2007fr}%
  \BibitemOpen
  \bibfield  {author} {\bibinfo {author} {\bibfnamefont {M.}~\bibnamefont
  {Csanad}}, \bibinfo {author} {\bibfnamefont {T.}~\bibnamefont {Csorgo}},\
  and\ \bibinfo {author} {\bibfnamefont {M.}~\bibnamefont {Nagy}},\ }\bibfield
  {title} {\bibinfo {title} {{Anomalous diffusion of pions at RHIC}},\ }\href
  {https://doi.org/10.1590/S0103-97332007000600018} {\bibfield  {journal}
  {\bibinfo  {journal} {Braz. J. Phys.}\ }\textbf {\bibinfo {volume} {37}},\
  \bibinfo {pages} {1002} (\bibinfo {year} {2007})},\ \Eprint
  {https://arxiv.org/abs/hep-ph/0702032} {arXiv:hep-ph/0702032} \BibitemShut
  {NoStop}%
\bibitem [{\citenamefont {Kincses}\ \emph {et~al.}(2025)\citenamefont
  {Kincses}, \citenamefont {Nagy},\ and\ \citenamefont
  {Csan{\'a}d}}]{Kincses:2024lnv}%
  \BibitemOpen
  \bibfield  {author} {\bibinfo {author} {\bibfnamefont {D.}~\bibnamefont
  {Kincses}}, \bibinfo {author} {\bibfnamefont {M.}~\bibnamefont {Nagy}},\ and\
  \bibinfo {author} {\bibfnamefont {M.}~\bibnamefont {Csan{\'a}d}},\ }\bibfield
   {title} {\bibinfo {title} {{L{\'e}vy walk of pions in heavy-ion
  collisions}},\ }\href {https://doi.org/10.1038/s42005-025-01973-x} {\bibfield
   {journal} {\bibinfo  {journal} {Commun. Phys.}\ }\textbf {\bibinfo {volume}
  {8}},\ \bibinfo {pages} {55} (\bibinfo {year} {2025})},\ \Eprint
  {https://arxiv.org/abs/2409.10373} {arXiv:2409.10373 [nucl-th]} \BibitemShut
  {NoStop}%
\bibitem [{\citenamefont {Bass}\ \emph {et~al.}(1998)\citenamefont {Bass} \emph
  {et~al.}}]{Bass:1998ca}%
  \BibitemOpen
  \bibfield  {author} {\bibinfo {author} {\bibfnamefont {S.~A.}\ \bibnamefont
  {Bass}} \emph {et~al.},\ }\bibfield  {title} {\bibinfo {title} {{Microscopic
  models for ultrarelativistic heavy ion collisions}},\ }\href
  {https://doi.org/10.1016/S0146-6410(98)00058-1} {\bibfield  {journal}
  {\bibinfo  {journal} {Prog. Part. Nucl. Phys.}\ }\textbf {\bibinfo {volume}
  {41}},\ \bibinfo {pages} {255} (\bibinfo {year} {1998})},\ \Eprint
  {https://arxiv.org/abs/nucl-th/9803035} {arXiv:nucl-th/9803035} \BibitemShut
  {NoStop}%
\bibitem [{\citenamefont {Bleicher}\ \emph {et~al.}(1999)\citenamefont
  {Bleicher} \emph {et~al.}}]{Bleicher:1999xi}%
  \BibitemOpen
  \bibfield  {author} {\bibinfo {author} {\bibfnamefont {M.}~\bibnamefont
  {Bleicher}} \emph {et~al.},\ }\bibfield  {title} {\bibinfo {title}
  {{Relativistic hadron hadron collisions in the ultrarelativistic quantum
  molecular dynamics model}},\ }\href
  {https://doi.org/10.1088/0954-3899/25/9/308} {\bibfield  {journal} {\bibinfo
  {journal} {J. Phys. G}\ }\textbf {\bibinfo {volume} {25}},\ \bibinfo {pages}
  {1859} (\bibinfo {year} {1999})},\ \Eprint
  {https://arxiv.org/abs/hep-ph/9909407} {arXiv:hep-ph/9909407} \BibitemShut
  {NoStop}%
\bibitem [{\citenamefont {Hartnack}\ \emph {et~al.}(2012)\citenamefont
  {Hartnack}, \citenamefont {Oeschler}, \citenamefont {Leifels}, \citenamefont
  {Bratkovskaya},\ and\ \citenamefont {Aichelin}}]{Hartnack:2011cn}%
  \BibitemOpen
  \bibfield  {author} {\bibinfo {author} {\bibfnamefont {C.}~\bibnamefont
  {Hartnack}}, \bibinfo {author} {\bibfnamefont {H.}~\bibnamefont {Oeschler}},
  \bibinfo {author} {\bibfnamefont {Y.}~\bibnamefont {Leifels}}, \bibinfo
  {author} {\bibfnamefont {E.~L.}\ \bibnamefont {Bratkovskaya}},\ and\ \bibinfo
  {author} {\bibfnamefont {J.}~\bibnamefont {Aichelin}},\ }\bibfield  {title}
  {\bibinfo {title} {{Strangeness Production close to Threshold in
  Proton-Nucleus and Heavy-Ion Collisions}},\ }\href
  {https://doi.org/10.1016/j.physrep.2011.08.004} {\bibfield  {journal}
  {\bibinfo  {journal} {Phys. Rept.}\ }\textbf {\bibinfo {volume} {510}},\
  \bibinfo {pages} {119} (\bibinfo {year} {2012})},\ \Eprint
  {https://arxiv.org/abs/1106.2083} {arXiv:1106.2083 [nucl-th]} \BibitemShut
  {NoStop}%
\bibitem [{\citenamefont {Fuchs}(2006)}]{Fuchs:2005zg}%
  \BibitemOpen
  \bibfield  {author} {\bibinfo {author} {\bibfnamefont {C.}~\bibnamefont
  {Fuchs}},\ }\bibfield  {title} {\bibinfo {title} {{Kaon production in heavy
  ion reactions at intermediate energies}},\ }\href
  {https://doi.org/10.1016/j.ppnp.2005.07.004} {\bibfield  {journal} {\bibinfo
  {journal} {Prog. Part. Nucl. Phys.}\ }\textbf {\bibinfo {volume} {56}},\
  \bibinfo {pages} {1} (\bibinfo {year} {2006})},\ \Eprint
  {https://arxiv.org/abs/nucl-th/0507017} {arXiv:nucl-th/0507017} \BibitemShut
  {NoStop}%
\end{thebibliography}%

\end{document}